\documentclass[12pt]{report}
\usepackage{amssymb}

\usepackage{graphicx}

\begin{document}

\title{Dual superconductor models of color confinement }
\author{Georges Ripka \and ECT*, Villa Tambosi, I-38050 Villazano (Trento), ITALY
\and Service de Physique Theorique, \and Centre d'Etudes de Saclay, F-91191
\and Gif-sur-Yvette Cedex, FRANCE \and ripka@cea.fr}
\maketitle
\tableofcontents

\chapter{Introduction}

These lectures were delivered at the ECT*\footnote{%
The ECT* is the European Centre for Theoretical Studies in Nuclear Physics
and Related Areas.} in Trento (Italy) in 2002 and 2003. They are addressed
to physicists who wish to acquire a minimal background to understand present
day attempts to model the confinement of QCD\footnote{%
QCD: quantum chromodynamics.} in
terms of dual superconductors. The lectures focus more on the models than on
attempts to derive them from QCD.

It is speculated that the QCD vacuum can be described in terms of a
Landau-Ginzburg model of a dual superconductor. Particle physicists often
refer to it as the Dual Abelian Higgs model. A dual superconductor is a
superconductor in which the roles of the electric and magnetic fields are
exchanged. Whereas, in usual superconductors, electric charges are condensed
(in the form of Cooper pairs, for example), in a dual superconductor,
magnetic charges are condensed. Whereas no QED\footnote{%
QED: quantum electrodynamics.} magnetic charges have as yet been observed,
the occurrence of color-magnetic charges in QCD,  
and the contention that their condensation
would lead to the confinement of quarks was speculated by various authors in
the early seventies, namely in the pioneering 1973 paper Nielsen and Olesen 
\cite{Nielsen1973}, the 1974 papers of Nambu \cite{Nambu74} and Creutz \cite
{Creutz1974}, the 1975 papers of 't Hooft \cite{tHooft75}, Parisi \cite
{Parisi1975}, Jevicki and Senjanovic \cite{Senjanovic1975}, and the 1976
paper of Mandelstam \cite{Mandelstam76}. Qualitatively, the confinement of
quarks embedded in a dual superconductor can be understood as follows.\ The
quarks carry color charge (see App.\ref{sec:colorcharges}). Consider a
static quark-antiquark $\left( q\bar{q}\right) $ configuration in which the
particles are separated by a distance $R$. The quark and anti-quark have
opposite color-charges so that they create a static color-electric field.\
The field lines stem from the positively charged particle and terminate on
the negatively charged particle. If the $q\bar{q}$ pair were embedded in a
normal (non-superconducting) medium, the color-electric field would be
described by a Coulomb potential and the energy of the system would vary as $%
-e^{2}/R$ where $e$ is the color-electric charge of the quarks. However, if
the $q\bar{q}$ pair is embedded in a dual superconductor, the Meissner
effect will attempt to eliminate the color-electric field.\ (Recall that, in
usual superconductors, the Meissner effect expels the magnetic field.) In
the presence of the color-electric charges of the quarks, Gauss' law
prevents the color-electric field from disappearing completely because the
flux of the electric field must carry the color-electric charge from the
quark (antiquark) to the antiquark (quark). The best the Meissner effect can
do is to compress the color-electric field lines into a minimal space,
thereby creating a thin flux tube which joins the quark and the antiquark in
a straight line. As the distance between the quark and antiquark increases,
the flux tube becomes longer but it maintains its minimal thickness. The
color-electric field runs parallel to the flux tube and maintains a constant
profile in the perpendicular direction. The mere geometry of the flux tube
ensures that the energy increases linearly with $R$ thereby creating a
linearly confining potential between the quark and the antiquark. This
qualitative description hides, in fact, many problems, which relate to
abelian projection (described in Chap.\ref{sec:abelgauge}), Casimir scaling,
etc. As a result, attempts to model quark confinement in terms of dual
superconductors are still speculative and somewhat ill defined.

The dual superconductor is described by the Landau-Ginzburg (Dual Abelian
Higgs) model \cite{Ginzburg50}. Because the roles of the electric and
magnetic fields are exchanged in a dual superconductor, it is natural to
express the lagrangian of the model in terms of a gauge potential $B^{\mu }$
associated to the \emph{dual} field tensor $\bar{F}^{\mu \nu }=\partial
^{\mu }B^{\nu }-\partial ^{\nu }B^{\mu }$. Indeed, when the field tensor $F$
is expressed in terms of the electric and magnetic fields , as in Eq.(\ref
{fmunu}), the corresponding expression (\ref{fmunudual}) of the dual field
tensor $\bar{F}^{\mu \nu }=\varepsilon ^{\mu \nu \alpha \beta }F_{\alpha
\beta }$ is obtained by the exchange $\vec{E}\rightarrow \vec{H}$ and $\vec{H%
}\rightarrow -\vec{E}$ of the electric and magnetic fields. (Recall that
electrodynamics is usually expressed in terms of the gauge potential $A^{\mu
}$ associated to the field tensor $F^{\mu \nu }=\partial ^{\mu }A^{\nu
}-\partial ^{\nu }A^{\mu }$.) When a system is described by the gauge
potential $B^{\mu }$, associated to the dual field tensor $\bar{F}$, the
coupling of electric charges (such as quarks) to the gauge field $B^{\mu }$
is analogous to the problem of coupling of magnetic charges in QED to the
gauge potential $A^{\mu }$. Such a coupling was formulated by Dirac in 1931\
and 1948 \cite{Dirac1948} and it requires the use of a Dirac string. The
Dirac theory of magnetic monopoles is reviewed in Chap.\ref{sec:dualsymmetry}%
. In Sect.\ref{sec:bdynamics}, it is applied to the coupling of electric
charges to the gauge field $B^{\mu }$ associated to the dual field tensor $%
\bar{F}$.

For a system consisting of a $q\bar{q}$ pair, the Dirac string stems from
the quark (or antiquark) and terminates on the antiquark (or quark). The
string should not, however, be confused with the flux tube which joins the
two particles in a straight line. Indeed, as explained in Sect.\ref
{sec:defdirstr}, the Dirac string can be deformed at will by a gauge
transformation.\ The latter does not modify the flux tube, because it is
formed by the electric field and the magnetic current, both of which are
gauge invariant. We refer here to the residual $U\left( 1\right) $ symmetry
which remains after the abelian gauge fixing (or projection), discussed in
Chap.\ref{sec:abelgauge}. However, as explained in Chapt.\ref
{sec:dualsupercond}, there is one gauge, the so-called unitary gauge, in
which the flux tube forms around the Dirac string. Calculations of flux
tubes have all been performed in this gauge.

One attempt \cite{Suzuki89} to apply the dual superconductor model to a
system of three quarks is discussed in Chap.\ref{sec:su3model}. Ultimately,
this is the goal aimed at by these lectures. We would like to formulate a
workable model of baryons and mesons, which would incorporate both
confinement and spontaneous chiral symmetry breaking and which could be
confronted to \emph{bona fide} experimental data and not only to lattice
data. Presently available models of hadrons incorporate either confinement
or chiral symmetry, but not both. It is likely that models, such as the one
described in Chap.\ref{sec:su3model}, will have to be implemented by an
interaction between quarks and a scalar chiral field, for which there is
also lattice evidence \cite{Faccioli2003}.\ 

The model is inspired by (but not derived from) several observations made in
lattice calculations. The first is the so-called abelian dominance, which is
the observation that, in lattice calculations performed in the maximal
abelian gauge, the confining string tension $\sigma $, which defines the
asymptotic confining potential $\sigma R$, can be extracted from the Abelian
link variables alone \cite{Suzuki1990-2,Suzuki1991},\cite{Shiba1994},\cite
{Stack1994},\cite{Bali1996},\cite{Polikarpov2003-3},\cite{Chernodub2003-3}.
Abelian gauge fixing is discussed in Chap.\ref{sec:abelgauge} both in the
continuum and on the lattice. \qquad \qquad

The second observation, made in lattice calculations, is that the confining
phase of the $SU\left( N\right) $ theory is related to the condensation of
monopoles \cite{DiGiacomo2000-1,DiGiacomo2000-2,DiGiacomo2001-3}, \cite
{Cosmai2000,Cosmai2001}.\ Such a statement can only be expressed in terms of
an abelian gauge projection. The condensation of monopoles and confinement
are found to disappear at the same temperature and it does not depend on the
chosen abelian projection \cite{DiGiacomo2003,Cosmai2000}. However,
confinement may well depend on the choice of the abelian gauge.\ In the
abelian Polyakov gauge, for example, monopole condensation is observed but
not confinement \cite{Chernodub2003-1}. In Chap.\ref{sec:abelgauge}, we show
how monopoles can be formed in the process of abelian projection. It is
often difficult to assess the reliability and the relevance of lattice
data.\ For example, on the lattice, even the free $U\left( 1\right) $ gauge
theory displays a confining phase in which magnetic monopoles are condensed 
\cite{Wiese1991-1,Wiese1991-2}.\ This confining phase disappears in the
continuum limit \cite{Guth1980} as it should, since a $U\left( 1\right) $
gauge theory describes a system of free photons. However, non-abelian gauge
theory is better behaved than $U\left( 1\right) $ gauge theory (it is free
of Landau poles) and lattice calculations point to the fact that, in the
non-abelian theory, the confining phase, detected by the area law of a
Wilson loop, survives even in the continuum limit.\ 

The third observation, which favors, although perhaps not exclusively, the
dual superconductor model, is the lattice measurement of the electric field
and the magnetic current, which form the flux tube joining two equal and
opposite static color-charges, in the maximal abelian gauge \cite
{Haymaker1993},\cite{Cosmai1995},
\cite{Bali1998-1,Bali1998-3,Bali1998-4},\cite{Polikarpov1999}%
.\ They are nicely fitted by the flux tube calculated with the
Landau-Ginzburg (Abelian Higgs) model, as discussed in Sect.\ref{sec:balifit}%
.

The model is, however, easily criticized and it has obvious failures.\ For
example, it confines color charges, in particular quarks, which form the
fundamental representation of the $SU\left( N\right) $ group and therefore
carry non-vanishing color-charge. However, it does not confine every color
source in the adjoint representation: for example, it would not confine
abelian gluons.\ (Color charges of quarks and gluons are listed in App.\ref
{sec:colorcharges}.) Because it is expressed in an abelian gauge, the model
also predicts the existence of particles, with masses the order of $1-2\;GeV$%
, which are not color singlets.

In addition, there is lattice evidence for competing scenarios of color
confinement, which involve the use of the maximal center gauge and center
projection, described in Sect.\ref{sec:centerproj}. They are usefully
reviewed in the 1998\ and 2003\ papers of Greensite \cite
{Greensite1998-2,Greensite2003}. They account for the full asymptotic string
tension as well as Casimir scaling. In fact, both the monopole and center
vortex mechanisms of the confinement are supported by the results of lattice
simulations. They are related in the sense that the main part of the
monopole trajectories lie on center projected vortices \cite{Greensite2000}, 
\cite{Polikarpov2003-1}. We do not describe the center-vortex model of
confinement in these lectures because it does not, as yet, lead to a
classical model, such as the Landau-Ginzburg (Abelian Higgs) model.\
Instead, it describes confinement in terms of (quasi) randomly distributed
magnetic fluxes in the vacuum. It is however, numerically simpler on the
lattice and flux tubes formed by both static $q\bar{q}$ and $2q2\bar{q}$
have been computed \cite{Polikarpov2003-2}. Further scenarios, such as the
Gribov coulomb gauge scenario developed by Zwanziger, Cucchieri \cite
{Zwanziger1997,Zwanziger1997-2} and Swanson \cite{Swanson2002}, and the
gluon chain model of Greensite and Thorn \cite{Greensite2002,Greensite2003-2}
are not covered by these lectures.

The relevant mathematical identities are listed in the appendices.

\chapter{The symmetry of electromagnetism with respect to electric and
magnetic charges}

\label{sec:dualsymmetry}

The possible existence of magnetic charges and the corresponding
electromagnetic theory was investigated by Dirac in 1931 and 1948 \cite
{Dirac1931,Dirac1948}. The reading of his 1948 paper is certainly
recommended. A useful introduction to the electromagnetic properties of
magnetic monopoles can be found in Sect.6.12 and 6.13 of Jackson's Classical
Electrodynamics \cite{Jackson1975}. The Dirac theory of magnetic monopoles,
which is briefly sketched in this chapter, will be incorporated into the
Landau-Ginzburg model of a dual superconductor, in order to couple electric
charges, which ultimately become confined. This will be done in Chapt.\ref
{sec:dualsupercond}.

\section{The symmetry between electric and magnetic charges at the level of
the Maxwell equations}

''The field equations of electrodynamics are symmetrical between electric
and magnetic forces. The symmetry between electricity and magnetism is,
however, disturbed by the fact that a single electric charge may occur on a
particle, while a single magnetic pole has not been observed on a particle.
In the present paper a theory will be developed in which a single magnetic
pole can occur on a particle, and the dissymmetry between electricity and
magnetism will consist only in the smallest pole which can occur, being much
greater than the smallest charge.'' This is how Dirac begins his 1948\ paper 
\cite{Dirac1948}.

The electric and magnetic fields $\vec{E}$ and $\vec{H}$ can be expressed as
components of the field tensor $F^{\mu \nu }$: 
\begin{equation}
F^{\mu \nu }=\left( 
\begin{array}{cccc}
0 & -E_{x} & -E_{y} & -E_{z} \\ 
E_{x} & 0 & -H_{z} & H_{y} \\ 
E_{y} & H_{z} & 0 & -H_{x} \\ 
E_{z} & -H_{y} & H_{x} & 0
\end{array}
\right)  \label{fmunu}
\end{equation}
\qquad They may equally well be expressed as the components of the \emph{dual%
} field tensor $\bar{F}^{\mu \nu }$: 
\begin{equation}
\overline{F}^{\mu \nu }=\frac{1}{2}\varepsilon ^{\mu \nu \alpha \beta
}F_{\alpha \beta }=\left( 
\begin{array}{cccc}
0 & -H_{x} & -H_{y} & -H_{z} \\ 
H_{x} & 0 & E_{z} & -E_{y} \\ 
H_{y} & -E_{z} & 0 & E_{x} \\ 
H_{z} & E_{y} & -E_{x} & 0
\end{array}
\right)  \label{fmunudual}
\end{equation}
where $\varepsilon ^{\mu \nu \alpha \beta }$ is the antisymmetric tensor
with $\varepsilon ^{0123}=1$. Thus, the cartesian components of the electric
and magnetic fields can be expressed as components of either the field
tensor $F$ or its dual $\bar{F}$: 
\begin{equation}
E^{i}=-F^{0i}=\frac{1}{2}\varepsilon ^{0ijk}\bar{F}_{jk}\;\;\;\;\;\;\;%
\;H^{i}=-\bar{F}^{0i}=-\frac{1}{2}\varepsilon ^{0ijk}F_{jk}  \label{heff}
\end{equation}
The appendix \ref{ap:apvt} summarizes the properties of vectors, tensors and
their dual forms. In the duality transformation $F\rightarrow \overline{F}$,
the electric and magnetic fields are interchanged as follows: 
\begin{equation}
F\rightarrow \overline{F}\;\;\;\;\;\;\vec{E}\rightarrow \vec{H}\;\;\;\;\;%
\vec{H}\rightarrow -\vec{E}  \label{dualsym}
\end{equation}

The electric charge $\rho $ and the electric current $\vec{j}$ are
components of the 4-vector $j^{\mu }$: 
\begin{equation}
j^{\mu }=\left( \rho ,\vec{j}\right)
\end{equation}
Similarly, the \emph{magnetic} charge $\rho _{mag}$ and the magnetic current 
$\vec{j}_{mag}$ are components of the 4-vector $j_{mag}^{\mu }$: 
\begin{equation}
j_{mag}^{\mu }=\left( \rho _{mag},\vec{j}_{mag}\right)
\end{equation}

At the level the Maxwell equations, there is a complete symmetry between
electric and magnetic currents and the coexistence of electric and magnetic
charges does not raise problems. The equations of motion for the electric
and magnetic fields $\vec{E}$ and $\vec{H}$ are the \emph{Maxwell equations}
which may be cast into the symmetric form: 
\begin{equation}
\partial _{\nu }F^{\nu \mu }=j^{\mu }\;\;\;\;\;\;\;\;\;\;\partial _{\nu }%
\overline{F}^{\nu \mu }=j_{mag}^{\mu }  \label{max}
\end{equation}
It is this symmetry which impressed Dirac, who probably found it upsetting
that the usual Maxwell equations are obtained by setting the magnetic
current $j_{mag}^{\mu }$ to zero. The Maxwell equations can also be
expressed in terms of the electric and magnetic fields $\vec{E}$ and $\vec{H}
$.\ Indeed, if we use the definitions (\ref{fmunu}) and (\ref{fmunudual}),
the Maxwell equations (\ref{max}) read:

\[
\partial _\nu F^{\nu \mu }=j^\mu \;\;\rightarrow \;\;\vec{\nabla}\cdot \vec{E%
}=\rho \;\;\;\;\;\;-\partial _t\vec{E}+\vec{\nabla}\times \vec{H}=\vec{j} 
\]
\begin{equation}
\partial _\nu \overline{F}^{\nu \mu }=j_{mag}^\mu \;\;\rightarrow \;\;\vec{%
\nabla}\cdot \vec{H}=\rho _{mag}\;\;\;\;\;\;\;-\partial _t\vec{H}-\vec{\nabla%
}\times \vec{E}=\vec{j}_{mag}  \label{maxeh}
\end{equation}

\section{Electromagnetism expressed in terms of the gauge field $A^{\mu }$
associated to the field tensor $F^{\mu \nu }$}

\label{sec:adynamics}

So far so good. Problems however begin to appear when we attempt to express
the theory in terms of vector potentials, alias gauge potentials. Why should
we?\ In the very words of Dirac \cite{Dirac1948}: ''To get a theory which
can be transferred to quantum mechanics, we need to put the equations of
motion into a form of an action principle, and for this purpose we require
the electromagnetic potentials.''

This is usually done by expressing the field tensor $F^{\mu \nu }$ in terms
of a vector potential $A^{\mu }=\left( \phi ,\vec{A}\right) $: 
\begin{equation}
F^{\mu \nu }=\partial ^{\mu }A^{\nu }-\partial ^{\nu }A^{\mu }\;\;\;\;\;\;\;%
\vec{E}=-\partial _{t}\vec{A}-\vec{\nabla}\phi \;\;\;\;\;\;\;\vec{H}=\vec{%
\nabla}\times \vec{A}  \label{fa}
\end{equation}
However, this expression leads to the identity\footnote{%
The identity $\partial \cdot \overline{\partial \wedge A}=0$ is often
referred to in the literature as a Bianchi identity.} $\partial _{\nu }%
\overline{F}^{\nu \mu }=0$ which contradicts the second Maxwell equation $%
\partial _{\nu }\overline{F}^{\nu \mu }=j_{mag}^{\mu }$. The expression $%
F=\partial \wedge A$ therefore precludes the existence of magnetic currents
and charges. In electromagnetic theory, this is a bonus which comes for free
since no magnetic charges have ever been observed. Dirac, however, was
apparently more seduced by symmetry than by this experimental observation.

Let us begin to use the compact notation, defined in App.\ref{ap:apvt}, and
in which $\partial \wedge A$ represents the antisymmetric tensor $\left(
\partial \wedge A\right) ^{\mu \nu }=\partial ^{\mu }A^{\nu }-\partial ^{\nu
}A^{\mu }$. The reader is earnestly urged to familiarize himself with this
notation by checking the formulas given in the appendix \ref{ap:apvt}, lest
he become irretrievably entangled in endless and treacherous strings of
indices.

Dirac proposed to modify the expression $F=\partial \wedge A$ by adding a
term $-\bar{G}$: 
\begin{equation}
F=\partial \wedge A-\bar{G}\;\;\;\;\;\;\bar{F}=\overline{\partial \wedge A}+G
\label{ffstring}
\end{equation}
where $G^{\mu \nu }=-G^{\nu \mu }$ is an antisymmetric tensor field\footnote{%
Remember that the dual of $\bar{F}$ is $-F$ !}. The latter satisfies the
equation: 
\begin{equation}
\partial \cdot G=j_{mag}  \label{stringeq}
\end{equation}
The field tensor $F^{\mu \nu }$ then satisfies both Maxwell equations,
namely: $\partial \cdot F=j$ and $\partial \cdot \bar{F}=j_{mag}$. In the
expressions above, the bar above a tensor denotes the dual tensor. For
example, $\bar{G}_{\mu \nu }=\frac{1}{2}\varepsilon _{\mu \nu \alpha \beta
}G^{\alpha \beta }$ (see App.\ref{ap:apvt}). For reasons which will become
apparent in Sect. \ref{sec:statstring}, we shall refer to the antisymmetric
tensor $G_{\mu \nu }$ as a \emph{Dirac string term}.

The string term $G_{\mu \nu }$ is not a dynamical variable. It simply serves
to couple the magnetic current $j_{mag}^\mu $ to the system. It acts as a
source term.\ Note that both $G$ and the equation $\partial \cdot G=j_{mag}$
are independent of the gauge potential $A^\mu $.

An equation for $A^{\mu }$ is provided by the Maxwell equation $\partial
\cdot F=j$. When the field tensor $F$ has the form (\ref{ffstring}), the
equation reads: 
\begin{equation}
\partial \cdot \left( \partial \wedge A\right) -\partial \cdot \bar{G}=j
\label{eqajj}
\end{equation}
The Maxwell equation (\ref{eqajj}) may be obtained from an action principle.
Indeed, since the string term $G$ does not depend on the gauge field $A$,
the variation of the action: 
\begin{equation}
I_{j,j_{mag}}\left( A\right) =\int d^{4}x\left( -\frac{1}{2}F^{2}-j\cdot
A\right) =\int d^{4}x\left( -\frac{1}{2}\left( \left( \partial \wedge
A\right) -\bar{G}\right) ^{2}-j\cdot A\right)  \label{iajj}
\end{equation}
with respect to the gauge field $A^{\mu }$, leads to the equation (\ref
{eqajj}). The action (\ref{iajj})\ is invariant with respect to the gauge
transformation $A\rightarrow A+\left( \partial \alpha \right) $ provided
that $\partial \cdot j=0$.

The source term $G$ has to satisfy two conditions. The first is the equation 
$\partial \cdot G=j_{mag}$. The second is that $\partial \cdot \bar{G}\neq
0. $ If the second condition is not satisfied, the magnetic current
decouples from the system. This is the reason why $G$ cannot simply be
expressed as $G=\partial \wedge B$, in terms of another gauge potential $%
B^{\mu }$.\footnote{%
The Zwanziger formalism, discussed in Sect.\ref{sec:zwanziger}, does in fact
make use of two gauge potentials.} String solutions (see Sect.\ref
{sec:worldsheet}) of the equation $\partial \cdot G=j_{mag}$ are constructed
in order to satisfy the condition $\partial \cdot \bar{G}\neq 0$.

The string term $G^{\mu \nu }$ can be expressed in terms of two vectors,
which we call $\vec{E}_{st}$ and $\vec{H}_{st}$: 
\begin{equation}
\vec{H}_{st}^{i}=-G^{0i}=\frac{1}{2}\varepsilon ^{0ijk}\bar{G}%
_{jk}=\;\;\;\;\;\;\vec{E}_{st}^{i}=-\bar{G}^{0i}=-\frac{1}{2}\varepsilon
^{0ijk}G_{jk}  \label{ehstring}
\end{equation}
The equation $\partial \cdot \bar{F}=\partial \cdot G=j_{mag}$ then
translates to: 
\begin{equation}
\vec{\nabla}\cdot \vec{H}^{st}=\rho _{mag}\;\;\;\;\;\;-\partial _{t}\vec{H}%
^{st}+\vec{\nabla}\times \vec{E}^{st}=\vec{j}_{mag}  \label{delhe}
\end{equation}

Let us express the electric and magnetic fields $\vec{E}$ and $\vec{H}$ in
terms of the vector potential and the string term.\ We define: 
\begin{equation}
A^{\mu }=\left( \phi ,\vec{A}\right)
\end{equation}
When the field tensor $F$ has the form (\ref{ffstring}), the electric and
magnetic fields can be obtained from (\ref{heff}), with the result: 
\begin{equation}
\vec{E}=-\partial _{t}\vec{A}-\vec{\nabla}\phi +\vec{E}_{st}\;\;\;\;\;\;\vec{%
H}=\vec{\nabla}\times \vec{A}+\vec{H}_{st}  \label{aehstring}
\end{equation}
and we have: 
\begin{equation}
-\frac{1}{2}F^{2}=-\frac{1}{2}\left( \partial \wedge A-\bar{G}\right) ^{2}=%
\frac{1}{2}\left( -\partial _{t}\vec{A}-\vec{\nabla}\chi -\vec{E}%
_{st}\right) ^{2}-\frac{1}{2}\left( \vec{\nabla}\times \vec{A}+\vec{H}%
_{st}\right) ^{2}
\end{equation}

\begin{itemize}
\item  \textbf{Exercise}: Consider the following expression of the field
tensor $F$ : 
\begin{equation}
F=\partial \wedge A-\overline{\partial \wedge B}
\end{equation}
in terms of two potentials $A^\mu $ and $B^\mu $.\ Show that $F$ will
satisfy the Maxwell equations $\partial \cdot F=j$ and $\partial \cdot \bar{F%
}=j_{mag}$ provided that the two potentials $A$ and $B$ satisfy the
equations: 
\begin{equation}
\partial \cdot \left( \partial \wedge A\right) =j\;\;\;\;\;\;\partial \cdot
\left( \partial \wedge B\right) =j_{mag}
\end{equation}
Check that the variation of the action: 
\begin{equation}
I_{j,j_{mag}}\left( A,B\right) =\int d^4x\left( -\frac 12F^2-j\cdot
A+j_{mag}\cdot B\right)
\end{equation}
with respect to $A$ and $B$ leads to the correct Maxwell equations. What is
wrong with this suggestion? A possible expression of the field tensor in
terms of two potentials is given in a beautiful 1971 paper of Zwanziger \cite
{Zwanziger1971} (see Sect.\ref{sec:zwanziger}).
\end{itemize}

\section{The current and world line\newline
of a charged particle.}

When we describe the trajectory of a point particle in terms of a
time-dependent position $\vec{R}\left( t\right) $, the Lorentz covariance is
not explicit because $t$ and $\vec{R}$ are different components of a Lorentz
4-vector. The function $\vec{R}\left( t\right) $ describes the trajectory in
3-dimensional euclidean space. Lorentz covariance can be made explicit if we
embed the trajectory in a 4-dimensional Minkowski space, where it is
described by a \emph{world line} $Z^{\mu }\left( \tau \right) $, which is a
4-vector parametrized by a scalar parameter $\tau $. The parameter $\tau $
may, but needs not, be chosen to be the proper-time of the particle. This is
how Dirac describes trajectories of magnetic monopoles in his 1948 paper and
much of the subsequent work is cast in this language, which we briefly
sketch below.

\begin{figure}[htbp]
\begin{center}
\includegraphics[width=10cm]{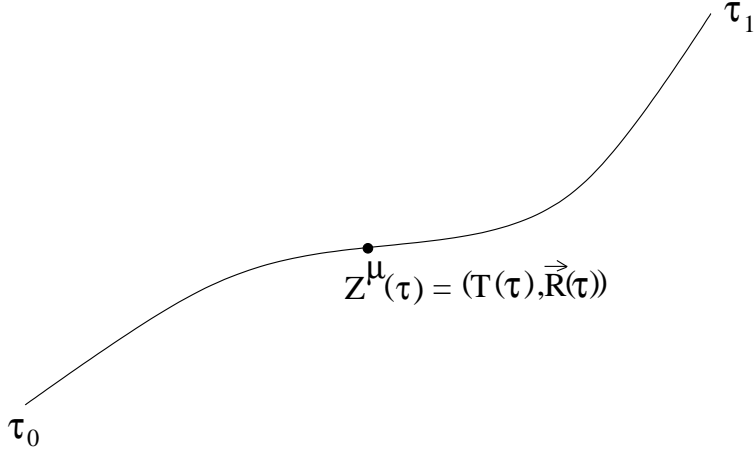}
\end{center}
\caption{The world line of a particle. For any value of $\tau $, the
4-vector $Z^{\mu }\left( \tau \right) $ indicates the position $\vec{R}%
\left( \tau \right) $ of the particle at the time $T\left( \tau \right) $.}
\label{fig:worldline}
\end{figure}

Let $Z^{\mu }\left( \tau \right) $ be the world line of a particle in
Minkowski space. A point $\tau $ on the world line $Z^{\mu }\left( \tau
\right) =\left( T\left( \tau \right) ,\vec{R}\left( \tau \right) \right) $
indicates the position $\vec{R}\left( \tau \right) $ of the particle at the
time $T\left( \tau \right) ,$as illustrated in Fig\ref{fig:worldline}. The
current $j^{\mu }\left( x\right) $ produced by a point particle with a
magnetic charge $g$ can be written in the form of a line integral: 
\begin{equation}
j^{\mu }\left( x\right) =g\int_{L}dZ^{\mu }\delta ^{4}\left( x-Z\right)
\end{equation}
along the world line of the particle. A more explicit form of the current
is: 
\begin{equation}
j^{\mu }\left( x\right) =g\int_{\tau _{0}}^{\tau _{1}}d\tau \frac{dZ^{\mu }}{%
d\tau }\delta ^{4}\left( x-Z\left( \tau \right) \right)  \label{jmu}
\end{equation}
where $\tau _{0}$ and $\tau _{1}$ denote the extremities of the world line,
which can, but need not, extend to infinity.

In order to exhibit the content of the current (\ref{jmu}), we express it in
terms of a density $\rho $ and a current $\vec{j}$: 
\begin{equation}
j^{\mu }=\left( \rho ,\vec{j}\right)
\end{equation}
Let $x^{\mu }=\left( t,\vec{r}\right) $. The current (\ref{jmu}), at the
position $\vec{r}$ and at the time $t$, has the more explicit form: 
\begin{equation}
j^{\mu }\left( t,\vec{r}\right) =g\int_{\tau _{0}}^{\tau _{1}}d\tau \frac{%
dZ^{\mu }}{d\tau }\delta \left( t-T\left( \tau \right) \right) \delta
_{3}\left( \vec{r}-\vec{R}\left( \tau \right) \right)
\end{equation}
The expression (\ref{jmu}) of the current is independent of the
parametrization $Z^{\mu }\left( \tau \right) $ which is chosen to describe
the world line. We can choose $\tau =T$. The density $\rho \left( t,\vec{r}%
\right) $ is then: 
\[
\rho \left( t,\vec{r}\right) =j^{0}\left( t,\vec{r}\right) =g\int_{\tau
_{0}}^{\tau _{1}}d\tau \frac{dT}{d\tau }\delta \left( t-T\left( \tau \right)
\right) \delta \left( \vec{r}-\vec{R}\left( \tau \right) \right) 
\]
\begin{equation}
=g\int_{\tau _{0}}^{\tau _{1}}d\tau \delta \left( t-\tau \right) \delta
\left( \vec{r}-\vec{R}\left( \tau \right) \right) =g\delta \left( \vec{r}-%
\vec{R}\left( t\right) \right)  \label{rhotr}
\end{equation}
and the current $\vec{j}\left( t,\vec{r}\right) $ is: 
\begin{equation}
\vec{j}\left( t,\vec{r}\right) =g\int_{\tau _{0}}^{\tau _{1}}d\tau \frac{d%
\vec{R}}{d\tau }\delta \left( t-\tau \right) \delta \left( \vec{r}-\vec{R}%
\left( \tau \right) \right) =g\frac{d\vec{R}}{dt}\delta \left( \vec{r}-\vec{R%
}\left( t\right) \right)  \label{jtr}
\end{equation}
The expressions (\ref{rhotr}) and (\ref{jtr}) are the familiar expressions
of the density and current produced by a point particle with magnetic charge 
$g$.

\section{The world sheet swept out by a Dirac string in Minkowski space}

\label{sec:worldsheet}

The Dirac string, which is added to the field tensor $F^{\mu \nu }$ in the
expression (\ref{ffstring}), is an antisymmetric tensor $G^{\mu \nu }\left(
x\right) $ which satisfies the equation: 
\begin{equation}
\partial \cdot G=j
\end{equation}
As stated above, not any solution of this equation will do. For example, if
we attempted to express the string term in terms of a potential $B^{\mu }$
by writing, for example, $G=\partial \wedge B$, we would have $\partial
\cdot \bar{G}=0$ and the string term would decouple from the action (\ref
{iajj}). For this reason, string solutions of the equation $\partial \cdot
G=j_{mag}$ have been proposed.

The string solution can be expressed as a surface integral over a \emph{%
world-sheet }$Z^{\mu }\left( \tau ,s\right) $: 
\begin{equation}
G^{\mu \nu }\left( x\right) =g\int d\tau ds\;\frac{\partial \left( Z_{\mu
},Z_{\nu }\right) }{\partial \left( s,\tau \right) }\delta ^{4}\left(
x-Z\right)  \label{gst}
\end{equation}
The world sheet $Z^{\mu }\left( \tau ,s\right) $ is parametrized by two
scalar parameters $\tau $ and $s$ and: 
\begin{equation}
\frac{\partial \left( Z_{\mu },Z_{\nu }\right) }{\partial \left( s,\tau
\right) }=\frac{\partial Z_{\mu }}{\partial s}\frac{\partial Z_{\nu }}{%
\partial \tau }-\frac{\partial Z_{\mu }}{\partial \tau }\frac{\partial
Z_{\nu }}{\partial s}
\end{equation}
is the Jacobian of the parametrization. A point $\left( \tau ,s\right) $ on
the world sheet $Z^{\mu }\left( \tau ,s\right) $ indicates the position $%
\vec{R}\left( \tau ,s\right) $ at the time $T\left( \tau ,s\right) $ of a
particle on the world sheet. The expression (\ref{gst}) for the string $G$
is independent of the parametrization of the world sheet $Z^{\mu }\left(
\tau ,s\right) $ and it can be written in a compact form as a surface
integral over the world sheet $Z$: 
\begin{equation}
G_{\mu \nu }\left( x\right) =g\int_{S}d\sigma _{\mu \nu }\;\delta \left(
x-Z\right)
\end{equation}
The surface element is: 
\begin{equation}
d\sigma _{\mu \nu }=d\tau ds\frac{\partial \left( Z_{\mu },Z_{\nu }\right) }{%
\partial \left( s,\tau \right) }=d\tau ds\left( \frac{\partial Z_{\mu }}{%
\partial s}\frac{\partial Z_{\nu }}{\partial \tau }-\frac{\partial Z_{\mu }}{%
\partial \tau }\frac{\partial Z_{\nu }}{\partial s}\right)
\end{equation}

\begin{figure}[htbp]
\begin{center}
\includegraphics[width=10cm]{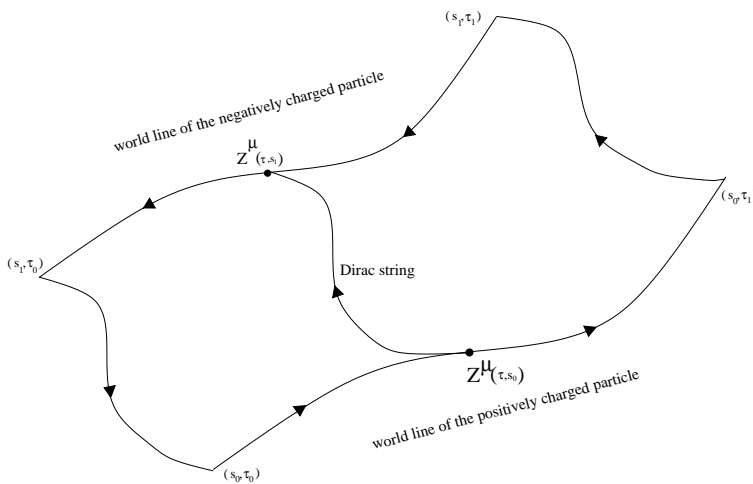}
\end{center}
\caption{The world sheet $Z^{\mu }\left( s,\tau \right) $ swept out by a
Dirac string, which stems from a particle with magnetic charge $g$ and
terminates on a particle with magnetic charge $-g$. The world line of the
positively charged particle is the segment joining the points $\left(
s_{0},\tau _{0}\right) $ and $\left( s_{0},\tau _{1}\right) $. The world
line of the negatively charged particle is the joins the points $\left(
s_{1},\tau _{0}\right) $ and $\left( s_{1},\tau _{1}\right) $. }
\label{fig:worldsheet}
\end{figure}

Figure \ref{fig:worldsheet} is an illustration of the world sheet which is
swept out by a Dirac string which stems from a particle with magnetic charge 
$g$ and terminates on a particle with magnetic charge $-g$. The word \emph{%
line} of the positively charged particle is the border of the world \emph{%
sheet} extending from the point $\left( s_{0},\tau _{0}\right) $ to the
point $\left( s_{0},\tau _{1}\right) $.\ The world line of the negatively
charged particle is the border of the world sheet extending from the point $%
\left( s_{1},\tau _{0}\right) $ to the point $\left( s_{1},\tau _{1}\right) $%
.\ For any value of $\tau $, we can view the string as a line on the world
sheet, which stems from the point $\left( s_{0},\tau \right) $ on the world
line of the positively charged particle to the point $\left( s_{1},\tau
\right) $ of the world line of the negatively charged particle (see Fig.\ref
{fig:worldline}). Often authors choose to work with a world sheet with a
minimal surface.\ This is equivalent to the use of straight line Dirac
strings. An observable, which is independent of the shape of the Dirac
string, is independent of the shape of the surface which defines the world
sheet. If the system is composed of a single magnetic monopole, that is, of
a single particle with magnetic charge $g$, then the attached string extends
to infinity.\ The corresponding world sheet has $s_{1}\rightarrow \infty $
and it becomes an infinite surface.

Let us check that the string form (\ref{gst}) satisfies the equation $%
\partial \cdot G=j$: 
\[
\partial _{\alpha }G^{\alpha \mu }\left( x\right) =g\int_{S}d\tau ds\;\left( 
\frac{\partial Z^{\alpha }}{\partial s}\frac{\partial Z^{\mu }}{\partial
\tau }-\frac{\partial Z^{\alpha }}{\partial \tau }\frac{\partial Z^{\mu }}{%
\partial s}\right) \frac{\partial }{\partial x^{\alpha }}\delta \left(
x-Z\right) 
\]
\begin{equation}
=-g\int_{S}d\tau ds\;\;\left( \frac{\partial Z^{\alpha }}{\partial s}\frac{%
\partial Z^{\mu }}{\partial \tau }-\frac{\partial Z^{\alpha }}{\partial \tau 
}\frac{\partial Z^{\mu }}{\partial s}\right) \frac{\partial }{\partial
Z^{\alpha }}\delta \left( x-Z\right)
\end{equation}
We have: 
\begin{equation}
\frac{\partial }{\partial \tau }\delta \left( x-Z\right) =\frac{\partial
Z^{\alpha }}{\partial \tau }\frac{\partial }{\partial Z^{\alpha }}\delta
\left( x-Z\right)
\end{equation}
and a similar expression holds for $\frac{\partial }{\partial s}\delta
\left( x-Z\right) $. We obtain thus: 
\begin{equation}
\partial _{\alpha }G^{\alpha \mu }\left( x\right) =-g\int_{S}d\tau
ds\;\left( \frac{\partial Z^{\mu }}{\partial \tau }\frac{\partial }{\partial
s}\delta \left( x-Z\right) -\frac{\partial Z^{\mu }}{\partial s}\frac{%
\partial }{\partial \tau }\delta \left( x-Z\right) \right)
\end{equation}
We can use Stoke's theorem which states that, for any two functions $U\left(
\tau ,s\right) $ and $V\left( \tau ,s\right) $, defined on the world sheet,
we have: 
\begin{equation}
\int_{S}\left( \frac{\partial U}{\partial \tau }\frac{\partial V}{\partial s}%
-\frac{\partial U}{\partial s}\frac{\partial V}{\partial \tau }\right)
=\oint_{C}U\left( \frac{\partial V}{\partial s}ds+\frac{\partial V}{\partial
\tau }d\tau \right)
\end{equation}
where the line integral is taken along the closed line $C$ which borders the
surface $S$. A more compact form of Stoke's theorem is: 
\begin{equation}
\int_{S}\frac{\partial \left( U,V\right) }{\partial \left( \tau ,s\right) }%
=\oint_{C}UdV
\end{equation}
We apply the theorem to the functions $U=\delta ^{4}\left( x-Z\right) $ and $%
V=Z^{\mu }$ so as to obtain: 
\begin{equation}
\partial _{\alpha }G^{\alpha \mu }\left( x\right) =g\oint_{C}\delta \left(
x-Z\right) \left( \frac{\partial Z^{\mu }}{\partial \tau }d\tau +\frac{%
\partial Z^{\mu }}{\partial s}ds\right) =g\oint_{C}dZ^{\mu }\delta \left(
x-Z\right)  \label{dgau}
\end{equation}
We can choose, for example, the world sheet to be such that the path $C$
begins at the point $\left( s_{0},\tau _{0}\right) $ and passes successively
through the points $\left( s_{0},\tau _{1}\right) $, $\left( s_{1},\tau
_{1}\right) $, $\left( s_{1},\tau _{0}\right) $ before returning to the
point $\left( s_{0},\tau _{0}\right) $. Then if the world line of the
charged particle begins at $\left( s_{0},\tau _{0}\right) $ and ends at $%
\left( s_{0},\tau _{1}\right) $, the other points being at infinity, the
expression (\ref{dgau}) reduces to: 
\begin{equation}
\partial _{\alpha }G^{\alpha \mu }\left( x\right) =g\int_{\tau _{0}}^{\tau
_{1}}d\tau \frac{dZ^{\mu }}{d\tau }\delta ^{4}\left( x-Z\left( \tau \right)
\right)
\end{equation}
which, in view of (\ref{jmu}), is the current $j^{\mu }\left( x\right) $
produced by the magnetically charged particle.

\section{The Dirac string joining equal and opposite magnetic charges}

\label{sec:statstring}

The string term $G^{\mu \nu }\left( x\right) $ can be expressed in terms of
the two vectors $\vec{H}_{st}$ and $\vec{E}_{st}$ defined in (\ref{ehstring}%
). We can use (\ref{gst}) to obtain an explicit expression for these
vectors. Thus: 
\[
H_{st}^i\left( t,\vec{r}\right) =G^{i0}\left( t,\vec{r}\right) =g\int_Sd\tau
ds\delta \left( t-T\right) \delta _3\left( \vec{r}-\vec{R}\right) \left( 
\frac{\partial \vec{R}_i}{\partial s}\frac{\partial T}{\partial \tau }-\frac{%
\partial \vec{R}_i}{\partial \tau }\frac{\partial T}{\partial s}\right) 
\]
For a given value of $s$, we can choose $\tau =T\left( \tau ,s\right) $ in
which case we have $\frac{\partial T}{\partial \tau }=1$ and $\frac{\delta
\tau }{\delta s}=0$. The string term $\vec{H}_{st}$ reduces to: 
\begin{equation}
\vec{H}_{st}\left( t,\vec{r}\right) =g\int_Lds\frac{\partial \vec{R}}{%
\partial s}\delta \left( \vec{r}-\vec{R}\left( t\right) \right) =g\int_Ld%
\vec{R}\delta \left( \vec{r}-\vec{R}\left( t\right) \right)  \label{hstrl}
\end{equation}
The expression for $\vec{E}_{st}$ is: 
\begin{equation}
E_{st}^i\left( t,\vec{r}\right) =\frac 12\varepsilon
^{0ijk}G_{jk}=\varepsilon ^{0ijk}g\int_Sd\tau ds\;\delta \left( t-T\right)
\delta _3\left( \vec{r}-\vec{R}\right) \left( \frac{\partial \vec{R}_j}{%
\partial s}\frac{\partial \vec{R}_k}{\partial \tau }\right)
\end{equation}
so that: 
\begin{equation}
\vec{E}_{st}\left( t,\vec{r}\right) =g\int_Lds\frac{\partial \vec{R}}{%
\partial s}\times \frac{\partial \vec{R}}{\partial t}\delta _3\left( \vec{r}-%
\vec{R}\left( t,\tau \right) \right) =g\int_Ld\vec{R}\times \frac{\partial 
\vec{R}}{\partial t}\delta \left( \vec{r}-\vec{R}\left( t\right) \right)
\end{equation}

The string terms $\vec{H}_{st}$ and $\vec{E}_{st}$ satisfy the equations (%
\ref{delhe}). Let us calculate : 
\begin{equation}
\vec{\nabla}_r\cdot \vec{H}_{st}\left( t,\vec{r}\right) =g\int_Ld\vec{R}%
\cdot \vec{\nabla}_r\delta \left( \vec{r}-\vec{R}\right) =-g\int_Ld\vec{R}%
\cdot \vec{\nabla}_R\delta \left( \vec{r}-\vec{R}\right)  \label{drr1}
\end{equation}
Now, for any function $f\left( \vec{R}\right) $ we have $d\vec{R}\cdot \vec{%
\nabla}_Rf\left( \vec{R}\right) =f\left( \vec{R}+\delta \vec{R}\right)
-f\left( \vec{R}\right) $. Let $\vec{R}_1\left( t\right) $ and $\vec{R}%
_2\left( t\right) $ be the points where the string $L$ originates and
terminates. We see that the expression (\ref{drr1}) is equal to: 
\begin{equation}
\vec{\nabla}_r\cdot \vec{H}_{st}\left( t,\vec{r}\right) =g\delta \left( \vec{%
r}-\vec{R}_1\left( t\right) \right) -g\delta \left( \vec{r}-\vec{R}_2\left(
t\right) \right)  \label{delhsstrl}
\end{equation}
The right hand side is equal to the magnetic density of a magnetic charge $g$
located at $\vec{R}_1$ and a magnetic charge $-g$ located at $\vec{R}_2$.
For such a system, we can choose a string which stems from the monopole $g$
and terminates at the monopole $-g$.

\section{Dirac strings with a constant orientation}

\label{sec:nstring}

Many calculations are made with the following solution to the equation $%
\partial \cdot G=j_{mag}$, namely: 
\begin{equation}
G=\frac 1{n\cdot \partial }n\wedge j_{mag}\;\;\;\;\;\;\;\;\;G^{\mu \nu
}=\frac 1{n\cdot \partial }\left( n^\mu j_{mag}^\nu -n^\nu j_{mag}^\mu
\right)  \label{straight}
\end{equation}
where $n^\mu $ is a given fixed vector and $n\cdot \partial =n_\mu \partial
^\mu $. We can check that this form also satisfies the equation $\partial
\cdot G=j_{mag}$: 
\begin{equation}
\partial _\alpha G^{\alpha \mu }=\frac 1{n\cdot \partial }\partial _\alpha
\left( n^\alpha j_{mag}^\mu -n^\mu j_{mag}^\alpha \right) =j_{mag}^\mu
\end{equation}
where we assumed that the current $j_{mag}$ is conserved: $\partial _\mu
j_{mag}^\mu =0$. The solution (\ref{straight}) is used in many applications
because it is simple and we shall call it a \emph{straight line string}.

Let us choose $n^\mu $ to be space-like: 
\begin{equation}
n^\mu =\left( 0,\vec{n}\right) \;\;\;\;\;\;n\cdot \partial =\vec{n}\cdot 
\vec{\nabla}
\end{equation}
The string terms $\vec{E}_{st}$ and $\vec{H}_{st}$, defined in (\ref
{ehstring}) are then: 
\begin{equation}
\vec{E}_{st}=-\frac 1{\vec{n}\cdot \vec{\nabla}}\vec{n}\times \vec{j}%
_{mag},\;\;\;\;\;\;\vec{H}_{st}=\frac{\vec{n}}{\vec{n}\cdot \vec{\nabla}}%
\rho _{mag}  \label{ehjmag}
\end{equation}

\begin{itemize}
\item  \textbf{Exercise}: Use (\ref{apvt:dab}) to check explicitly that the
form (\ref{ehjmag}) of the string terms satisfies the form (\ref{delhe}) of
the equation $\partial \cdot G=j_{mag}$.
\end{itemize}

Consider first the case of a single magnetic monopole sitting at the point $%
\vec{R}_1$. The monopole is described by the following magnetic current $%
j_{mag}^\mu $: 
\begin{equation}
j_{mag}^\mu =\left( \rho _{mag},\vec{j}_{mag}\right) \;\;\;\;\;\rho
_{mag}=g\delta \left( \vec{r}-\vec{R}_1\right) \;\;\;\;\;\vec{j}_{mag}=0
\label{magmon}
\end{equation}
The equations (\ref{ehjmag}) show that: 
\begin{equation}
\vec{\nabla}\cdot \vec{H}_{st}=\rho _{mag}=g\delta \left( \vec{r}-\vec{R}%
_1\right) \;\;\;\;\;\;\vec{E}_{st}=0  \label{gij}
\end{equation}
The Fourier transform of $\vec{H}_{st}$ is: 
\begin{equation}
\vec{H}_{st}\left( \vec{k}\right) =g\int d^3r\;e^{i\vec{k}\cdot \vec{r}}\;%
\frac{\vec{n}}{\vec{n}\cdot \vec{\nabla}}\delta \left( \vec{r}-\vec{R}%
_1\right) =ig\frac{\vec{n}}{\vec{n}\cdot \vec{k}}e^{i\vec{k}\cdot \vec{R}_1}
\end{equation}
Let us choose the $z$-axis parallel to $\vec{n}$ so that $\frac{\vec{n}}{%
\vec{n}\cdot \vec{k}}=\vec{e}_{\left( z\right) }\frac 1{k_z}$ where $\vec{e}%
_{\left( z\right) }$ is a unit vector pointing in the $z$ direction. We then
have $\vec{H}_{st}\left( \vec{k}\right) =ig\vec{e}_{\left( z\right) }\frac
1{k_z}e^{i\vec{k}\cdot \vec{R}_1}$. The inverse Fourier transform is: 
\begin{equation}
\vec{H}_{st}\left( \vec{r}\right) =\frac{ig}{\left( 2\pi \right) ^3}\vec{e}%
_{\left( z\right) }\int d^3k\;e^{-i\vec{k}\left( \cdot \vec{r}-\vec{R}%
_1\right) }\frac 1{k_z}=\frac g{\left( 2\pi \right) ^3}\vec{e}_{\left(
z\right) }\int d^3k\;e^{-i\vec{k}\cdot \left( \vec{r}-\vec{R}_1\right)
}\int_0^\infty dz^{\prime }e^{ik_zz^{\prime }}
\end{equation}
Let us define a vector $\vec{R}\left( z^{\prime }\right) =\left(
X_1,Y_1,Z_1+z^{\prime }\right) $. We have: 
\begin{equation}
\vec{k}\cdot \left( \vec{r}-\vec{R}_1\right) -k_zz^{\prime }=\vec{k}\cdot
\left( \vec{r}-\vec{R}\left( z^{\prime }\right) \right) \;\;\;\;\;\;d\vec{R}%
\left( z^{\prime }\right) =\vec{e}_{\left( z\right) }dz^{\prime }
\end{equation}
so that: 
\begin{equation}
\vec{H}_{st}\left( \vec{r}\right) =\vec{e}_{\left( z\right)
}g\;\int_0^\infty dz^{\prime }\;\delta \left( \vec{r}-\vec{R}\left(
z^{\prime }\right) \right) =g\int_Ld\vec{R}\;\delta \left( \vec{r}-\vec{R}%
\right)  \label{hzline}
\end{equation}
where the path $L$ starts at the point $\vec{R}_1$ and runs to infinity
parallel to the positive $z$-axis.\ 

Thus, when the density $\rho \left( \vec{r}\right) $ represents a single
monopole at the point $\vec{R}_{1}$, the straight line solution (\ref{ehjmag}%
)\ is identical to a Dirac string which stems from the monopole and
continues to infinity in a straight line parallel to the vector $\vec{n}$.
If the system consists of two equal and opposite magnetic charges, located
respectively at positions $\vec{R}_{1}$ and $\vec{R}_{2}$, then the straight
line solution (\ref{ehjmag}) represents two strings emanating from the
charges and running to infinity parallel to the $z$-axis.\ Straight line
strings, such as (\ref{ehjmag}) with a fixed vector $n_{\mu }$, are used in
the Zwanziger formalism discussed in Sect.\ref{sec:zwanziger}.

\section{The vector potential $\vec{A}$ in the vicinity of a magnetic
monopole}

\label{sec:vecfielda}

Let us calculate the vector potential in the presence of the magnetic
monopole. Since $j=0$, the equation (\ref{eqajj}) for the vector potential $%
A^\mu $ is: 
\begin{equation}
\partial \cdot \left( \partial \wedge A\right) -\partial \cdot \bar{G}=0
\label{astreq}
\end{equation}
Let us write: 
\begin{equation}
A^\mu =\left( \phi ,\vec{A}\right)
\end{equation}
The equation (\ref{astreq}) can then be broken down to: 
\[
\vec{\nabla}\cdot \left( -\vec{\nabla}\phi -\partial _t\vec{A}\right) =0 
\]
\begin{equation}
\partial _t\left( \partial _t\vec{A}+\vec{\nabla}\phi \right) +\vec{\nabla}%
\times \left( \vec{\nabla}\times \vec{A}\right) =\vec{\nabla}\times \vec{H}%
_{st}
\end{equation}
For a static monopole, it is natural to seek a static (time-independent)
solution. We can choose $\phi =0$. The equation for $\vec{A}$ reduces to: 
\begin{equation}
\vec{\nabla}\times \left( \vec{\nabla}\times \vec{A}\right) =\vec{\nabla}%
\times \vec{H}_{st}  \label{ddapath}
\end{equation}
Let us distinguish the longitudinal and transverse parts of the vector
potential, respectively $\vec{A}_L$ and $\vec{A}_T$: 
\begin{equation}
\vec{A}_L=\frac 1{\nabla ^2}\vec{\nabla}\left( \vec{\nabla}\cdot \vec{A}%
\right) \;\;\;\;\;\;\;\vec{A}_T=\vec{A}-\frac 1{\nabla ^2}\vec{\nabla}\left( 
\vec{\nabla}\cdot \vec{A}\right) \;\;\;\;\;\;\vec{\nabla}\times \vec{A}%
_L=0\;\;\;\;\;\;\vec{\nabla}\cdot \vec{A}_T=0  \label{alt}
\end{equation}
The equation (\ref{ddapath}) determines only the transverse part $\vec{A}_T$
of the vector potential $\vec{A}$ because $\vec{\nabla}\times \vec{A}=\vec{%
\nabla}\times \vec{A}_T$. It leaves the longitudinal part undetermined.
Since the transverse part has $\vec{\nabla}\cdot \vec{A}_T=0$, we have $\vec{%
\nabla}\times \left( \vec{\nabla}\times \vec{A}_T\right) =-\nabla ^2\vec{A}%
_T $. Substituting for $\vec{H}_{st}$ the string (\ref{hzline}), the
expression for $\vec{A}_T$ becomes: 
\begin{equation}
-\nabla ^2\vec{A}_T=g\vec{\nabla}\times \int_Ld\vec{R}\;\delta \left( \vec{r}%
-\vec{R}\right)
\end{equation}
At this point, a useful trick consists in using the identity (\ref{coulid})
to rewrite the $\delta $-function. We obtain thus: 
\begin{equation}
\nabla ^2\vec{A}_T=\frac g{4\pi }\vec{\nabla}\times \int_Ld\vec{R}\;\nabla
^2\frac 1{\left| \vec{r}-\vec{R}\right| }
\end{equation}
so that we can take: 
\begin{equation}
\vec{A}_T\left( \vec{r}\right) =\frac g{4\pi }\vec{\nabla}\times \int_Ld\vec{%
R}\;\frac 1{\left| \vec{r}-\vec{R}\right| }  \label{apath}
\end{equation}
In a gauge such that $\vec{\nabla}\cdot \vec{A}=0$, the expression above
becomes an expression for $\vec{A}$, but no matter. The expression (\ref
{apath}) gives the vector potential $\vec{A}_T$ for a Dirac string defined
by the path $L$.

An analytic expression for $\vec{A}_T$ may be obtained when the path $L$ is
a straight line, as, for example the straight line string (\ref{hzline})
which runs along the positive $z$-axis. In this case the expression (\ref
{apath}) reads: 
\begin{equation}
\vec{A}_T\left( \vec{r}\right) =\frac g{4\pi }\vec{\nabla}\times \vec{e}%
_{\left( z\right) }\int_0^\infty dz^{\prime }\;\frac 1{\sqrt{\rho ^2+\left(
z-z^{\prime }\right) ^2}}
\end{equation}
In cylindrical coordinates (Appendix \ref{apvt:cylcoord}), $\vec{A}_T$ can
be expressed in the form: 
\begin{equation}
\vec{A}_T\left( \vec{r}\right) =\vec{e}_{\left( \theta \right) }A\left( \rho
,z\right)  \label{atrz}
\end{equation}
and (\ref{apvt:cyl4}) shows that this form is consistent with $\vec{\nabla}%
\cdot \vec{A}_T=0$. Using again (\ref{apvt:cyl5}) we obtain: 
\begin{equation}
\vec{A}_T\left( \vec{r}\right) =-\frac g{4\pi }\vec{e}_{\left( \theta
\right) }\int_0^\infty dz^{\prime }\;\frac \partial {\partial \rho }\frac 1{%
\sqrt{\rho ^2+\left( z-z^{\prime }\right) ^2}}
\end{equation}
After performing the derivative with respect to $\rho $, the integral over $%
z^{\prime }$ becomes analytic and we obtain the vector potential in the
form: 
\begin{equation}
\vec{A}_T\left( \rho ,\theta ,z\right) =\vec{e}_{\left( \theta \right)
}\frac g{4\pi }\frac 1\rho \left( 1+\frac z{\sqrt{\left( \rho ^2+z^2\right) }%
}\right)  \label{apolar}
\end{equation}
In spherical coordinates (Appendix \ref{ap:spherical}), the vector potential
has the form: 
\begin{equation}
\vec{A}_T\left( \rho ,\theta ,z\right) =\vec{e}_{\left( \varphi \right)
}\frac g{4\pi }\frac{1+\cos \theta }{r\sin \theta }  \label{amonop}
\end{equation}
We shall see in Sect. \ref{sec:occmonop} that the abelian gluon field
acquires such a form in the vicinity of points where gauge fixing becomes
undetermined.

\begin{itemize}
\item  \textbf{Exercise}: Use (\ref{apvt:cyl4}) to check that (\ref{apolar})
is transverse: $\vec{\nabla}\cdot \vec{A}_T=0$. Use (\ref{apvt:cyl5}) to
calculate the magnetic field from (\ref{apolar}). Check that, at all points
which are not on the positive $z$-axis (where $\vec{H}_{st}=0$), the
magnetic field is equal to: 
\begin{equation}
\vec{H}=\frac g{4\pi r^2}\;\frac{\vec{r}}r
\end{equation}

\item  \textbf{Exercise}: Calculate the Coulomb potential produced by a
charge situated at the point $\vec{r}=\vec{R}$ and deduce the identity: 
\begin{equation}
\delta \left( \vec{r}-\vec{R}\right) =-\nabla ^2\frac 1{4\pi \left| \vec{r}-%
\vec{R}\right| }  \label{coulid}
\end{equation}
\end{itemize}

\section{The irrelevance of the shape of the Dirac string}

Let us calculate the electric and magnetic fields $\vec{E}$ and $\vec{H}$
generated by the monopole. They are given by the field tensor (\ref{fmunu}).
There are two ways to calculate them. The complicated, although instructive,
way consists in starting from the expression (\ref{ffstring}) of the field
tensor in terms of the vector potential and the string term. The dual string
term $\bar{G}$ is then: 
\begin{equation}
E_{st}^{i}=-\bar{G}^{0i}=0\;\;\;\;\;\;H_{st}^{i}=\varepsilon ^{0ijk}\bar{G}%
_{ij}
\end{equation}
If we use (\ref{fmunu}) , the electric and magnetic fields become: 
\begin{equation}
\vec{E}=0\;\;\;\;\;\;\vec{H}=\vec{\nabla}\times \vec{A}+\vec{H}_{st}
\label{ehdelg}
\end{equation}
Now $\vec{\nabla}\times \vec{A}$ can be calculated from (\ref{apath}): 
\begin{equation}
\vec{\nabla}_{r}\times \vec{A}=-\frac{g}{4\pi }\int_{L}\;\vec{\nabla}%
_{r}\times \left( \vec{\nabla}_{r}\times d\vec{R}\frac{1}{\left| \vec{r}-%
\vec{R}\right| }\right)
\end{equation}
We use $\vec{\nabla}\times \left( \vec{\nabla}\times \vec{a}\right) =\vec{%
\nabla}\left( \vec{\nabla}\cdot \vec{a}\right) -\nabla ^{2}\vec{a}$ to
calculate: 
\[
\vec{\nabla}_{r}\times \left( \vec{\nabla}_{r}\times d\vec{R}\frac{1}{\left| 
\vec{r}-\vec{R}\right| }\right) =\vec{\nabla}_{r}\left( \vec{\nabla}%
_{r}\cdot d\vec{R}\frac{1}{\left| \vec{r}-\vec{R}\right| }\right) -d\vec{R}%
\;\nabla _{r}^{2}\frac{1}{\left| \vec{r}-\vec{R}\right| } 
\]
\begin{equation}
=-\vec{\nabla}\left( \vec{\nabla}_{R}\cdot d\vec{R}\frac{1}{\left| \vec{r}-%
\vec{R}\right| }\right) +4\pi d\vec{R}\;\delta \left( \vec{r}-\vec{R}\right)
\end{equation}
Substituting back into the expression for $\vec{\nabla}\times \vec{A}$, we
obtain: 
\begin{equation}
\vec{\nabla}_{r}\times \vec{A}=-\frac{g}{4\pi }\vec{\nabla}\frac{1}{r}%
-g\int_{L}d\vec{R}\;\delta \left( \vec{r}-\vec{R}\right) =\frac{g}{4\pi r^{2}%
}\vec{e}_{\left( r\right) }-\vec{H}_{st}\;\;\;\;\;\;\left( \vec{e}_{\left(
r\right) }=\frac{\vec{r}}{r}\right)
\end{equation}
where we used (\ref{hzline}). Substituting these results into (\ref{ehdelg})
we find that the electric and magnetic fields are: 
\begin{equation}
\vec{E}=0\;\;\;\;\;\;\;\;\;\;\;\;\vec{H}=\vec{\nabla}\times \vec{A}+\vec{H}%
_{st}=\frac{g}{4\pi r^{2}}\vec{e}_{\left( r\right) }  \label{ehmax}
\end{equation}
\qquad

The fields (\ref{ehmax}) could, of course, also have been obtained by simply
solving the Maxwell equations (\ref{maxeh}) with the magnetic current (\ref
{magmon}), without appealing to the Dirac string. We have calculated them
the hard way in order to show that the string term, which breaks rotational
invariance, does not contribute to the electric and magnetic fields. As a
result, the trajectory of an electrically or magnetically charged particle,
flying by, will not feel be Dirac string. However, we shall see that the
string term can modify the phase of the wavefunction of, say, an electron
flying by, and this effect leads to Dirac's charge quantization.

\begin{itemize}
\item  \textbf{Exercise}: Start from the action (\ref{iajj}) and derive an
expression for the energy of the system.\ Show that is does not depend on
the Dirac string term.

\item  \textbf{Exercise}: Equation (\ref{ddapath}) states that $\vec{\nabla}%
\times \left( \vec{\nabla}\times \vec{A}\right) =-\vec{\nabla}\times \vec{H}%
_{st}$. What would we have missed if we had concluded that $\vec{\nabla}%
\times \vec{A}=-\vec{H}_{st}$ ?
\end{itemize}

\section{Deformations of Dirac strings and charge quantization}

\label{sec:defdirstr}

Although the electric and magnetic fields are independent of the string
term, the vector potential is not. What happens to the vector potential if
we deform the Dirac string? Let us show that a deformation of the Dirac
string is equivalent to a gauge transformation. This is, of course, why the
magnetic field is not affected by the string term.

\begin{figure}[htbp]
\begin{center}
\includegraphics[width=10cm]{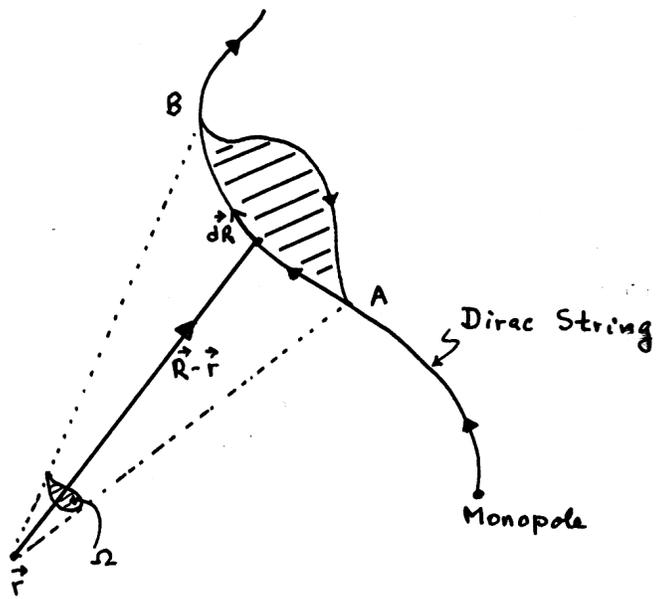}
\end{center}
\caption{The effect on the vector potential of deforming a Dirac string.}
\label{fig:string}
\end{figure}

The vector potential is given by (\ref{apath}). Let us deform only a segment
of the path, situated between two points $A$ and $B$ on the string. The
difference $\delta \vec{A}\left( \vec{r}\right) $ between the vector
potentials, calculated with the two different paths, is the contour
integral: 
\begin{equation}
\delta \vec{A}\left( \vec{r}\right) =-\frac{g}{4\pi }\vec{\nabla}\times
\int_{C}d\vec{R}\;\frac{1}{\left| \vec{r}-\vec{R}\right| }  \label{delac}
\end{equation}
where the contour $C$ follows the initial path from $A$ to $B$ and then
continues back from $B$ to $A$ along the deformed path, as shown on Fig.\ref
{fig:string}. Using the identity (\ref{apvt:stokes2}), we can transform the
contour integral into an integral over a surface $S$ whose boundary is the
path $C$: 
\begin{equation}
\delta \vec{A}\left( \vec{r}\right) =-\frac{g}{4\pi }\vec{\nabla}\times
\int_{S}d\vec{s}\times \vec{\nabla}_{R}\frac{1}{\left| \vec{r}-\vec{R}%
\right| }=-\frac{g}{4\pi }\vec{\nabla}\times \left( \vec{\nabla}\times
\int_{S}d\vec{s}\frac{1}{\left| \vec{r}-\vec{R}\right| }\right)
\end{equation}
Using the identity $\vec{\nabla}\times \left( \vec{\nabla}\times \vec{a}%
\right) =\vec{\nabla}\left( \vec{\nabla}\cdot \vec{a}\right) -\nabla ^{2}%
\vec{a}$, we obtain: 
\[
\delta \vec{A}\left( \vec{r}\right) =-\frac{g}{4\pi }\vec{\nabla}\left(
\int_{S}d\vec{s}\cdot \vec{\nabla}\frac{1}{\left| \vec{r}-\vec{R}\right| }%
\right) +\frac{g}{4\pi }\nabla ^{2}\left( \int_{S}d\vec{s}\frac{1}{\left| 
\vec{r}-\vec{R}\right| }\right) 
\]
\begin{equation}
=\frac{g}{4\pi }\vec{\nabla}\left( \int_{S}d\vec{s}\cdot \vec{\nabla}_{R}%
\frac{1}{\left| \vec{r}-\vec{R}\right| }\right) -g\int_{S}d\vec{s}\;\delta
\left( \vec{r}-\vec{R}\right)
\end{equation}

The second term vanishes at any point $\vec{r}$ not on the surface and can
be dropped.\ The first term is the gradient of the solid angle $\Omega $,
subtended by the surface $S$, when viewed from the point $\vec{r}$: 
\begin{equation}
\delta \vec{A}\left( \vec{r}\right) =\frac{g}{4\pi }\vec{\nabla}\left(
\int_{S}d\vec{s}\cdot \vec{\nabla}_{R}\frac{1}{\left| \vec{r}-\vec{R}\right| 
}\right) =-\frac{g}{4\pi }\vec{\nabla}\int_{S}\;d\vec{s}\cdot \left( \vec{r}-%
\vec{R}\right) \frac{1}{\left| \vec{r}-\vec{R}\right| ^{3}}=-\frac{g}{4\pi }%
\vec{\nabla}\Omega \left( \vec{r}\right)  \label{deltaa}
\end{equation}
To see why, consider first a very small surface $\delta \vec{s}$, such that $%
\left| \vec{r}-\vec{R}\right| $ remains essentially constant.\ Then $\delta 
\vec{s}\cdot \left( \vec{r}-\vec{R}\right) \frac{1}{\left| \vec{r}-\vec{R}%
\right| ^{3}}=\frac{1}{\left| \vec{r}-\vec{R}\right| ^{2}}\left( \delta \vec{%
s}\cdot \frac{\vec{r}-\vec{R}}{\left| \vec{r}-\vec{R}\right| }\right)
=\delta \Omega _{s}$. A finite surface $S$ can be decomposed into small
surfaces bounded by small contours which overlap (and therefore cancel each
other) everywhere except on the boundary of the surface, that is, on the
path $C$.\ The result (\ref{deltaa}) follows.

The expression (\ref{deltaa}) shows that $\delta \vec{A}$ is a gradient. The
deformation of the string therefore adds a gradient to the vector potential $%
\vec{A}$ and this corresponds to a gauge transformation.\ A deformation of
the Dirac string can therefore be compensated by a gauge transformation.

This is, however, only true at points which do not lie on the surface $S$.
Indeed, the solid angle $\Omega \left( \vec{r}\right) $ \emph{is a
discontinuous function} of $\vec{r}$.\ The vector $\vec{r}-\vec{R}$ changes
sign as the point $\vec{r}$ crosses the surface. If the point $\vec{r}$ lies
close to and on one side of the surface $S$ (the shaded area in Fig. \ref
{fig:string}), the solid angle $\Omega \left( \vec{r}\right) $ is equal to $%
2\pi $ (half a sphere).\ As soon as point $\vec{r}$ crosses the surface, the
solid angle switches to $-2\pi $. Thus the solid angle $\Omega \left( \vec{r}%
\right) $ undergoes a discontinuous variation of $4\pi $ as the point
crosses the surface $S$. As a result, the gauge transformation which
compensates a deformation of the Dirac string is a singular gauge
transformation. This point will be further discussed in Sect. \ref
{sec:chargequant}.

Consider the wavefunction $\psi \left( \vec{r}\right) $ of an electron.\ (We
consider an electron because it is a particle with the smallest observed
electric charge.) When the vector potential undergoes a gauge transformation 
$\vec{A}\rightarrow \vec{A}-\frac g{4\pi }\vec{\nabla}\Omega $, the electron
wavefunction undergoes the gauge transformation $\psi \rightarrow e^{ie\frac{%
g\Omega }{4\pi }}\psi $. This means that, on either side of the surface $S$,
the electron wavefunction differs by a phase $e^{ie\frac{g\Omega }{4\pi }%
}=e^{ieg}$. This would make the Dirac string observable, unless we impose
the condition: 
\begin{equation}
eg=2n\pi  \label{chargequant}
\end{equation}
where $n$ is an integer. The expression (\ref{chargequant}) is the charge
quantization condition proposed by Dirac. In his own words, ''the mere
existence of one [magnetic] pole of strength $g$ would require all electric
charges to be quantized in units of $2\pi n/g$ and similarly, the existence
of one [electric] charge would require all [magnetic] poles to quantized.
The quantization of electricity is one of the most fundamental and striking
features of atomic physics, and there seems to be no explanation of it apart
from the theory of poles. This provides some grounds for believing in the
existence of these poles''\footnote{%
In the 1948 paper of Dirac, the quantization condition is stated as $eg=%
\frac{1}{2}n\hbar c$. Excepting the use of units $\hbar =c=1$, the charge $e$
defined by Dirac is equal to $4\pi $ times the charge we use in this paper.
For example, Dirac writes the Maxwell equation $\partial \cdot F=4\pi j$
whereas we write it as $\partial \cdot F=j$. The quotation of Dirac's paper
is modified so as to take this difference into account.} \cite{Dirac1948}.
This was written in 1948. Dirac used a different argument to prove the
quantization rule (\ref{chargequant}). The proof given above is taken from
Chap. 6 of Jackson's Classical Electrodynamics \cite{Jackson1975}. There
have been many other derivations \cite{Zwanziger1971,Goldhaber1965}. For a
late 2002 reflection of Jackiw on the subject, see reference \cite
{Jackiw2002}.

The discontinuity of the solid angle $\Omega \left( \vec{r}\right) $ implies
that, on the surface $S$, we have $\vec{\nabla}\times \left( \vec{\nabla}%
\Omega \right) \neq 0$, so that, for example, $\left( \partial _x\partial
_y-\partial _y\partial _x\right) \Omega \neq 0$. To see this, consider the
line integral $\oint_L\left( \vec{\nabla}\Omega \right) \cdot d\vec{l}$
taken along a path $L$ which crosses the surface $S$ (the shaded area on
Fig. \ref{fig:string}). The integral is, of course, equal to the
discontinuity of $\Omega $ across the surface $S$.\ It is therefore equal to 
$4\pi $. However, in view of Stoke's theorem (\ref{apvt:stokes1}, we have: 
\begin{equation}
\oint_L\left( \vec{\nabla}\Omega \right) \cdot d\vec{l}=\int_{S\left(
L\right) }d\vec{s}\cdot \left( \vec{\nabla}\times \left( \vec{\nabla}\Omega
\right) \right) =4\pi  \label{curldiv}
\end{equation}
where $S\left( L\right) $ is the surface bounded by the path $L$. It follows
that $\vec{\nabla}\times \left( \vec{\nabla}\Omega \right) \neq 0$ on the
surface $S$.

\section{The way Dirac originally argued for the string}

\label{sec:diracstring}

In his 1948\ paper \cite{Dirac1948}, Dirac had an elegant way of conceiving
the string term in order to accommodate magnetic monopoles. He first noted
that the relation $F_{\mu \nu }=\partial _\mu A_\nu -\partial _\nu A_\mu $
implied the absence of magnetic charges. In an attempt to preserve this
relation as far as possible, he argued as follows. If the field tensor had
the form $F_{\mu \nu }=\partial _\mu A_\nu -\partial _\nu A_\mu $, then the
magnetic field would be $\vec{H}=\vec{\nabla}\times \vec{A}$. The flux of
the magnetic field through any closed surface $S$ would then vanish because
of the divergence theorem (\ref{apvt:divergence}): 
\begin{equation}
\int_S\vec{H}.d\vec{s}=\int_S\left( \vec{\nabla}\times \vec{A}\right) .d\vec{%
s}=\int_Vd^3r\;\vec{\nabla}.\left( \vec{\nabla}\times \vec{A}\right) =0
\end{equation}
where $V$ is the volume enclosed by the surface $S$ and $d\vec{s}$ a surface
element directed outward normal to the surface. However, if the surface $S$
encloses a magnetic monopole of charge $g$, then the Maxwell equation $\vec{%
\nabla}.\vec{H}=g\delta \left( \vec{r}\right) $ states that the total
magnetic flux crossing the surface $S$ should equal the magnetic charge $g$
of the monopole: 
\begin{equation}
\int_S\vec{H}.d\vec{s}=\int_Vd^3r\;\vec{\nabla}.\vec{H}=\int_Vd^3r\;g\delta
\left( \vec{r}\right) =g
\end{equation}
Dirac concluded that ''the equation $F_{\mu \nu }=\partial _\mu A_\nu
-\partial _\nu A_\mu $ must then fail somewhere on the surface $S$'' and he
assumed that it fails at only one point on the surface $S$. ''The equation
will then fail at one point on every closed surface surrounding the magnetic
monopole, so that it will fail on a line of points'', which he called a 
\emph{string}.\ ''The string may be any curved line, extending from the pole
to infinity or ending at another monopole of equal and opposite strength.\
Every magnetic monopole must be at the end of such a string.'' Dirac went on
to show that the strings are unphysical variables which do not influence
physical phenomena and that they must not pass through electric charges. He
therefore replaced the expression $\vec{H}=\vec{\nabla}\times \vec{A}$, by
the modified expression: 
\begin{equation}
\vec{H}=\vec{\nabla}\times \vec{A}+\vec{H}_{st}
\end{equation}

\begin{itemize}
\item  \textbf{Exercise}: Try the formulate the theory in terms of two
strings, each one carrying a fraction of the flux of the magnetic field.
Under what conditions can a monopole be attached to two strings?
\end{itemize}

\section{Electromagnetism expressed in terms of the gauge field $B^{\mu }$
associated to the dual field tensor $\bar{F}^{\mu \nu }$}

\label{sec:bdynamics}

The scheme developed in Sect. \ref{sec:adynamics} is useful when a model
action is expressed in terms of the vector potential $A^\mu $ related to the
field tensor $F=\partial \wedge A-\bar{G}$ and when magnetic charges and
currents are present. We shall however be interested in the Landau-Ginzburg
model of a dual superconductor, which is expressed in terms of the potential 
$B^\mu $ associated to the \emph{dual} tensor $\bar{F}^{\mu \nu }$ and in
which (color) electric charges are present. We cannot write the dual field
tensor in the form $\bar{F}=\partial \wedge B$ because that would imply that 
$\partial \cdot F=-\partial \cdot \overline{\partial \wedge B}=0$, which
would preclude the existence of electric charges. The way out, of course, is
to modify the expression of $\bar{F}^{\mu \nu }$ by adding a string term $%
\bar{G}^{\mu \nu }$ and writing the dual field tensor in the form: 
\begin{equation}
\bar{F}=\partial \wedge B+\bar{G}\;\;\;\;\;\;F=-\overline{\partial \wedge B}%
+G  \label{fbg}
\end{equation}
We require the string term $G$ to be independent of $B$ and to satisfy the
equation: 
\begin{equation}
\partial \cdot G=j  \label{dotgj}
\end{equation}

Note that, in the expression (\ref{fbg}), the string term $G$ is added to
the field tensor $F$, whereas, in the Dirac formulation (\ref{ffstring}),
the string term is added to the field tensor $\bar{F}$. The roles of $F$ and 
$\bar{F}$ are indeed interchanged when we express electromagnetism in terms
of the gauge field $B^{\mu }$.

This way, the first Maxwell equation $\partial \cdot F=\partial \cdot G=j$
is satisfied independently of the field $B^\mu $. The latter is determined
by the second Maxwell equation $\partial \cdot \bar{F}=j_{mag}$, namely: 
\begin{equation}
\partial \cdot \left( \partial \wedge B\right) +\partial \cdot \bar{G}%
=j_{mag}
\end{equation}
where $j_{mag}^\mu $ is a magnetic current, which in the dual
Landau-Ginzburg model, is provided by a gauged complex scalar field. The
equation for $B^\mu $ may be obtained from the variation of the action: 
\begin{equation}
I_{j,j_{mag}}\left( B\right) =\int d^4x\left( -\frac 12\bar{F}%
^2-j_{mag}\cdot B\right) =\int d^4x\left( -\frac 12\left( \partial \wedge B+%
\bar{G}\right) ^2-j_{mag}\cdot B\right)  \label{ibjj}
\end{equation}
with respect to the gauge field $B^\mu $. The action (\ref{ibjj}) is
invariant under the gauge transformation $B\rightarrow B+\left( \partial
\beta \right) $ provided that $\partial \cdot j_{mag}=0$.

The source term $G$ has to satisfy two conditions. The first is the equation 
$\partial \cdot G=j$. The second is: $\partial \cdot \bar{G}\neq 0$.
Otherwise, the electric current decouples from the system. String solutions
satisfy the second condition, whereas a form, such as $G=\partial \wedge A$
does not.

We define: 
\begin{equation}
B^\mu =\left( \vec{B},\chi \right) \;\;\;\;\;\;\partial ^\mu =\frac \partial
{\partial x_\mu }=\left( \partial _t,-\vec{\nabla}\right) \;\;\;\;\;\;j^\mu
=\left( \rho ,\vec{j}\right)
\end{equation}
We can express the string term $G^{\mu \nu }$ and its dual $\overline{G}%
^{\mu \nu }$ in terms of two euclidean 3-vectors $\vec{E}_{st}$ and $\vec{H}%
_{st}$ in the same way as the field tensor $F^{\mu \nu }$ is expressed in
terms of the electric and magnetic fields.\ In analogy with (\ref{heff}) we
define: 
\begin{equation}
E_{st}^i=-G^{0i}=\frac 12\varepsilon ^{0ijk}\bar{G}_{jk}\;\;\;\;\;%
\;H_{st}^i=-\bar{G}^{0i}=-\frac 12\varepsilon ^{0ijk}G_{jk}
\label{ehgstring}
\end{equation}
Be careful not to confuse these definitions with the definitions (\ref
{ehstring})!

With a field tensor of the form (\ref{fbg}), the electric and magnetic
fields can be obtained from (\ref{heff}) with the result: 
\begin{equation}
\vec{E}=-\vec{\nabla}\times \vec{B}+\vec{E}_{st}\;\;\;\;\;\;\vec{H}%
=-\partial _t\vec{B}-\vec{\nabla}\chi +\vec{H}_{st}  \label{ehbg}
\end{equation}
The equation $\partial \cdot G=j$ translates to: 
\begin{equation}
\vec{\nabla}\cdot \vec{E}_{st}=\rho \;\;\;\;\;\;-\partial _t\vec{E}_{st}+%
\vec{\nabla}\times \vec{H}_{st}=\vec{j}  \label{gkeq}
\end{equation}
and we have: 
\begin{equation}
-\frac 12\bar{F}^2=-\frac 12\left( \partial \wedge B+\bar{G}\right) ^2=\frac
12\left( -\partial _t\vec{B}-\vec{\nabla}\chi +\vec{H}_{st}\right) ^2-\frac
12\left( -\vec{\nabla}\times \vec{B}+\vec{E}_{st}\right) ^2
\end{equation}

The source term $\vec{E}^{st}$ has to satisfy two conditions. The first is
the equation $\vec{\nabla}\cdot \vec{E}_{st}=\rho $. The second is that $%
\vec{\nabla}\times \vec{E}^{st}\neq 0$.

The magnetic charge density and current are given by (\ref{maxeh}): 
\[
\rho _{mag}=\vec{\nabla}\cdot \vec{H}=\vec{\nabla}\cdot \left( -\partial _t%
\vec{B}-\vec{\nabla}\chi +\vec{H}_{st}\right) 
\]
\begin{equation}
\vec{j}_{mag}=-\partial _t\vec{H}-\vec{\nabla}\times \vec{E}=-\partial
_t\left( -\partial _t\vec{B}-\vec{\nabla}\chi +\vec{H}_{st}\right) -\vec{%
\nabla}\times \left( -\vec{\nabla}\times \vec{B}+\vec{E}_{st}\right)
\label{gkmag}
\end{equation}

Consider the case where the system consists of a static point electric
charge $e$ at the position $\vec{R}_1$ and a static electric charge $-e$ at
the position $\vec{R}_2$. The charge density is then: 
\begin{equation}
\rho \left( \vec{r}\right) =e\delta \left( \vec{r}-\vec{R}_1\right) -e\delta
\left( \vec{r}-\vec{R}_2\right)
\end{equation}
and the electric current $\vec{j}$ vanishes. In that case, we can use a
string term of the form: 
\begin{equation}
\vec{E}_{st}\left( \vec{r}\right) =e\int_Ld\vec{R}\,\,\delta \left( \vec{r}-%
\vec{R}\right) \;\;\;\;\;\;\;\;\;\vec{H}_{st}\left( \vec{r}\right) =0
\label{estpath}
\end{equation}
where the line integral follows a path $L$ (which is the string), which
stems from the charge $e$ at the point $\vec{R}_1$ and terminates at the
charge $-e$ at the point $\vec{R}_2$. A point on the path $L$ can be
parametrized by a function $\vec{R}\left( s\right) $ such that: 
\begin{equation}
\vec{R}_1=\vec{R}\left( s_1\right) \;\;\;\;\;\;\;\vec{R}_2=\vec{R}\left(
s_2\right) \;\;\;\;\;\;\;\;d\vec{R}=ds\,\frac{d\vec{R}}{ds}
\end{equation}
in which case the line integral (\ref{estpath}) acquires the more explicit
form: 
\begin{equation}
\vec{E}_{st}\left( \vec{r}\right) =e\int_{s_1}^{s_2}ds\frac{d\vec{R}}{ds}%
\,\,\delta \left( \vec{r}-\vec{R}\left( s\right) \right)
\end{equation}
The argument which follows equation (\ref{drr1}) can be repeated here with
the result: 
\begin{equation}
\vec{\nabla}_r\cdot \vec{E}_{st}=e\int_Ld\vec{R}\cdot \vec{\nabla}%
_r\,\,\delta \left( \vec{r}-\vec{R}\right) =-e\int_Ld\vec{R}\cdot \vec{\nabla%
}_R\,\,\delta \left( \vec{r}-\vec{R}\right) =e\delta \left( \vec{r}-\vec{R}%
_1\right) -e\delta \left( \vec{r}-\vec{R}_2\right)
\end{equation}
If we had a single electric charge $e$ at the point $\vec{R}_1$ the string
would extend out to infinity.

Many calculations are performed with \emph{straight line strings}, discussed
in Sect. \ref{sec:nstring}, and which is the following solution of the
equation $\partial \cdot G=j$: 
\begin{equation}
G=\frac{1}{n\cdot \partial }n\wedge j\;\;\;\;\;\;\;\;\;G^{\mu \nu }=\frac{1}{%
n\cdot \partial }\left( n^{\mu }j^{\nu }-n^{\nu }j^{\mu }\right)  \label{gnj}
\end{equation}
where $n^{\mu }$ is a given fixed vector and $n\cdot \partial =n_{\mu
}\partial ^{\mu }$. This form solves the equation $\partial \cdot G=j$ if $%
\partial _{\mu }j^{\mu }=0$, that is, if the electric current is conserved.
If we choose $n^{\mu }$ to be space-like, the string terms $\vec{E}_{st}$
and $\vec{H}_{st}$ are given by: 
\begin{equation}
n^{\mu }=\left( 0,\vec{n}\right) ,\;\;\;\;\;\;n\cdot \partial =\vec{n}\cdot 
\vec{\nabla},\;\;\;\;\;\;\vec{E}_{st}=\frac{\vec{n}}{\vec{n}\cdot \vec{\nabla%
}}\rho ,\;\;\;\;\;\;\vec{H}_{st}=-\frac{1}{\vec{n}\cdot \vec{\nabla}}\vec{n}%
\times \vec{j}  \label{nmuest}
\end{equation}
Consider the Fourier transform of the source term $\vec{E}_{st}$: 
\begin{equation}
\vec{E}_{st}\left( \vec{k}\right) =\int d^{3}r\;e^{i\vec{k}\cdot \vec{r}}%
\frac{\vec{n}}{\vec{n}\cdot \vec{\nabla}}\rho \left( \vec{r}\right) =-\frac{%
\vec{n}}{i\vec{n}\cdot \vec{k}}\;\rho \left( \vec{k}\right)
\end{equation}
The inverse Fourier transform is: 
\begin{equation}
\vec{E}_{st}\left( \vec{r}\right) =-\frac{1}{\left( 2\pi \right) ^{3}}\int
d^{3}k\;e^{-i\vec{k}\cdot \vec{r}}\frac{\vec{n}}{i\vec{n}\cdot \vec{k}}%
\;\int d^{3}r^{\prime }e^{-i\vec{k}\cdot \vec{r}^{\prime }}\rho \left( \vec{r%
}^{\prime }\right)
\end{equation}
Let us choose the $z$-axis to be parallel to $\vec{n}$, so that $\frac{\vec{n%
}\cdot \vec{k}}{n}=k_{z}$ and $\frac{\vec{n}}{n}=\vec{e}_{\left( z\right) }$
where $\vec{e}_{\left( z\right) }$ is the unit vector pointing in the $z$%
-direction. We obtain: 
\begin{equation}
\vec{E}_{st}\left( \vec{r}\right) =-\vec{e}_{\left( z\right) }\frac{1}{%
\left( 2\pi \right) ^{3}}\int d^{3}k\;e^{-i\vec{k}\cdot \vec{r}}\frac{1}{%
ik_{z}}\;\int d^{3}r^{\prime }e^{-i\vec{k}\cdot \vec{r}^{\prime }}\rho
\left( \vec{r}^{\prime }\right)
\end{equation}
We obtain: 
\begin{equation}
\vec{E}_{st}\left( \vec{r}\right) =\vec{e}_{\left( z\right) }\int
d^{3}r^{\prime }\delta \left( x-x^{\prime }\right) \delta \left( y-y^{\prime
}\right) \theta \left( z-z^{\prime }\right) \rho \left( \vec{r}^{\prime
}\right) =\vec{e}_{\left( z\right) }\int_{0}^{\infty }dz^{\prime }\;\rho
\left( x,y,z-z^{\prime }\right)
\end{equation}
Let 
\begin{equation}
\vec{R}\left( z^{\prime }\right) =\left( 0,0,z^{\prime }\right) \;\;\;\;\;\;d%
\vec{R}=\vec{e}_{\left( z\right) }dz^{\prime }\;\;\;\;\;\;\vec{r}-\vec{R}%
=\left( x,y,z-z^{\prime }\right)
\end{equation}
We obtain: 
\begin{equation}
\vec{E}_{st}\left( \vec{r}\right) =\int_{0}^{\infty }dz^{\prime }\frac{%
dR\left( z^{\prime }\right) }{dz^{\prime }}\;\rho \left( \vec{r}-\vec{R}%
\left( z^{\prime }\right) \right) =\int_{L}d\vec{R}\;\rho \left( \vec{r}-%
\vec{R}\right)  \label{estrho}
\end{equation}
where the path $L$ is a straight line, starting at the origin and running
parallel to the $z$-axis. The expression (\ref{estrho}) provides for a
determination of the operator $\frac{\vec{n}}{\vec{n}\cdot \vec{\nabla}}$
used in (\ref{nmuest}). For example, if the system has a single charge $e$
at the point $\vec{R}_{1}$, the density is $\rho \left( \vec{r}\right)
=e\delta \left( \vec{r}-\vec{R}_{1}\right) $ and the expression yields a
string term $\vec{E}_{st}\left( \vec{r}\right) $ which stems from the point $%
\vec{R}_{1}$ and extends to infinity parallel to the $z$-axis.\ If there is
an additional charge $-e$ located at the point $\vec{R}_{2}$, then the
expression will yield an additional parallel string, stemming from the
charge $-e$ and extending to infinity parallel to the $z$-axis.\ It will 
\emph{not} be a string joining the two charges, unless the two charges
happen to be located along the $z$-axis. Of course, if the strings can be
deformed so as to merge at some point, then the two strings become
equivalent to a single string joining the equal and opposite charges.

\chapter{The Landau-Ginzburg model of a dual superconductor}

\label{sec:dualsupercond}

We shall describe color confinement in the QCD\ ground state in terms of a
dual superconductor, which differs from usual metallic superconductors in
that the roles of the electric and magnetic fields are exchanged. The dual
superconductor will be described in terms of a suitably adapted
Landau-Ginzburg model of superconductivity. The original model was developed
in 1950 by Ginzburg and Landau \cite{Ginzburg50}. Particle physicists refer
to it today as the Dual Abelian Higgs model. The crucial property of the
dual superconductor will be the Meissner effect \cite{Meissner1933}, which
expels the electric field (instead of the magnetic field, as in a usual
superconductors). As a result, the color-electric field which is produced,
for example, by a quark-antiquark pair embedded in the dual superconductor,
acquires the shape of a color flux tube, thereby generating an
asymptotically linear confining potential.

It is easy to formulate a model in which the Meissner effect applies to the
electric field. All we need to do is to formulate the Landau-Ginzburg theory
in terms of a vector potential $B^{\mu }$ associated to the \emph{dual}
field tensor $\bar{F}^{\mu \nu }$, as in Sect.\ref{sec:bdynamics}. For an
early review and a historical background, see the 1975\ paper of Jevicki and
Senjanovic \cite{Senjanovic1975}. The presentation given below owes a lot to
the illuminating account of superconductivity given in Sect.21.6 of vol.2 of
Steven Weinberg's \emph{Quantum Theory of Fields} \cite{Weinberg1996}. We
first study the dual Landau-Ginzburg model with no reference to the color
degrees of freedom. The way the latter are incorporated is discussed in Chap.%
\ref{sec:su3model}.

\section{The Landau-Ginzburg action of a dual superconductor}

\label{sec:landact}

The Landau-Ginzburg (Abelian Higgs) model is expressed in terms of a gauged
complex scalar field $\psi $, which, presumably, represents a magnetic
charge condensate. The model action is: 
\begin{equation}
I_{j}\left( B,\psi ,\psi ^{\ast }\right) =\int d^{4}x\left( -\frac{1}{4}%
\overline{F}_{\mu \nu }\overline{F}^{\mu \nu }+\frac{1}{2}\left( D_{\mu
}\psi \right) \left( D_{\mu }\psi \right) ^{\ast }-\frac{1}{2}b\left( \psi
\psi ^{\ast }-v^{2}\right) ^{2}\right)  \label{landginz}
\end{equation}
where $\psi $ is the complex scalar field and $\bar{F}^{\mu \nu }$ the dual
field tensor. The covariant derivative is $D_{\mu }=\partial _{\mu
}+igB_{\mu }$, where $B^{\mu }$ is a vector potential, and: 
\begin{equation}
\left( D_{\mu }\psi \right) =\left( \partial _{\mu }\psi +igB_{\mu }\psi
\right) \;\;\;\;\;\left( D_{\mu }\psi \right) ^{\ast }=\left( \partial _{\mu
}\psi ^{\ast }-igB_{\mu }\psi ^{\ast }\right)  \label{covu1}
\end{equation}
The dimensionless constant $g$ can be viewed as a \emph{magnetic} charge. As
explained in Sect.\ref{sec:bdynamics}, the presence of \emph{electric}
charges and currents can be taken into account by adding a string term $\bar{%
G}$ to the dual field strength tensor, which is: 
\begin{equation}
\bar{F}^{\mu \nu }=\left( \partial \wedge B\right) ^{\mu \nu }+\bar{G}^{\mu
\nu }\;\;\;\;\;\;F^{\mu \nu }=-\left( \overline{\partial \wedge B}\right)
^{\mu \nu }+G^{\mu \nu }  \label{dotjbis}
\end{equation}
The string term is related to the electric current by the equation: 
\begin{equation}
\partial _{\alpha }G^{\alpha \mu }=j^{\mu }  \label{dgjmu}
\end{equation}
The Landau-Ginzburg action (\ref{landginz}) is invariant under the abelian
gauge transformation: 
\[
\Omega \left( x\right) =e^{-ig\beta \left( x\right) }\;\;\;\;\;\;\;D_{\mu
}\rightarrow \Omega D_{\mu }\Omega ^{\dagger } 
\]
\begin{equation}
B_{\mu }\rightarrow B_{\mu }+\left( \partial _{\mu }\beta \right)
\;\;\;\;\;\;\psi \rightarrow e^{-ig\beta }\psi  \label{gagpsi}
\end{equation}

The last term of the action is a potential which drives the scalar field to
a non-vanishing expectation value $\psi \psi ^{*}=v^2$ in the ground state
of the system. The superconducting phase occurs when $\psi \psi ^{*}=v^2\neq
0$, and it will model the color-confined phase of QCD. The normal phase
occurs when $\psi \psi ^{*}=0$ and it represents the perturbative phase of
QCD.\ The model parameter $v$ may be temperature and density dependent, and
its variation can drive the system to the normal phase. Of course, other
processes may also contribute to the phase transition. Note that, when $\psi
\psi ^{*}\neq 0$, the action (\ref{landsphi}) is not invariant under the
gauge transformation $B\rightarrow B+\left( \partial \beta \right) $ of the
field $B^\mu $ alone because, loosely speaking, the gauge field $B^\mu $
acquires a squared mass $g^2\psi \psi ^{*}$. In the compact notation
described in App.\ref{ap:apvt}, the action reads: 
\begin{equation}
I_j\left( B,\psi ,\psi ^{*}\right) =\int d^4x\left( -\frac 12\left( \partial
\wedge B+\bar{G}\right) ^2+\frac 12\left| \partial \psi +igB\psi \right|
^2-\frac 12b\left( \psi \psi ^{*}-v^2\right) ^2\right)  \label{lgpsipsi}
\end{equation}

The physical content of the model is often more transparent in a polar
representation of the complex field $\psi $: 
\begin{equation}
\psi \left( x\right) =S\left( x\right) e^{ig\varphi \left( x\right)
}\;\;\;\;\;\;\psi ^{*}\left( x\right) =S\left( x\right) e^{-ig\varphi \left(
x\right) }  \label{salpha}
\end{equation}
The Landau-Ginzburg action (\ref{landginz}) can be expressed in terms of the
real fields $S$ and $\varphi $: 
\begin{equation}
I_j\left( B,\varphi ,S\right) =\int d^4x\left( -\frac 12\left( \partial
\wedge B+\bar{G}\right) ^2+\frac{g^2S^2}2\left( B+\partial \varphi \right)
^2+\frac 12\left( \partial S\right) ^2-\frac 12b\left( S^2-v^2\right)
^2\right)  \label{landsphi}
\end{equation}
The action (\ref{landsphi}) is invariant under the gauge transformation: 
\begin{equation}
B\rightarrow B+\left( \partial \beta \right) \;\;\;\;\;\;\varphi \rightarrow
\varphi -\beta \;\;\;\;\;\;S\rightarrow S  \label{gaginland}
\end{equation}
\qquad

In the ground state of the system, $S=v$, and fluctuations of the scalar
field $S$ describe a scalar particle with a mass: 
\begin{equation}
m_{H}=2v\sqrt{b}  \label{mh}
\end{equation}
Particle physicists like to refer to $S$ as a Higgs field and to $m_{H}$ as
a Higgs mass. The field $\varphi $ remains massless and is sometimes
referred to as a Goldstone field. The gauge field develops a mass: 
\begin{equation}
m_{V}=gv  \label{mv}
\end{equation}
\qquad \qquad Properties of superconductors are often described in terms of
a \emph{penetration depth} $\lambda $ and a \emph{correlation length} $\xi $%
, which are equal to the inverse vector and Higgs masses: 
\begin{equation}
\lambda =\frac{1}{m_{V}}\;\;\;\;\;\;\xi =\frac{1}{m_{H}}  \label{lengths}
\end{equation}
In usual metallic superconductors, the penetration length is the distance
within which an externally applied magnetic field disappears inside the
superconductor. In our dual superconductor, the penetration length $\lambda $
will measure the distance within which the electric field and the magnetic
current vanish outside the flux-tube which develops, for example, between a
quark and an antiquark. The correlation length is related to the distance
within which the scalar field acquires its vacuum value $S=v$. It is also a
measure of the energy difference, per unit volume, of the normal and
superconducting phase, usually referred to as the bag constant: 
\begin{equation}
\mathcal{B}=\frac{1}{8}m_{H}^{2}v^{2}=\frac{v^{2}}{8\xi ^{2}}=\frac{m_{V}^{2}%
}{8g^{2}\xi ^{2}}
\end{equation}
In type I superconductors (pure metals except niobium) $\xi >\lambda $ and $%
m_{V}>m_{H}$. In type II superconductors (alloys and niobium) $\lambda >\xi $
and $m_{H}>m_{V}$. In Sect. \ref{sec:balifit} we shall see that the dual
superconductors which model the confinement of color charge have $%
m_{H}\preceq m_{V}$. They are close to the boundary which separates type I
and type II\ superconductors. The London limit (Sect. \ref{sec:london}), in
which it is assumed that $b\rightarrow \infty $ so that $m_{H}\gg m_{V}$, is
an extreme example of a type II superconductor. In type II superconductors,
the only stable vortex lines are those with minimum flux. In type I
superconductors, vortices attract each other whereas they repel each other
in type II superconductors. Useful reviews of these properties can be found
in Chap.21.6 (volume 2) of Weinberg's ''Quantum Theory of Fields'' \cite
{Weinberg1996} and in Chap.4.3 of Vilenkin and Shellard's ''Cosmic Strings
and Other Topological Defects'' \cite{Vilenkin1994}.

\section{The Landau-Ginzburg action in terms of euclidean fields}

Let us write: 
\begin{equation}
B^\mu =\left( \chi ,\vec{B}\right)
\end{equation}
and let us express the antisymmetric source term $G^{\mu \nu }$ in terms of
the two vectors $\vec{E}^{st}$ and $\vec{H}^{st}$ as in (\ref{ehgstring}).
The action (\ref{landginz}) can then be broken down to the form: 
\[
I_j\left( \psi ,\psi ^{*},\vec{B},\chi \right) =\int d^4x 
\]
\[
\left[ \frac 12\left( -\partial _t\vec{B}-\vec{\nabla}\chi +\vec{H}%
_{st}\right) ^2-\frac 12\left( -\vec{\nabla}\times \vec{B}+\vec{E}
_{st}\right) ^2+\frac 12\left( \partial _t\psi +ig\chi \psi \right) \left(
\partial _t\psi ^{*}-ig\chi \psi ^{*}\right) \right. 
\]
\begin{equation}
\left. -\frac 12\left( \vec{\nabla}\psi -ig\vec{B}\psi \right) \left( \vec{%
\nabla}\psi ^{*}+ig\vec{B}\psi ^{*}\right) -\frac 12b\left( \psi \psi
^{*}-v^2\right) ^2\right]  \label{actpsi}
\end{equation}
Since no time derivative acts on the field $\chi $, it acts as the
constraint $\frac{\delta I}{\delta \chi }=0$, namely: 
\begin{equation}
\vec{\nabla}\cdot \left( -\partial _t\vec{B}-\vec{\nabla}\chi +\vec{H}%
_{st}\right) +\frac{ig}2\left( \psi \partial _t\psi ^{*}-\psi ^{*}\partial
_t\psi \right) +g^2\chi \psi \psi ^{*}=0  \label{constrpsi}
\end{equation}
The Eq.(\ref{dgjmu}) which relates the source terms to the electric charge
density and current reads: 
\begin{equation}
\vec{\nabla}\cdot \vec{E}_{st}=\rho \;\;\;\;\;\;-\partial _t\vec{E}_{st}+%
\vec{\nabla}\times \vec{H}_{st}=\vec{j}  \label{gkeqbis}
\end{equation}

\section{The flux tube joining two equal and opposite electric charges}

\label{sec:fluxtube}

Consider a system composed of two static equal and opposite electric charges 
$\pm e$ placed on the $z$-axis at equal distances from the origin and
separated by a distance $R$.\ The charge density is then: 
\begin{equation}
\rho \left( \vec{r}\right) =e\delta \left( \vec{r}-\vec{R}_{1}\right)
-e\delta \left( \vec{r}-\vec{R}_{2}\right) \;\;\;\;\;\;\vec{R}_{1}=\left(
0,0,-\frac{R}{2}\right) \;\;\;\;\;\;\vec{R}_{2}=\left( 0,0,\frac{R}{2}\right)
\label{rest}
\end{equation}
The electric current is then $j^{\mu }=\delta ^{\mu 0}\rho $ with $\vec{j}=0$%
. As shown in Sect.\ref{sec:bdynamics}, the string terms satisfy the
equations: 
\begin{equation}
\vec{\nabla}\cdot \vec{E}^{st}=\rho \;\;\;\;\;\;\;\vec{\nabla}\times \vec{E}%
^{st}\neq 0\;\;\;\;\;\;\vec{H}_{st}\left( \vec{r}\right) =0
\end{equation}
Note the condition $\vec{\nabla}\times \vec{E}^{st}\neq 0$. If this
condition is not satisfied, the electric density decouples from the system,
as can be seen on the expression (\ref{epsi}) of the energy. String
solutions are designed to avoid this.

The string term $\vec{E}^{st}$ has the form (\ref{estpath}): 
\begin{equation}
\vec{E}_{st}\left( \vec{r}\right) =e\int_{\vec{R}_{1}}^{\vec{R}_{2}}d\vec{Z}%
\,\,\delta \left( \vec{r}-\vec{Z}\right)  \label{este}
\end{equation}
where the integral follows a path $L$ (the string), which stems from the
point $\vec{R}_{1}$ and terminates at the point $\vec{R}_{2}$. Following the
steps described in Sect. \ref{sec:statstring}, we can easily check that the
form (\ref{este}) satisfies the equation $\vec{\nabla}\cdot \vec{E}%
_{st}=\rho $ with $\rho $ given by (\ref{rest}).

When the fields are time-independent, the energy density is equal to \emph{%
minus} the action density given by (\ref{actpsi}). The energy of the system
is thus: 
\[
\mathcal{E}_\rho \left( \psi ,\psi ^{*},\vec{B},\chi \right) =\int
d^3r\left[ -\frac 12\left( \vec{\nabla}\chi \right) ^2+\frac 12\left( -\vec{%
\nabla}\times \vec{B}+\vec{E}_{st}\right) ^2-\frac 12g^2\chi ^2\psi \psi
^{*}\right. 
\]
\begin{equation}
\left. +\frac 12\left( \vec{\nabla}\psi -ig\vec{B}\psi \right) \left( \vec{%
\nabla}\psi ^{*}+ig\vec{B}\psi ^{*}\right) +\frac 12b\left( \psi \psi
^{*}-v^2\right) ^2\right]
\end{equation}
The constraint (\ref{constrpsi}) is satisfied with $\chi =0$. The energy
becomes the following sum of positive terms: 
\[
\mathcal{E}_\rho \left( \psi ,\psi ^{*},\vec{B}\right) =\int d^3r\left[
\frac 12\left( -\vec{\nabla}\times \vec{B}+\vec{E}_{st}\right) ^2\right. 
\]
\begin{equation}
\left. +\frac 12\left( \vec{\nabla}\psi -ig\vec{B}\psi \right) \left( \vec{%
\nabla}\psi ^{*}+ig\vec{B}\psi ^{*}\right) +\frac 12b\left( \psi \psi
^{*}-v^2\right) ^2\right]  \label{epsi}
\end{equation}
This expression can also be derived from the classical energy (\ref{hcps})
by making the energy stationary with respect to the conjugate momenta $\vec{H%
}$ and $P$, as becomes time-independent fields.

\subsection{The Ball-Caticha expression of the string term}

\label{sec:ballcati}

The string term (\ref{este}) does not depend on the fields $B,\psi $ and $%
\psi ^{*}$.\ A useful trick, introduced by Ball and Caticha \cite{Ball1988},
and used in all subsequent work, consists in expressing the string term $%
\vec{E}^{st}$ in terms of the electric field $\vec{E}^0$ and the dual vector
potential $\vec{B}^0$, which are produced by the electric charges when they
are embedded \emph{in the normal vacuum} where $\psi =0$. It is a way to
express the longitudinal and transverse parts of the string term $\vec{E}%
^{st}$ in terms of the known fields $\vec{E}^0$ and $\vec{B}^0$.

The electric field $\vec{E}^0$ produced by the electric charges embedded in
the normal vacuum is the well known Coulomb field: 
\begin{equation}
\vec{E}^0\left( \vec{r}\right) =-\frac e{4\pi }\vec{\nabla}\left( \frac
1{\left| \vec{r}-\vec{R}_1\right| }-\frac 1{\left| \vec{r}-\vec{R}_2\right|
}\right) =\frac e{4\pi }\left( \frac{\vec{r}-\vec{R}_1}{\left| \vec{r}-\vec{R%
}_1\right| ^3}-\frac{\vec{r}-\vec{R}_2}{\left| \vec{r}-\vec{R}_2\right| ^3}%
\right)  \label{ezeror}
\end{equation}
This electric field $\vec{E}^0$ can, however, also be expressed in terms of
the dual potential $\vec{B}^0$ and the string term $\vec{E}^{st}$, using the
expression (\ref{ehbg}): 
\begin{equation}
\vec{E}^0=-\vec{\nabla}\times \vec{B}^0+\vec{E}^{st}  \label{estringbo}
\end{equation}
The idea is to use this equation in order to express the string term $\vec{E}%
^{st}$ in terms of $\vec{E}^0$ and $\vec{B}^0$.

We can calculate the dual vector potential $\vec{B}^0$ as in Sect. \ref
{sec:vecfielda}. Since no magnetic current $\vec{j}_{mag}$ occurs in the
normal vacuum, the field $\vec{B}^0$ is given by (\ref{gkmag}): 
\begin{equation}
\vec{\nabla}\times \left( \vec{\nabla}\times \vec{B}^0\right) =\vec{\nabla}%
\times \vec{E}^{st}
\end{equation}
The string term $\vec{E}^{st}$ is given by the line integral (\ref{este}) so
that: 
\begin{equation}
\vec{\nabla}\times \left( \vec{\nabla}\times \vec{B}^0\right) =e\vec{\nabla}%
_r\times \int_Ld\vec{Z}\,\,\delta \left( \vec{r}-\vec{Z}\right)  \label{ddb0}
\end{equation}
This is an equation for the transverse part $\vec{B}_T^0$ which is the only
part we need.\ We have $\vec{\nabla}\times \left( \vec{\nabla}\times \vec{B}%
^0\right) =-\nabla ^2\vec{B}_T^0$ and we can use (\ref{coulid}) to write the
equation above in the form:

\begin{equation}
-\nabla _r^2\vec{B}_T^0=-\frac e{4\pi }\vec{\nabla}_r\times \int_Ld\vec{Z}%
\;\nabla _r^2\frac 1{\left| \vec{r}-\vec{Z}\right| }
\end{equation}
so that we can take: 
\begin{equation}
\vec{B}^0\left( \vec{r}\right) =\frac e{4\pi }\vec{\nabla}_r\times \int_Ld%
\vec{Z}\;\frac 1{\left| \vec{r}-\vec{Z}\right| }  \label{botpath}
\end{equation}

From (\ref{estringbo}) we see that we can write the string term in the form $%
\vec{E}^{st}=\vec{E}^0+\vec{\nabla}\times \vec{B}^0$. We substitute this
expression into the energy (\ref{epsi}), with the result: 
\[
\mathcal{E}_\rho \left( \psi ,\psi ^{*},\vec{B}\right) =\int d^3r\left[
\frac 12\left( -\vec{\nabla}\times \vec{B}+\vec{E}^0+\vec{\nabla}\times \vec{%
B}^0\right) ^2\right. 
\]
\begin{equation}
\left. +\frac 12\left( \vec{\nabla}\psi -ig\vec{B}\psi \right) \left( \vec{%
\nabla}\psi ^{*}+ig\vec{B}\psi ^{*}\right) +\frac 12b\left( \psi \psi
^{*}-v^2\right) ^2\right]
\end{equation}
Since $\vec{E}^0$ is a gradient, the mixed term $\vec{E}^0\cdot \left( \vec{%
\nabla}\times \left( -\vec{B}+\vec{B}^0\right) \right) $ vanishes. The field 
$\vec{E}^0$ contributes a simple Coulomb term to the energy: 
\begin{equation}
\int d^3r\frac 12\;\vec{E}_0^2=-\frac{e^2}{4\pi R}\;+\;\left(
terms\;independent\;of\;R\right)
\end{equation}
In the following, we neglect the (albeit infinite) self-energy terms which
are independent of $R$. The energy can thus be written in the form: 
\[
\mathcal{E}_\rho \left( \psi ,\psi ^{*},\vec{B}\right) =-\frac{e^2}{4\pi R}%
\;+\int d^3r\left[ \frac 12\left( -\vec{\nabla}\times \vec{B}+\vec{\nabla}%
\times \vec{B}^0\right) ^2\right. 
\]
\begin{equation}
\left. +\frac 12\left( \vec{\nabla}\psi -ig\vec{B}\psi \right) \left( \vec{%
\nabla}\psi ^{*}+ig\vec{B}\psi ^{*}\right) +\frac 12b\left( \psi \psi
^{*}-v^2\right) ^2\right]  \label{ebpsi}
\end{equation}
where $\vec{B}^0$ is given by (\ref{botpath}).

\subsection{Deformations of the string and charge quantization}

\label{sec:chargequant}

Consider the effect of deforming the string, that is, the path $L$ which
defines the vector $\vec{B}^{0}\left( \vec{r}\right) $ in the expression (%
\ref{botpath}). Let us deform a segment of the path, situated between two
points $A$ and $B$ on the path. The corresponding change of $\vec{B}^{0}$
is: 
\begin{equation}
\vec{B}^{0}\left( \vec{r}\right) \rightarrow \vec{B}^{0}\left( \vec{r}%
\right) +\frac{e}{4\pi }\vec{\nabla}_{r}\times \int_{C}d\vec{Z}\;\frac{1}{%
\left| \vec{r}-\vec{Z}\right| }  \label{belac}
\end{equation}
where the contour $C$ follows the initial path from $A$ to $B$ and then
continues back from $B$ to $A$ along the modified path, as illustrated on
Fig.\ref{fig:string}. The expression (\ref{belac}) is the same as the
expression (\ref{delac}) and we can therefore repeat the argument given in
Sect.\ref{sec:defdirstr} to show that the deformation of the string $L$ adds
a gradient to $\vec{B}^{0}$: 
\begin{equation}
\vec{B}^{0}\left( \vec{r}\right) \rightarrow \vec{B}^{0}\left( \vec{r}%
\right) +\frac{e}{4\pi }\vec{\nabla}\Omega  \label{delbo}
\end{equation}
where $\Omega $ is the solid angle subtended by the surface $S$, bounded by
the contour $C$, viewed from the point $\vec{r}$. The energy (\ref{ebpsi})
is changed to: 
\[
\mathcal{E}_{\rho }\left( \psi ,\psi ^{\ast },\vec{B}\right) =-\frac{e^{2}}{%
4\pi R}\;+\int d^{3}r\left[ \frac{1}{2}\left( -\vec{\nabla}\times \vec{B}+%
\vec{\nabla}\times \vec{B}^{0}+\frac{e}{4\pi }\vec{\nabla}\times \left( \vec{%
\nabla}\Omega \right) \right) ^{2}\right. 
\]
\begin{equation}
\left. +\frac{1}{2}\left( \vec{\nabla}\psi -ig\vec{B}\psi \right) \left( 
\vec{\nabla}\psi ^{\ast }+ig\vec{B}\psi ^{\ast }\right) +\frac{1}{2}b\left(
\psi \psi ^{\ast }-v^{2}\right) ^{2}\right]  \label{egauged}
\end{equation}
We have purposely not set to zero the term $\vec{\nabla}\times \left( \vec{%
\nabla}\Omega \right) $ because, as shown in Sect.\ref{sec:defdirstr}, the
solid angle $\Omega \left( \vec{r}\right) $ is a discontinuous function of $%
\vec{r}.$ It undergoes a sudden change of $4\pi $ as $\vec{r}$ crosses the
surface $S$ bordered by the path $C$ (the shaded area in Fig. \ref
{fig:string}). The Eq. (\ref{curldiv}) shows that $\vec{\nabla}\times \left( 
\vec{\nabla}\Omega \right) $ is non-vanishing on the surface $S$. We can,
however, compensate for the extra term $\frac{e}{4\pi }\vec{\nabla}\times
\left( \vec{\nabla}\Omega \right) $ in the energy (\ref{egauged}) by
performing the following \emph{singular} gauge transformation: 
\begin{equation}
\vec{B}\rightarrow \vec{B}+\frac{e}{4\pi }\left( \vec{\nabla}\Omega \right)
\;\;\;\;\;\;\psi \rightarrow e^{ige\frac{\Omega }{4\pi }}\psi  \label{singaz}
\end{equation}
which reduces the energy (\ref{egauged}) to its original form (\ref{ebpsi}%
).\ The energy (\ref{ebpsi}) becomes thus independent of the shape of the
string. The gauge transformation (\ref{singaz}) is not well defined because
the transformed field $e^{-ige\frac{\Omega }{4\pi }}\psi $ is a
discontinuous function of $\vec{r}$ thereby making the gradient $\vec{\nabla}%
e^{-ige\frac{\Omega }{4\pi }}\psi $ ill defined. However, we can impose the
condition: 
\begin{equation}
eg=2n\pi  \label{egquant}
\end{equation}
which makes the field $e^{-ige\frac{\Omega }{4\pi }}\psi $ a continuous and
differentiable function of $\vec{r}$. We recover the Dirac quantization
condition (\ref{chargequant}). Thus, deformations of the string can be
compensated by singular gauge transformations.

\subsection{The relation between the Dirac string\newline
and the flux tube in the unitary gauge}

\label{sec:choiceunit}

It is convenient to express the energy (\ref{ebpsi}) in terms of the polar
representation (\ref{salpha}) of the complex scalar field: 
\[
\mathcal{E}_{\rho }\left( \vec{B},\varphi ,S\right) =-\frac{e^{2}}{4\pi R}%
\;+\int d^{3}r\left[ \frac{1}{2}\left( -\vec{\nabla}\times \vec{B}+\vec{%
\nabla}\times \vec{B}^{0}\right) ^{2}\right. 
\]
\begin{equation}
\left. +\frac{g^{2}S^{2}}{2}\left( \vec{B}-\vec{\nabla}\varphi \right) ^{2}+%
\frac{1}{2}\left( \vec{\nabla}S\right) ^{2}+\frac{1}{2}b\left(
S^{2}-v^{2}\right) ^{2}\right]  \label{ebbo}
\end{equation}
As shown in Sect.\ref{sec:chargequant}, a modification of the shape of the
string, which defines $\vec{B}_{0}$ in Eq.(\ref{botpath}), adds the gradient 
$\frac{e}{4\pi }\vec{\nabla}\Omega $ to $\vec{B}_{0}$. The corresponding
modification of the energy can be compensated by the gauge transformation 
\begin{equation}
\vec{B}\rightarrow \vec{B}+\frac{e}{4\pi }\left( \vec{\nabla}\Omega \right)
\;\;\;\;\;\;\varphi \rightarrow \varphi +ige\frac{\Omega }{4\pi }
\end{equation}

Because the energy (\ref{ebbo}) is invariant under the gauge transformation: 
\begin{equation}
\vec{B}\rightarrow \vec{B}-\vec{\nabla}\beta \;\;\;\;\;\;\varphi \rightarrow
\varphi -\beta
\end{equation}
we can choose the gauge $\beta =\varphi $, which is usually referred to as
the unitary gauge. In this gauge, the field $\varphi $ vanishes and the
energy (\ref{ebbo}) is equal to: 
\[
\mathcal{E}_{\rho }\left( \vec{B},S\right) =-\frac{e^{2}}{4\pi R}\;+\int
d^{3}r\left[ \frac{1}{2}\left( -\vec{\nabla}\times \vec{B}+\vec{\nabla}%
\times \vec{B}^{0}\right) ^{2}\right. 
\]
\begin{equation}
\left. +\frac{g^{2}S^{2}}{2}\vec{B}^{2}+\frac{1}{2}\left( \vec{\nabla}%
S\right) ^{2}+\frac{1}{2}b\left( S^{2}-v^{2}\right) ^{2}\right]
\label{egunit}
\end{equation}

The energy (\ref{egunit}), expressed in the unitary gauge, is \emph{not}
independent of the shape of the string which defines the field $\vec{B}^0$,
nor should it be, because modifications of the shape of the string are
compensated by modifications of the phase $\varphi $. When, for example,
flux tubes joining electric charges are calculated by minimizing the energy (%
\ref{egunit}) expressed in the unitary gauge, the flux tubes follow and
develop around the Dirac strings.\ For example, in Sect.\ref{sec:elecmagcur}%
, we shall see that the string term represents the longitudinal part of the
electric field. In the unitary gauge, the shapes of the Dirac strings can be
chosen so as to minimize the energy. They can subsequently be deformed, by
re-introducing the field $\varphi $, as in the expression (\ref{ebbo}) for
example.

\subsection{The flux tube calculated in the unitary gauge}

\label{sec:flux}

When the system consists of two static equal and opposite charges, the
charge density is given by (\ref{rest}) and the Dirac string is a straight
line joining the two charges. The field $\vec{B}^{0}\left( \vec{r}\right) $
is given by the expression (\ref{botpath}) with $d\vec{Z}=\vec{e}_{\left(
z\right) }dz$, where $\vec{e}_{\left( z\right) }$ is a unit vector parallel
to the $z$-axis. The explicit expression can be written in cylindrical
coordinates (App.\ref{apvt:cylcoord}): \ 
\begin{equation}
\vec{B}^{0}\left( \vec{r}\right) =\frac{e}{4\pi }\vec{\nabla}_{r}\times \vec{%
e}_{\left( z\right) }\int_{-\frac{R}{2}}^{\frac{R}{2}}dz^{\prime }\;\frac{1}{%
\sqrt{\rho ^{2}+\left( z-z^{\prime }\right) ^{2}}}
\end{equation}
Use (\ref{apvt:cyl5}) to get: 
\[
\vec{B}^{0}\left( \vec{r}\right) =\frac{e}{4\pi }\vec{e}_{\left( \theta
\right) }\int_{-\frac{R}{2}}^{\frac{R}{2}}dz^{\prime }\frac{\rho }{\left(
\rho ^{2}+\left( z-z^{\prime }\right) ^{2}\right) ^{\frac{3}{2}}} 
\]
\begin{equation}
=-\vec{e}_{\left( \theta \right) }\frac{e}{4\pi }\frac{1}{\rho }\left( \frac{%
z-\frac{R}{2}}{\sqrt{\left( \rho ^{2}+\left( z-\frac{R}{2}\right)
^{2}\right) }}-\frac{z+\frac{R}{2}}{\sqrt{\left( \rho ^{2}+\left( z+\frac{R}{%
2}\right) ^{2}\right) }}\right) \equiv \vec{e}_{\left( \theta \right)
}B^{0}\left( \rho ,z\right)  \label{bot}
\end{equation}
This is an analytic expression for the field $\vec{B}^{0}$ in cylindrical
coordinates. From (\ref{apvt:cyl4}) we can check directly that $\vec{\nabla}.%
\vec{B}^{0}=0$ so that $\vec{B}^{0}$ is transverse. Near the $z$-axis, where 
$\rho $ is small, the field $B^{0}\left( \rho ,z\right) $ becomes singular: 
\[
B^{0}\left( \rho ,z\right) \mathrel{\mathop{\rightarrow }\limits_{\rho
\rightarrow 0}}=-\frac{e}{4\pi }\frac{1}{\rho }\left( \frac{z-\frac{1}{2}R}{%
\left| z-\frac{1}{2}R\right| }-\frac{z+\frac{1}{2}R}{\left| z+\frac{1}{2}%
R\right| }\right) 
\]
\begin{equation}
\; 
\begin{array}{l}
=\frac{e}{2\pi \rho }\;\;\;when\;\;-\frac{1}{2}R<z<\frac{R}{2} \\ 
=0\;\;\;\;when\;\;z<-\frac{R}{2}\;\;\;and\;\;\;z>\frac{R}{2}
\end{array}
\label{sbot}
\end{equation}

The fields which make the energy (\ref{egunit}) stationary satisfy the
equations: 
\[
\vec{\nabla}\times \left( \vec{\nabla}\times \vec{B}\right) -\vec{\nabla}%
\times \left( \vec{\nabla}\times \vec{B}^0\right) +g^2S^2\vec{B}=0 
\]
\begin{equation}
\left[ -\vec{\nabla}^2+2b\left( S^2-v^2\right) +g^2\vec{B}^2\right] S=0
\label{fieldeq}
\end{equation}
When $\vec{B}^0$ has the form (\ref{bot}), a solution exists, in cylindrical
coordinates, in which the fields $S$ and $\vec{B}$ have the form: 
\begin{equation}
S\left( \vec{r}\right) =S\left( \rho ,z\right) \;\;\;\;\;\;\;\;\;\;\vec{B}%
\left( \vec{r}\right) =\vec{e}_{\left( \theta \right) }B\left( \rho ,z\right)
\label{sbphi}
\end{equation}
The field equations (\ref{fieldeq}) reduce to the following set of coupled
equations for the functions $B\left( \rho ,z\right) $ and $S\left( \rho
,z\right) $: 
\[
-\frac{\partial ^2\left( B-B_0\right) }{\partial z^2}-\frac \partial
{\partial \rho }\frac 1\rho \left( \frac \partial {\partial \rho }\rho
\left( B-B_0\right) \right) +g^2S^2B=0 
\]
\begin{equation}
-\frac 1\rho \frac \partial {\partial \rho }\left( \rho \frac \partial
{\partial \rho }S\right) +2b\left( S^2-v^2\right) S+g^2C^2B=0
\end{equation}

Using (\ref{apvt:cyl5}), we can express the energy (\ref{egunit}${}$){\ }in
terms of the functions $S\left( \rho ,z\right) $, $B\left( \rho ,z\right) $
and $B^0\left( \rho ,z\right) $: 
\[
\mathcal{E}_\rho \left( \vec{B},S\right) =-\frac{e^2}{4\pi R}\;+2\pi
\int_0^\infty \rho d\rho \int_{-\infty }^\infty dz\left[ \frac 12\left( 
\frac{\partial \left( B-B^0\right) }{\partial z}\right) ^2+\frac 1{2\rho
^2}\left( \frac \partial {\partial \rho }\rho \left( B-B^0\right) \right)
^2\right. 
\]
\begin{equation}
\left. +\frac{g^2S^2}2B^2+\frac 12\left( \frac{\partial S}{\partial \rho }%
\right) ^2+\frac 12\left( \frac{\partial S}{\partial z}\right) ^2+\frac
12b\left( S^2-v^2\right) ^2\right]  \label{elcyl}
\end{equation}

We require that, far from the sources ($\rho \rightarrow \infty
,\;z\rightarrow \pm \infty $), the fields should recover their ground state
values $S=v$ and $B=0$. Close to the $z$-axis, the field $B\left( \rho
,z\right) \approx \frac{1}{\rho }$ thereby making the electric field finite
in this region. However, such a behavior would make the contribution of the
term $\frac{g^{2}S^{2}}{2}B^{2}$ diverge, \emph{unless} $S\rightarrow 0$ in
the vicinity of the string. This is reason why the energy (as well as the
string tension), calculated in the London limit, has an ultraviolet
divergence (see Sect.\ref{sec:abrikosov}).

The role of the model parameters is made more explicit, if we work with the
following dimensionless fields and distances: 
\[
B\left( \rho ,z\right) -B^{0}\left( \rho ,z\right) =vl\left( x,y\right)
\;\;\;\;\;\;S\left( \rho ,z\right) =vs\left( x,y\right)
\;\;\;\;\;\;B^{0}\left( \rho ,z\right) =vb_{0}\left( x,y\right) 
\]
\begin{equation}
x=gv\rho =m_{V}\rho \;\;\;\;\;\;\;\;y=gvz=m_{V}z  \label{nodim}
\end{equation}
where $m_{H}$ and $m_{V}$ are the Higgs and vector masses (\ref{mh}) and \ref
{mv}), respectively equal to the inverse penetration and correlation lengths
(\ref{lengths}). The energy acquires the form: 
\[
\mathcal{E}_{R}\left( l,s\right) =-\frac{e^{2}}{4\pi R}+\frac{m_{V}}{g^{2}}%
2\pi \int_{0}^{\infty }xdx\int_{-\infty }^{\infty }dy\left[ \frac{1}{2}%
\left( \frac{\partial l}{\partial y}\right) ^{2}+\frac{1}{2x^{2}}\left( 
\frac{\partial }{\partial x}xl\right) ^{2}+\frac{1}{2}s^{2}\left(
c-b_{0}\right) ^{2}\right. 
\]
\begin{equation}
\left. +\frac{1}{2}\left( \frac{\partial s}{\partial x}\right) ^{2}+\frac{1}{%
2}\left( \frac{\partial s}{\partial y}\right) ^{2}+\frac{m_{H}^{2}}{%
8m_{V}^{2}}\left( s^{2}-1\right) ^{2}\right]  \label{ered}
\end{equation}
where $\frac{m_{V}}{g^{2}}=\frac{v}{g}$.

\begin{figure}[htbp]
\begin{center}
\includegraphics[width=10cm]{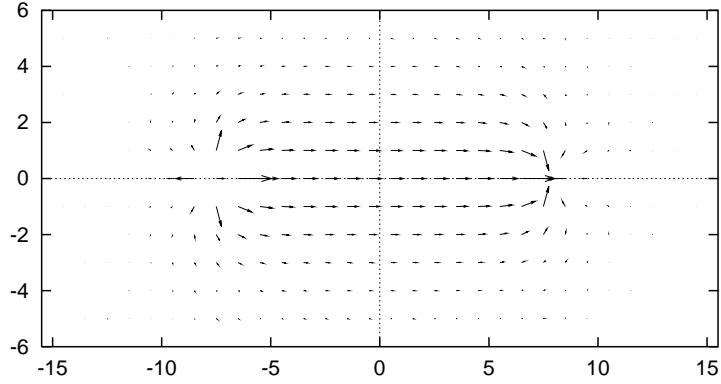}
\end{center}
\caption{The lines of force of the electric field between two static
color-electric $SU\left( 2\right) $ sources obtained from a lattice
calculation in the maximal abelian projection \protect\cite{Bali1998-1}.}
\label{fig:bali4}
\end{figure}

The flux tube obtained by minimizing the energy (\ref{ered}) is calculated
and displayed in the 1990 paper of Maedan, Matsubara and Suzuki \cite
{Suzuki1990}. The flux tube is very similar to the one obtained in lattice
calculations, illustrated in Fig.\ref{fig:bali4}.\ A more detailed
comparison to lattice data is made in Sect.\ref{sec:balifit}. More recently,
progress towards analytic forms for the solutions has been reported in the
1998 paper of Baker, Brambilla, Dosch and Vairo \cite{Baker1998}.

\subsection{The electric field and the magnetic current}

\label{sec:elecmagcur}

The electric and magnetic fields are given by (\ref{ehbg}): 
\begin{equation}
\vec{E}=-\vec{\nabla}\times \vec{B}+\vec{E}^{st}\;\;\;\;\;\;\;\vec{H}=0
\label{ehg}
\end{equation}
The longitudinal part of the electric field is the string term $\vec{E}^{st}$
and it is given by: 
\begin{equation}
\vec{\nabla}\cdot \vec{E}=\vec{\nabla}\cdot \vec{E}^{st}=\rho
\end{equation}
A magnetic current (\ref{gkmag}) is produced by the transverse part of the
electric field: 
\begin{equation}
\vec{j}_{mag}=-\vec{\nabla}\times \vec{E}  \label{ecjmag}
\end{equation}
The relation $\vec{j}_{mag}=-\vec{\nabla}\times \vec{E}$ is sometimes
referred to as the ''Ampere law''.

When the fields have the cylindrical symmetry (\ref{sbphi}), the electric
field $\vec{E}$ and the magnetic current $\vec{j}_{mag}$, given by (\ref{ehg}%
) and (\ref{ecjmag}), are: 
\[
\vec{E}\left( \rho ,z\right) =-\vec{\nabla}\times \vec{B}+\vec{\nabla}\times 
\vec{B}^0+\vec{E}^0=\vec{e}_{\left( \rho \right) }\frac{\partial B}{\partial
z}-\vec{e}_{\left( z\right) }\frac 1\rho \left( \frac \partial {\partial
\rho }\rho B\right) +\vec{E}^0 
\]
\[
\vec{j}_{mag}\left( \rho ,z\right) =\vec{\nabla}\times \left( \vec{\nabla}%
\times \vec{B}\right) -\vec{\nabla}\times \left( \vec{\nabla}\times \vec{B}%
_0\right) 
\]
\begin{equation}
=-\vec{e}_{\left( \theta \right) }\left( \frac{\partial ^2\left(
B-B^0\right) }{\partial z^2}+\frac \partial {\partial \rho }\frac 1\rho
\frac \partial {\partial \rho }\rho \left( B-B^0\right) \right)  \label{ej}
\end{equation}
We see that the magnetic current circulates around the $z$-axis.\ This is
why the flux tube is often called a \emph{vortex}.

\subsection{\protect\bigskip The Abrikosov-Nielsen-Olesen vortex}

\label{sec:abrikosov}

When $m_VR\gg 1$ and $m_HR\gg 1$, that is, when the distance which separates
the electric charges is much larger than the width of the flux tube, in
regions of space where both $\rho $ and $z$ are much smaller than $R$, the
source term (\ref{bot}) reduces to $\vec{B}^0=-\vec{e}_{\left( \theta
\right) }\frac e{2\pi \rho }$, and this in turn implies that $\vec{\nabla}%
\times \vec{B}^0=0$. In this region of space, the fields (\ref{sbphi}),
which are solutions of the equations (\ref{fieldeq}), become independent of $%
z$ and they acquire the simpler form: 
\begin{equation}
S\left( \vec{r}\right) =S\left( \rho \right) \;\;\;\;\;\;\;\;\;\;\vec{B}%
\left( \vec{r}\right) =\vec{e}_{\left( \theta \right) }B\left( \rho \right)
\end{equation}
The electric field (\ref{ej}) points in the $z$-direction: 
\begin{equation}
\vec{E}\left( \vec{r}\right) =-\vec{\nabla}\times \vec{B}=-\vec{e}_{\left(
z\right) }\frac 1\rho \frac \partial {\partial \rho }\left( \rho B\right)
\label{ezb}
\end{equation}
and a magnetic current $\vec{j}_{mag}$ circulates around the $z$-axis: 
\begin{equation}
\vec{j}_{mag}=-\vec{\nabla}\times \vec{E}=-\vec{e}_{\left( \theta \right)
}\frac \partial {\partial \rho }\frac 1\rho \frac \partial {\partial \rho
}\rho B  \label{jmagb}
\end{equation}
A flux tube is formed which is referred to as the Abrikosov-Nielsen-Olesen
vortex, studied in the 1973 paper of Nielsen and Olesen \cite{Nielsen1973}.
It is the analogue of the vortex lines, which were predicted to occur in
superconductors by Abrikosov \cite{Abrikosov1957}.\ Being particularly
simple, the Abrikosov-Nielsen-Olesen vortex has been extensively studied. In
Sect. \ref{sec:balifit}, we shall see that lattice simulations confirm the
formation of an Abrikosov-Nielsen-Olesen vortex between equal and opposite
charges.

The energy density becomes independent of both $z$ and $\theta $, so that,
in terms of the dimensionless fields and distances (\ref{nodim}), the energy
per unit length along the $z$-axis reduces to: 
\begin{equation}
\frac{\partial \mathcal{E}_R\left( b,s\right) }{\partial z}=2\pi
v^2\int_0^\infty xdx\left( \frac 1{2x^2}\left( \frac \partial {\partial
x}xb\right) ^2+\frac 12s^2b^2+\frac 12\left( \frac{\partial s}{\partial x}%
\right) ^2+\frac{m_H^2}{8m_V^2}\left( s^2-1\right) ^2\right)  \label{dedz}
\end{equation}
where $b\left( x\right) =vB\left( z\right) $ and where $v^2=\frac{m_V^2}{g^2}%
.$ The fields $s\left( x\right) $ and $b\left( x\right) $ which make the
energy stationary are the solutions of the equations: 
\[
-\frac d{dx}\frac 1x\frac d{dx}\left( xb\right) +s^2b=0 
\]
\begin{equation}
-\frac 1x\frac d{dx}x\frac d{dx}s+s^2b+\frac{m_H^2}{2m_V^2}\left(
s^2-1\right) s=0  \label{redeq}
\end{equation}
The boundary conditions are: 
\begin{equation}
b\mathrel{\mathop{\rightarrow }\limits_{x\rightarrow 0}}\frac ax\;\;\;\;\;\;s%
\mathrel{\mathop{\rightarrow }\limits_{x\rightarrow 0}}0\;\;\;\;\;\;b%
\mathrel{\mathop{\rightarrow }\limits_{x\rightarrow \infty }}0\;\;\;\;\;\;s%
\mathrel{\mathop{\rightarrow }\limits_{x\rightarrow 0}}1
\end{equation}
The constant $a$ can be determined from the flux of the electric field
crossing a surface normal to the vortex.

The expression (\ref{dedz}) is interpreted as the string tension, that is,
the coefficient $\sigma $ of the asymptotically linear potential $\sigma R$
which develops between static electric charges embedded in the dual
superconductor. The string tension depends on the two parameters $v$ and $%
\frac{m_{H}}{m_{V}}$, the ratio of the Higgs and vector masses. When $%
m_{H}=m_{V}$, that is, when the system is on the borderline between a type I
and II superconductor, analytic solutions of the equations of motion have
been found by Bogomolnyi \cite{Bogomolnyi1976}. In this Bogomolnyi limit,
the string tension is given by the expression: 
\begin{equation}
\sqrt{\sigma }=\sqrt{\frac{\pi }{g^{2}}m^{2}}\;\;\;\;\;\;m=m_{H}=m_{V}
\end{equation}
The stability and extensions to supersymmetry have also been investigated.
For a review, see the Sect.4.1 of the useful book by Vilenkin and Shellard 
\cite{Vilenkin1994}. The interaction between vortices is discussed in
Sect.4.3 of that book. When $m_{H}\gg m_{V}$ (type II superconductors),
vortices repel each other. When $m_{H}\ll m_{V}$ (type I superconductors) an
attraction between vortices occurs.

\begin{itemize}
\item  \textbf{Exercise}: Consider two equal and opposite electric charges $%
\pm e$ embedded in a dual superconductor.\ Assume that they are separated by
a large distance $L$, such that a flux tube is formed, the energy of which
is proportional to its length $L$. We can then write the energy of the flux
tube in the form $\mathcal{E}=\alpha e^2L$. Study how the energy of the
system varies as a function of the electric charge $e$, for a fixed value of 
$L$. Show that, on the average, the energy increases linearly (and not
quadratically) with $e$, because several flux tubes can form.
\end{itemize}

\subsection{Divergencies of the London limit}

\label{sec:divlondon}

\begin{figure}[htbp]
\begin{center}
\includegraphics[width=10cm]{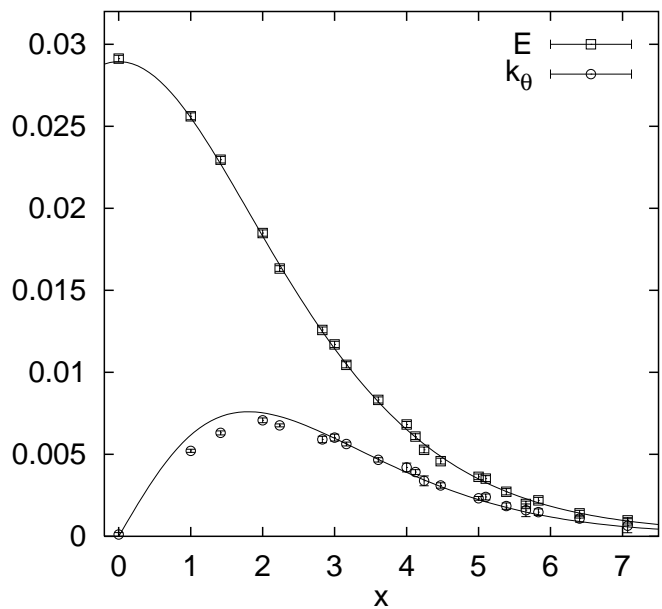}
\end{center}
\caption{The profiles of the electric field $E$ and of the magnetic current,
marked $k_\protect\theta $ on the figure, are plotted as a function of the
the distance $x$ from the center of the flux tube. The results are obtained
from a pure gauge $SU\left( 2\right) $ lattice calculation in the maximal
abelian gauge \protect\cite{Bali1998-4}. The curves are obtained from the
Landau-Ginzburg model of a dual superconductor.}
\label{fig:fit3}
\end{figure}

The London limit is the extreme case $m_H\gg m_V$ of a type II
superconductor.\ In this limit, the energy (\ref{dedz}) is be minimized when
the field $s\left( x\right) $ maintains its ground state value $s=1$ ($S=v$)
for all values of $x$. The equation for $b\left( x\right) $ reduces then to: 
\[
-\frac d{dx}\frac 1x\frac d{dx}\left( xb\right) +b=0 
\]
which is equivalent to the following equation for the field $C\left(
z\right) :$%
\begin{equation}
-\frac d{d\rho }\frac 1\rho \frac d{d\rho }\left( \rho B\right) +m_V^2B=0
\end{equation}
The solution, which vanishes far from the vortex is: 
\begin{equation}
b\left( x\right) =aK_1\left( x\right) \;\;\;or\;\;\;B\left( z\right) =\frac
avK_1\left( m_V\rho \right)
\end{equation}
However, for small values of $x$, we have: 
\begin{equation}
x\left( K_1\left( x\right) \right) ^2=\allowbreak x^{-1}+\left( -\ln \frac
1x-\ln 2+\gamma-\frac 12\right) x+O\left( x^3\right)
\end{equation}
so that the integral of $\frac 12s^2b^2$, in the string tension (\ref{dedz}%
), produces a logarithmic divergence at small $x$. This is the origin of the
divergence obtained in the analytic expression (\ref{vlr1}) of the string
tension in the London limit. The electric field, far from the sources, is $%
\vec{E}=-\vec{\nabla}\times \vec{B}$ and it has the singular behavior $\vec{E%
}\left( \vec{r}\right) \simeq -\vec{e}_{\left( z\right) }\frac av\ln \left(
m_V\rho \right) $ when $\rho \rightarrow 0$. This singular behavior does not
occur in the Landau-Ginzburg model and Fig. \ref{fig:fit3} shows that it is
also not observed in lattice simulations.

\section{Comparison of the Landau-Ginzburg model with lattice data}

\label{sec:balifit}

\begin{figure}[htbp]
\begin{center}
\includegraphics[width=10cm]{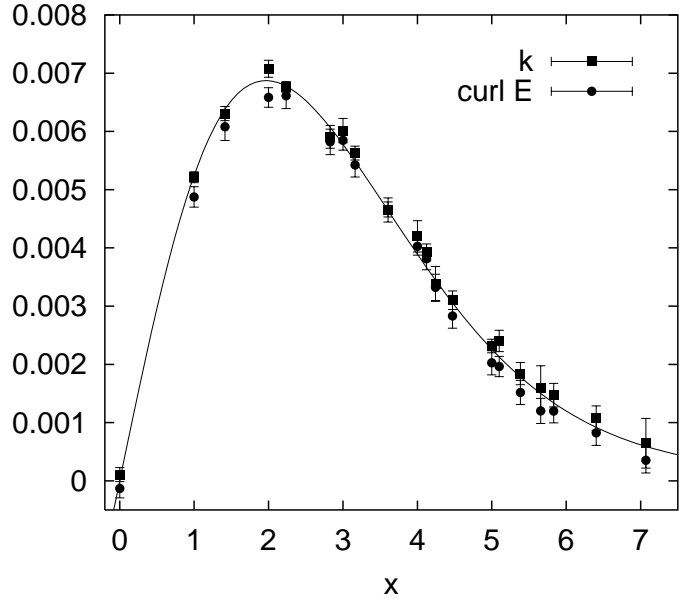}
\end{center}
\caption{The figure is a lattice confirmation of the ''Ampere law'' $\vec{j}%
_{mag}=-\vec{\nabla}\times \vec{E} $. The profiles of the magnetic current,
marked $k$ on the figure and of $\vec{\nabla}\times \vec{E}$, marked $curlE$
on the figure, are plotted as a function of the distance $x$ from the center
of the flux tube. The results are obtained from a pure gauge $SU\left(
2\right) $ lattice calculation in the maximal abelian gauge \protect\cite
{Bali1998-4}.}
\label{fig:curle}
\end{figure}

In 1998, Bali, Schlichter and Schilling \cite{Bali1998-4} compared the flux
tube formed by two static sources in a $SU\left( 2\right) $ lattice
calculation in the maximal abelian gauge (see Chapt.\ref{sec:abelgauge})
with the Abrikosov-Nielsen-Olesen vortex formed in the Landau-Ginzburg model
of a dual superconductor. They measured both the electric field and magnetic
current. The flux tube is depicted in Fig.\ref{fig:bali4}. Figure \ref
{fig:curle} shows that the relation $\vec{j}_{mag}=-\vec{\nabla}\times \vec{E%
}$ between the electric field (\ref{ezb}) and the magnetic current (\ref
{jmagb}) is satisfied, and they checked that a magnetic current circulates
around the flux tube. They also measured the profiles of the electric field
and of the magnetic current and compared it to the corresponding expressions
(\ref{ezb}) and (\ref{jmagb}) obtained for the Abrikosov-Nielsen-Olesen
vortex. Figure \ref{fig:fit3} shows the fit obtained with the parameters: 
\begin{equation}
m_{V}=gv=1.23\;GeV,\;\;\;\;m_{H}=2v\sqrt{b}=1.04\;GeV  \label{balifit}
\end{equation}
The values of $m_{V}$ and $m_{H}$ show that system is a type I
superconductor, but close to the border separating type I and type II
superconductors. A similar conclusion was reached by Matsubara, Ejiri and
Suzuki in an earlier 1994 paper \cite{Suzuki1994} for both color $SU\left(
2\right) $ and $SU\left( 3\right) $. The fit appears to be good enough to be
significant. Note that the electric field behaves quite regularly when $%
x\rightarrow 0$, that is, close to the $z$-axis.\ This is in contradiction
with the electric field calculated in the London limit. In a 1999 paper \cite
{Polikarpov1999}, Gubarev, Ilgenfritz, Polikarpov and Suzuki fitted the same
lattice data with the parameters: 
\begin{equation}
g=5.\,827\pm 0.004\;\;\;\;\;\;m_{V}=1.31\pm
0.07\;GeV\;\;\;\;\;\;m_{H}=1.36\pm 0.01\;GeV
\end{equation}
The data are taken with a lattice spacing fitted to the observed string
tension $\sqrt{\sigma _{SU\left( 2\right) }}=440\;MeV$.\ The Landau-Ginzburg
dual superconductor gives a string tension equal to: 
\begin{equation}
\sqrt{\sigma }=400.1\pm 53.0\;MeV\approx 0.91\sqrt{\sigma _{SU\left(
2\right) }}
\end{equation}
A more recent fit reported in the 2003\ paper by Koma, Ilgenfritz and Suzuki 
\cite{Koma2003} yields the following parameters: 
\begin{equation}
m_{V}=953(20)\;MeV\;\;\;\;\;\;\;m_{H}=1091(7)\;MeV
\end{equation}
In this paper, the authors find a small dependence of the fitted coupling
constant $g$ on the distance separating the quark and antiquark, which is
compatible with antiscreening of the effective QCD\ coupling constant,
derived from the Dirac quantization condition $e_{eff}=\frac{4\pi }{g}$.

Suganuma and Toki \cite{Suganuma2000-2} argue in favor of a different set of
model parameters, namely: 
\begin{equation}
m_{V}=0.5\;GeV\;\;\;\;\;\;\;\;\;m_{H}=1.26\;MeV
\end{equation}
which would make the QCD\ ground state is a type II\ superconductor. They do
not, however, fit the profiles of the Abrikosov-Nielsen-Olesen vortex.

The scalar field $S=\left| \psi \right| $, which acts as an order parameter
in the Landau-Ginzburg model, does not, as such, specify the nature of the
monopole condensation, assumed to occur in the QCD\ ground state.\
Similarly, the Landau-Ginzburg model of usual superconductors was not direct
evidence of electron-pair condensation, which was postulated and discovered
six years later \cite{Cooper1956,Bardeen1957}.

\section{The dielectric function of the color-dielectric model}

The Abrikosov-Nielsen-Olesen vortex may also be described in terms of the
color dielectric model of T.D.Lee \cite{Lee81}. The model is described by
the action: 
\begin{equation}
I_j\left( A,\sigma \right) =\int d^4x\left( -\frac 12\kappa \left( \sigma
\right) \left( \partial \wedge A\right) ^2+\frac 12\left( \partial \sigma
\right) ^2-\frac b2\left( \sigma ^2-v^2\right) ^2-j\cdot A\right)
\label{colordi}
\end{equation}
where $\kappa \left( \sigma \right) $ is a dielectric constant chosen such
that is decreases regularly from $1$ to $0$ as $\sigma $ varies from zero to 
$v$ and where $j^\mu $ is a source for electric charges and currents. The
equations of motion are: 
\[
\partial \cdot \kappa \left( \partial \wedge A\right) =j 
\]
\begin{equation}
-\partial ^2\sigma -\frac 12\left( \partial \wedge A\right) ^2\kappa
^{\prime }\left( \sigma \right) -2b\left( \sigma ^2-v^2\right) \sigma
=0\;\;\;\;\;\;\left( \kappa ^{\prime }\left( x\right) =\frac{\delta \kappa
\left( \sigma \right) }{\delta \sigma \left( x\right) }\right)
\end{equation}

If we define the field strength tensor to be $F^{\mu \nu }=\kappa \left(
\sigma \right) \left( \partial \wedge A\right) $ then the equation of motion
for $A^{\mu }$ becomes equivalent to the Maxwell equation $\partial \cdot
F=j $. The system also develops a magnetic current\footnote{%
I am indebted to Gunnar Martens for showing this to me prior to the
publication of his calculations.}, given by: 
\begin{equation}
j_{mag}=\partial \cdot \bar{F}=\left( \partial \kappa \right) \cdot 
\overline{\partial \wedge A}
\end{equation}
Let us write: 
\begin{equation}
A^{\mu }=\left( \phi ,\vec{A}\right) \;\;\;\;\;\;j^{\mu }=\left( \rho ,\vec{j%
}\right)
\end{equation}
The action can be broken down to: 
\[
I_{j}\left( A,\phi ,\sigma \right) =\int d^{4}x\left\{ \frac{1}{2}\kappa
\left( -\partial _{t}\vec{A}-\vec{\nabla}\phi \right) ^{2}-\frac{1}{2}\kappa
\left( \vec{\nabla}\times A\right) ^{2}\right. 
\]
\begin{equation}
\left. +\frac{1}{2}\left( \partial _{t}\sigma \right) ^{2}-\frac{1}{2}\left( 
\vec{\nabla}\sigma \right) ^{2}-\frac{b}{2}\left( \sigma ^{2}-v^{2}\right)
^{2}-\rho \phi +\vec{j}\cdot \vec{A}\right\}
\end{equation}
The field $\phi $ imposes the constraint: 
\begin{equation}
\vec{\nabla}\cdot \kappa \left( -\partial _{t}\vec{A}-\vec{\nabla}\phi
\right) =\rho
\end{equation}
In the presence of static charges: 
\[
\rho \left( x\right) =\rho \left( \vec{r}\right) \,\,\,\,\,\;\;\;\;\;\vec{j}%
\left( x\right) =0 
\]
the source $j^{\mu }$ is time independent and the fields $A^{\mu }$ and $%
\sigma $ may also be assumed to be time independent. The energy in the
presence of static sources is: 
\begin{equation}
\mathcal{E}_{\rho }\left( A,\phi ,\sigma \right) =\int d^{3}r\left( -\frac{1%
}{2}\kappa \left( \vec{\nabla}\phi \right) ^{2}+\frac{1}{2}\kappa \left( 
\vec{\nabla}\times A\right) ^{2}+\frac{1}{2}\left( \vec{\nabla}\sigma
\right) ^{2}+\frac{b}{2}\left( \sigma ^{2}-v^{2}\right) ^{2}-\rho \phi
\right)
\end{equation}
and the constraint is: 
\begin{equation}
-\kappa ^{\prime }\left( \vec{\nabla}\sigma \right) \cdot \left( \vec{\nabla}%
\phi \right) -\kappa \nabla ^{2}\phi =\rho  \label{phieq}
\end{equation}
The energy is minimized when: 
\begin{equation}
\vec{\nabla}\times \vec{A}=0
\end{equation}
in which case the energy reduces to: 
\begin{equation}
\mathcal{E}_{\rho }\left( \phi ,\sigma \right) =\int d^{3}r\left( -\frac{1}{2%
}\kappa \left( \vec{\nabla}\phi \right) ^{2}+\frac{1}{2}\left( \vec{\nabla}%
\sigma \right) ^{2}+\frac{b}{2}\left( \sigma ^{2}-v^{2}\right) ^{2}-\rho
\phi \right)
\end{equation}
The field $\sigma $ which minimizes the energy satisfies the equation: 
\begin{equation}
-\nabla ^{2}\sigma +2b\left( \sigma ^{2}-v^{2}\right) \sigma -\frac{1}{2}%
\kappa ^{\prime }\left( \vec{\nabla}\phi \right) ^{2}=0  \label{sigeq}
\end{equation}

Consider two static charges on the $z$-axis, placed symmetrically with
respect to the origin: 
\begin{equation}
\rho \left( \vec{r}\right) =-e\delta \left( \vec{r}-\vec{R}_1\right)
+e\delta \left( \vec{r}-\vec{R}_2\right)
\end{equation}
If the charges are far apart, then, close to the $x-y$ plane, we expect the
electric field to be constant and directed along $\vec{e}_{\left( z\right) }$%
. A solution of equations (\ref{phieq}) and (\ref{sigeq}) exists in
cylindrical coordinates, of the form: 
\begin{equation}
\sigma \left( \vec{r}\right) =\sigma \left( \rho \right) \;\;\;\;\;\;\phi
\left( \vec{r}\right) =az
\end{equation}
and the function $\sigma \left( \rho \right) $ is the solution of the
equation: 
\begin{equation}
-\frac 1\rho \frac d{d\rho }\left( \rho \frac{d\sigma }{d\rho }\right)
+2b\left( \sigma ^2-v^2\right) \sigma -\frac 12a^2\kappa ^{\prime }=0
\end{equation}
In this region, the electric and magnetic fields are: 
\begin{equation}
\vec{E}\left( \rho \right) =-\kappa \vec{\nabla}\phi =-\vec{e}_{\left(
z\right) }a\kappa \left( \rho \right) \;\;\;\;\;\;\;\;\;\vec{H}=0
\end{equation}
In this sense, lattice calculations, as well as the Landau-Ginzburg model
which agrees with them, determine the function $\kappa \left[ \sigma \left(
\rho \right) \right] $ of the dielectric model. The magnetic current $%
j_{mag}^\mu =\left( \rho _{mag},\vec{j}_{mag}\right) $ is such that $\rho
_{mag}=0$ and: 
\begin{equation}
\;\vec{j}_{mag}=-\vec{\nabla}\times \vec{E}=\left( \vec{\nabla}\kappa
\right) \times \left( \vec{\nabla}\phi \right) =\kappa ^{\prime }\left( \vec{%
\nabla}\sigma \right) \times \left( \vec{\nabla}\phi \right) =-\vec{e}%
_{\left( \theta \right) }a\kappa ^{\prime }\frac{d\sigma }{d\rho }
\end{equation}
so that a magnetic current flows around the vortex which may be viewed as an
Abrikosov-Nielsen-Olesen vortex.

There is however an important difference between the color-dielectric model
described by the action (\ref{colordi}) and the Landau-Ginzburg model. If
the system, described by the color-dielectric model, is perturbed or raised
to a finite temperature, the expectation value of the field $\sigma $ may be
shifted to, typically, a lower value and the color dielectric function no
longer vanishes, in spite of having been fine-tuned to do so in the vacuum
at zero temperature. The model therefore describes the physical vacuum in
terms of an unstable phase, which is not the case of the Landau-Ginzburg
model which describes the superconducting phase within a finite range of
temperatures.

\section{The London limit of the Landau-Ginzburg model}

\label{sec:london}

The London limit of the Landau Ginzburg model is obtained by letting $%
b\rightarrow \infty $, which means that the Higgs mass $m_{H}$ is much
larger than the vector meson mass $m_{V}.$ In that limit, the scalar field
maintains its vacuum value $S=v$ and the model action (\ref{landsphi})
reduces to: 
\begin{equation}
I_{j}\left( B,\varphi \right) =\int d^{4}x\left( -\frac{1}{2}\left( \partial
\wedge B+\bar{G}\right) ^{2}+\frac{m_{V}^{2}}{2}\left( B+\partial \varphi
\right) ^{2}\right)  \label{london}
\end{equation}
where $m_{V}$, the mass of the vector boson, is equal to the inverse
penetration depth (\ref{lengths}), and where the source term $G$ is related
to the electric current $j$ by the equation: 
\begin{equation}
\partial \cdot G=j  \label{geq}
\end{equation}

Because of its simplicity (the action is a quadratic form of the single
field $B^\mu $), the London limit has been extensively studied.\ It yields
simple analytic forms for the confining potential, field strength
correlators, and more. However, as explained in Sect. \ref{sec:flux}, the
linear confining potential develops an ultraviolet divergence because, in
the London limit, the scalar field $S$ is not allowed to vanish in the
center of the flux tube.\ For the same reason, the electric field develops a
singularity in the center of the flux tube and this is not observed in
lattice simulations (see Sect. \ref{sec:balifit}).

We shall study the London limit for a system consisting of two static
electric charges. The electric current is then: 
\[
j^{\mu }=\left( \rho \left( \vec{r}\right) ,0,0,0\right)
\;\;\;\;\;\;\partial ^{2}j^{\mu }=-\delta _{0}^{\mu }\vec{\nabla}^{2}\rho 
\]
\begin{equation}
\rho \left( \vec{r}\right) =e\delta \left( \vec{r}-\vec{r}_{1}\right)
-e\delta \left( \vec{r}-\vec{r}_{2}\right)  \label{jstat}
\end{equation}
We work in the unitary gauge and we assume a straight line Dirac string,
running from one charge to the other. In this case, we can use the form (\ref
{gnj}): 
\begin{equation}
G=\frac{1}{n\cdot \partial }n\wedge j  \label{gnjd}
\end{equation}
with a space-like vector $n^{\mu }=\left( 0,\vec{n}\right) $ and with $\vec{n%
}$ parallel to the line joining the charges: 
\[
\vec{n}=\vec{r}_{2}-\vec{r}_{1} 
\]
\begin{equation}
n^{\mu }=\left( 0,\vec{n}\right) \;\;\;\;\;\;n^{2}=-\vec{n}%
^{2}\;\;\;\;\;\;n\cdot j=0\;\;\;\;\left( n\cdot \partial \right) j=\vec{n}%
\cdot \vec{\nabla}\rho  \label{nspatial}
\end{equation}
The equation of motion for the field $B^{\mu }$ is: 
\begin{equation}
\partial \cdot \left( \partial \wedge B+\overline{G}\right) +m_{V}^{2}B=0
\end{equation}
If we decompose $B^{\mu }$ into longitudinal and transverse parts using the
projectors (\ref{apvt:tldd}), we find that $B^{\mu }$ is transverse and
equal to: 
\begin{equation}
B=-\frac{1}{\partial ^{2}+m_{V}^{2}}\partial \cdot \overline{G}
\end{equation}
We can then eliminate the field $B^{\mu }$ from the action, which reduces
to: 
\begin{equation}
I_{j}=\int d^{4}x\left( -\frac{1}{2}\left( \partial \cdot \bar{G}\right) 
\frac{1}{\partial ^{2}+m_{V}^{2}}\left( \partial \cdot \bar{G}\right) -\frac{%
1}{2}\bar{G}^{2}\right)  \label{ijgg}
\end{equation}

\subsection{The gluon propagator}

\label{sec:glueprop}

When $G$ has the form (\ref{gnjd}), we have: 
\begin{equation}
\left( \partial \cdot \overline{G}\right) ^\mu =\frac 1{n\cdot \partial
}\left( \partial \cdot \overline{n\wedge j}\right) ^\mu =-\frac 1{n\cdot
\partial }\varepsilon ^{\mu \alpha \beta \gamma }\partial _\alpha n_\beta
j_\gamma \;\;\;\;\;\;\overline{G}^2=-\left( \frac 1{n\cdot \partial }\left(
n\wedge j\right) \right) ^2
\end{equation}
Substituting, the action (\ref{ijgg}) becomes: 
\[
I_j=\int d^4x\;\left( \frac 12j_3\left( \varepsilon ^{\mu 123}\partial
_1n_2\right) \frac 1{\left( n\cdot \partial \right) ^2}\frac 1{\partial
^2+m_V^2}\left( \varepsilon _{\mu 456}\partial ^4n^5j^6\right) -\frac
12\left( n\wedge j\right) \frac 1{\left( n\cdot \partial \right) ^2}\left(
n\wedge j\right) \right) 
\]
We can use (\ref{apvt:epscontr}) to evaluate $\varepsilon ^{\mu
123}\varepsilon _{\mu 456}$. Remembering that $\partial _\mu j^\mu =0$, a
straightforward, albeit risky calculation yields the action: 
\begin{equation}
I_j=\int d^4x\left( \frac 12j\;\frac 1{\left( n\cdot \partial \right)
^2}\left( \partial ^2n^2-\left( n\cdot \partial \right) ^2\right) \frac
1{\partial ^2+m_V^2}j-\frac 12\left( n\cdot j\right) \frac{\partial ^2}{%
\left( n\cdot \partial \right) ^2}\frac 1{\partial ^2+m_V^2}\left( n\cdot
j\right) \right)
\end{equation}
\begin{equation}
-\int d^4x\left( -\frac 12\left( n\cdot j\right) \;\frac 1{\left( n\cdot
\partial \right) ^2}\left( n\cdot j\right) \right)
\end{equation}
We group together the terms which depend on $n\cdot \partial $ and those
which do not, to get: 
\begin{equation}
I_j=\int d^4x\;j_\mu \left( -\frac 12\frac{g^{\mu \nu }}{\partial ^2+m_V^2}%
-\frac 12\frac{n^2}{\left( n\cdot \partial \right) ^2}\left( \frac{m_V^2}{%
\partial ^2+m_V^2}\right) \left( g^{\mu \nu }-\frac{n^\mu n^\nu }{n^2}%
\right) \right) j_\nu  \label{ijlondon}
\end{equation}
Since $j^\mu $ is a source term for the gluon field $A_\mu $, the London
limit of the gluon propagator in the dual superconductor can be read off the
expression (\ref{ijlondon}): 
\begin{equation}
D^{\mu \nu }=-\frac 12\frac{g^{\mu \nu }}{\partial ^2+m_V^2}-\frac 12\frac{%
n^2}{\left( n\cdot \partial \right) ^2}\left( \frac{m_V^2}{\partial ^2+m_V^2}%
\right) \left( g^{\mu \nu }-\frac{n^\mu n^\nu }{n^2}\right)  \label{glueprop}
\end{equation}

\subsection{The energy in the presence of static electric charges in the
London limit}

\label{sec:lonpot}

When the system consists of static electric charges, the fields are
time-independent and the energy density is equal to minus the lagrangian.\
The energy obtained from the action \ref{ijlondon} is thus: 
\[
\mathcal{E}_\rho =\int d^3r\;j_\mu \left( \frac 12\frac{g^{\mu \nu }}{%
\partial ^2+m_V^2}+\frac 12\frac{n^2}{\left( n\cdot \partial \right) ^2}%
\left( \frac{m_V^2}{\partial ^2+m_V^2}\right) \left( g^{\mu \nu }-\frac{%
n^\mu n^\nu }{n^2}\right) \right) j_\nu 
\]
\[
=\int d^3r\frac 12\rho \left( \;\frac 1{-\vec{\nabla}^2+m_V^2}-\;\frac{\vec{n%
}^2}{\left( \vec{n}\cdot \vec{\nabla}\rho \right) ^2}\left( \frac{m_V^2}{-%
\vec{\nabla}^2+m_V^2}\right) \right) \rho 
\]
\begin{equation}
=\frac 1{\left( 2\pi \right) ^3}\int d^3k\;\frac 12\rho _{\vec{k}}\left(
\frac 1{k^2+m_V^2}+\frac{m_V^2}{\left( k^2+m_V^2\right) }\frac{\vec{n}^2}{%
\left( \vec{n}\cdot \vec{k}\right) ^2}\right) \rho _{-\vec{k}}  \label{erhok}
\end{equation}
If we substitute the form (\ref{jstat}) of $\rho $ into the energy (\ref
{erhok}), we obtain: 
\begin{equation}
\mathcal{E}_\rho =\frac 12\int d^3r_1d^3r_2\rho \left( \vec{r}_1\right)
v\left( \vec{r}_1-\vec{r}_2\right) \rho \left( \vec{r}_2\right)
\label{erpot}
\end{equation}
with: 
\begin{equation}
v\left( \vec{r}\right) =\frac{-e^2}{\left( 2\pi \right) ^3}\int d^3ke^{-i%
\vec{k}\cdot \vec{r}}\left( \frac 1{k^2+m_V^2}+\frac{m_V^2}{\left(
k^2+m_V^2\right) }\frac{r^2}{\left( \vec{r}\cdot \vec{k}\right) ^2}\right)
\label{vor}
\end{equation}
where we set $\vec{n}=\vec{r}$ in accordance with (\ref{nspatial}).

\subsection{The confining potential in the London limit}

\label{sec:londonvr}

The first term of the potential (\ref{vor}) is a short ranged Yukawa
potential: 
\begin{equation}
v_{SR}\left( \vec{r}\right) =\frac{-e^{2}}{\left( 2\pi \right) ^{3}}\int
d^{3}ke^{-i\vec{k}\cdot \vec{r}}\frac{1}{k^{2}+m_{V}^{2}}=\frac{-e^{2}}{4\pi
r}e^{-m_{V}r}  \label{vsrr}
\end{equation}
Consider the second (long-range) term: 
\begin{equation}
v_{LR}\left( \vec{r}\right) =\frac{-e^{2}}{\left( 2\pi \right) ^{3}}\int
d^{3}ke^{-i\vec{k}\cdot \vec{r}}\frac{m_{V}^{2}}{\left(
k^{2}+m_{V}^{2}\right) }\frac{r^{2}}{\left( \vec{r}\cdot \vec{k}\right) ^{2}}
\end{equation}
We write $\vec{k}\cdot \vec{r}=kr\cos \theta $ and the long range potential
becomes: 
\[
v_{LR}\left( \vec{r}\right) =\frac{-e^{2}}{\left( 2\pi \right) ^{3}}\int
d^{3}ke^{-ikr\cos \theta }\frac{m_{V}^{2}}{k^{2}\left(
k^{2}+m_{V}^{2}\right) \cos ^{2}\theta } 
\]
$\allowbreak $%
\begin{equation}
=\frac{4\pi }{\left( 2\pi \right) ^{3}}\int_{0}^{\infty }dk\frac{m_{V}^{2}}{%
\left( k^{2}+m_{V}^{2}\right) }\int_{0}^{1}dx\frac{\cos \left( krx\right) }{%
x^{2}}
\end{equation}
The integral diverges both at small $x$ and at large $k$. The divergence at
small $x$ contributes an infinite term which is independent of $r$. This can
be seen by making a subtraction, that is, by evaluating $v_{LR}\left( \vec{r}%
\right) -v_{LR}\left( \vec{r}_{0}\right) $. Indeed, we have: 
\begin{equation}
\int_{0}^{1}\frac{\cos \left( krx\right) -\cos \left( kr_{0}x\right) }{x^{2}}%
dx=-\cos kr-krSi\left( kr\right) +\cos kr_{0}+kr_{0}Si\left( kr_{0}\right)
\end{equation}
so that: 
\begin{equation}
v_{LR}\left( \vec{r}\right) =v_{LR}\left( \vec{r}_{0}\right) -\frac{4\pi
e^{2}}{\left( 2\pi \right) ^{3}}m_{V}^{2}r\int_{0}^{\Lambda r}dy\frac{1}{%
\left( y^{2}+m_{V}^{2}r^{2}\right) }\left( -\cos y-ySi\left( y\right) \right)
\end{equation}
Except for regions where $m_{V}r\lesssim 1$, we can approximate the function 
$\cos y+ySi\left( y\right) $ by its asymptotic value $\frac{\pi }{2}y$ and,
adding the short range part (\ref{vsrr}), the potential becomes: 
\begin{equation}
v_{LR}\left( \vec{r}\right) =v_{LR}\left( \vec{r}_{0}\right) -\frac{e^{2}}{%
4\pi r}e^{-m_{V}r}+\frac{e^{2}m_{V}^{2}}{8\pi }\;r\;\ln \frac{\Lambda
^{2}+m_{V}^{2}}{m_{V}^{2}}  \label{vlr1}
\end{equation}
where a sharp cut-off $\Lambda $ was introduced to make the $k$-integral
converge at large $k$. As discussed in Sects.\ref{sec:divlondon} and \ref
{sec:london}, the ultraviolet divergence is an artifact of the London limit.

Equal and opposite electric charges are thus confined by a linearly rising
potential.\ Let us define the string tension $\sigma $ by writing the
potential in the form $V\left( r\right) =-\frac{e^2}{4\pi }\frac{e^{-m_Vr}}%
r+\sigma r$. The London limit of the Landau-Ginzburg model produces a string
tension equal to: 
\begin{equation}
\sigma =\frac{e^2m_V^2}{8\pi }\ln \frac{\Lambda ^2+m_V^2}{m_V^2}=\frac{%
n^2\pi }2v^2\ln \frac{\Lambda ^2+m_V^2}{m_V^2}  \label{stringten}
\end{equation}
where we used the relation $eg=2n\pi $ between the magnetic and electric
charges. The string tension may be compared to the prediction (\ref{dedz})
obtained in the Landau-Ginzburg model, without taking the London limit,
namely: 
\begin{equation}
\frac{\partial \mathcal{E}_R\left( c,s\right) }{\partial z}=2\pi
v^2\int_0^\infty xdx\left( \frac 1{2x^2}\left( \frac \partial {\partial
x}xc\right) ^2+\frac 12s^2c^2+\frac 12\left( \frac{\partial s}{\partial x}%
\right) ^2+\frac{m_H^2}{8m_V^2}\left( s^2-1\right) ^2\right)
\end{equation}
The latter depends on two parameters, namely $v$ and the ratio $\frac{m_H}{%
m_V}$ of the vector and Higgs masses. The cut-off in the expression obtained
in the London limit mimics the missing vanishing of the scalar field in the
vicinity of the vortex.

\subsection{Chiral symmetry breaking}

The gluon propagator (\ref{glueprop}) has been used by Suganuma and Toki as
an input for a Schwinger-Dyson calculation, in which the quark propagator is
dressed by a sum of rainbow diagrams in the Landau gauge \cite{Suganuma2000}%
. The quark gluon propagator was found to be strong enough to produce
spontaneous chiral symmetry breaking. The dependence of the gluon propagator
(\ref{glueprop}) on the direction of the vector $n^\mu $ was averaged out.
It was further modified at large euclidean momenta so as to make it merge
with the perturbative QCD\ value, and at small momenta so as eliminate its
divergence as $k\rightarrow 0$. As a result, the statement that the
Landau-Ginzburg model explains the observed chiral symmetry breaking is only
qualitative. However, they did observe that, when the vector mass $m_V=gv$
was small enough, both confinement and chiral symmetry vanished. More
recently, the relation between monopoles and instantons has been studied in
Abelian projected QCD, both analytically and by lattice simulations \cite
{Suganuma1995-2,Teper1996,Brower1997,Diakonov1997}. The topic is reviewed by
Toki and Suganuma \cite{Suganuma2000-2}.

\section{The field-strength correlator}

\label{sec:brambilla}

In these lectures, we have expressed 4-vectors and tensors in a Minkowski
space with the metric $g^{\mu \nu }=diag\left( 1,-1,-1,-1\right) $, such
that $\det g=-1$.\ Such ''Minkowski actions'' are expressed in terms of
vectors and tensors in Minkowski space, and they are suitable for canonical
quantization and for various representations of the evolution operator $%
e^{iHt}$, by means of path integrals for example. However, present day
lattice calculations are limited to evaluations of the partition function $%
Tre^{-\beta H}$, which is expressed in terms of a functional integral of a
Euclidean action.\ The latter is written in terms of vectors and tensors in
a Euclidean space with a metric $g^{\mu \nu }=diag\left( 1,1,1,1\right)
=\delta _{\mu \nu }$, such that $\det g=+1$. A Euclidean action is obtained
when the functional integral for the partition function is derived from the
Hamiltonian of the system. Crudely speaking, the Euclidean action which is
thus obtained, is related to the Minkowski action by making the $\mu =0$
components of vectors and tensors imaginary. This is summarized in the table
(\ref{apme:trans}) of App.\ref{ap:minkeucl}. The reader who is not familiar
with the use of Euclidean actions, is referred to standard textbooks \cite
{Zinn89,Ripka86,Negele1988}.

The Euclidean form of the Landau-Ginzburg action (\ref{landsphi}) in the
unitary gauge, as given in App.\ref{ap:minkeucl}, has the form: 
\begin{equation}
I_{j}\left( B,S\right) =\int d^{4}x\left( \frac{1}{2}\left( \partial \wedge
B+\bar{G}\right) ^{2}+\frac{g^{2}S^{2}}{2}B^{2}+\frac{1}{2}\left( \partial
S\right) ^{2}+\frac{1}{2}b\left( S^{2}-v^{2}\right) ^{2}\right)
\label{euclidlg}
\end{equation}

In a 1998\ paper \cite{Baker1998}, Baker, Brambilla, Dosch and Vairo
suggested to use this Landau-Ginzburg action in order to model the field
strength correlator which is observed in lattice calculations, in the
absence of quarks. Formally, the source term $\bar{G}$ appearing in the
action, can be used as a source term for the dual field tensor $\bar{F}%
=\partial \wedge B$.\ 

Let us illustrate the method by restricting ourselves to the London limit
(Sect. \ref{sec:london}) in which the field $S$ maintains its ground state
value $S=v$. The euclidean action reduces then to: 
\begin{equation}
I_{j}\left( B\right) =\int d^{4}x\left( \frac{1}{2}\left( \partial \wedge B+%
\bar{G}\right) ^{2}+\frac{m_{V}^{2}}{2}B^{2}\right)  \label{euclid}
\end{equation}
The equation of motion of the field $B^{\mu }$ is: 
\begin{equation}
-\partial \cdot \left( \partial \wedge B+\bar{G}\right) +m_{V}^{2}B=0
\end{equation}
By considering the longitudinal and transverse parts, defined in App. \ref
{apvt:vectors}), we can solve this equation in the form: 
\begin{equation}
B=\frac{1}{-\partial ^{2}+m_{V}^{2}}\partial \cdot \bar{G}
\end{equation}
Substituting back into the action, we obtain: 
\begin{equation}
I_{j}=\int d^{4}x\left( -\frac{1}{2}\left( \partial \cdot \bar{G}\right) 
\frac{1}{-\partial ^{2}+m_{V}^{2}}\left( \partial \cdot \bar{G}\right) +%
\frac{1}{2}\bar{G}^{2}\right)
\end{equation}
We may drop the second term which is simply due to the fact that the source
term appears quadratically in the action. Use (\ref{apvt:k}) to write the
action in the form: 
\begin{equation}
I_{j}=\int d^{4}x\frac{1}{2}\bar{G}\left( \frac{\partial ^{2}K}{-\partial
^{2}+m_{V}^{2}}\right) \bar{G}=\frac{1}{2\left( 2\pi \right) ^{4}}\int
d^{4}k\;\bar{G}\left( k\right) \left( \frac{-k^{2}K}{k^{2}+m_{V}^{2}}\right) 
\bar{G}\left( -k\right)  \label{threpee}
\end{equation}
where $K_{\mu \nu ,\alpha \beta }$ is the differential operator defined in (%
\ref{apvt:kedef}). In the Euclidean metric, it has the form: 
\begin{equation}
K_{\mu \nu ,\alpha \beta }=\frac{1}{\partial ^{2}}\left( \delta _{\mu \alpha
}\partial _{\nu }\partial _{\beta }-\delta _{\nu \alpha }\partial _{\mu
}\partial _{\beta }+\delta _{\nu \beta }\partial _{\mu }\partial _{\alpha
}-\delta _{\mu \beta }\partial _{\nu }\partial _{\alpha }\right)
\label{kform}
\end{equation}
and its properties are listed in (\ref{apme:ke}). From the form (\ref{euclid}%
) of the action, we see that $\bar{G}$ may be considered as a source term
for the dual field strength $\bar{F}=\partial \wedge B$. The Fourier
transform of the dual field strength propagator can be read off (\ref
{threpee}): 
\begin{equation}
\int d^{4}xe^{-ik\cdot x}\left\langle T\left[ \bar{F}\left( x\right) \bar{F}%
\left( 0\right) \right] \right\rangle =\frac{-k^{2}K}{k^{2}+m_{V}^{2}}
\end{equation}
Let us define the function: 
\[
L\left( x\right) =\frac{1}{\left( 2\pi \right) ^{4}}\int d^{4}k\;e^{ik\cdot
x}\frac{1}{k^{2}+m_{V}^{2}}=\frac{1}{\left( 2\pi \right) ^{4}}\frac{4\pi ^{2}%
}{x}\int_{0}^{\infty }\frac{k^{2}}{k^{2}+m_{V}^{2}}J_{1}\left( kx\right) dk 
\]
\begin{equation}
=\frac{m_{V}^{2}}{\left( 2\pi \right) ^{2}}\frac{K_{1}\left( m_{V}x\right) }{%
m_{V}x}\;\;\;\;\;\;\left( x\equiv \left| x\right| \right)
\end{equation}
The dual field strength propagator can be written as: 
\begin{equation}
\left\langle T\left[ \bar{F}\left( x\right) \bar{F}\left( 0\right) \right]
\right\rangle =\frac{1}{\left( 2\pi \right) ^{4}}\int d^{4}k\frac{-k^{2}K}{%
k^{2}+m_{V}^{2}}=K\partial ^{2}L\left( x\right)
\end{equation}
Since $\bar{F}\left( x\right) =\varepsilon F\left( x\right) $ and since, in
the euclidean metric, we have $\varepsilon K\varepsilon =E=G-K$, the field
strength propagator is: 
\begin{equation}
\left\langle T\left[ F\left( x\right) F\left( 0\right) \right] \right\rangle
=\left( G-K\right) \partial ^{2}L\left( x\right)  \label{fprop1}
\end{equation}
where $G_{\mu \nu ,\alpha \beta }=\delta _{\mu \alpha }\delta _{\nu \beta
}-\delta _{\mu \beta }\delta _{\nu \alpha }$.

In order to conform to the conventional notation found in the literature, we
note that $L$ is a function of $x^2=x_\mu x_\mu $ so that $\partial _\mu
L=2x_\mu \frac{dL}{dx^2}$. Substituting the form (\ref{kform}) into the
expression (\ref{fprop1}), we find that the field strength propagator can be
written in the form: 
\[
\left\langle T\left[ F_{\mu \nu }\left( x\right) F_{\alpha \beta }\left(
0\right) \right] \right\rangle =\left( \delta _{\mu \alpha }\delta _{\nu
\beta }-\delta _{\mu \beta }\delta _{\nu \alpha }\right) \partial ^2L 
\]
\begin{equation}
-2\left( \delta _{\mu \alpha }\partial _\nu x_\beta -\delta _{\nu \alpha
}\partial _\mu x_\beta +\delta _{\nu \beta }\partial _\mu x_\alpha -\delta
_{\mu \beta }\partial _\nu x_\alpha \right) \frac{dL}{dx^2}  \label{fprop2}
\end{equation}

The contention of Baker, Brambilla, Dosch and Vairo \cite{Baker1998} is that
this field strength propagator can be compared to the gauge invariant field
strength correlator $\left\langle g^2F_{\mu \nu }\left( x\right) U\left(
x,0\right) F_{\alpha \beta }\left( 0\right) U\left( 0,x\right) \right\rangle 
$ which is measured on the lattice \cite{DiGiacomo1997-2,DiGiacomo1997-3}
and which is usually parametrized in terms of two functions $D\left(
x^2\right) $ and $D_1\left( x^2\right) $: 
\[
\left\langle g^2F_{\mu \nu }\left( x\right) U\left( x,0\right) F_{\alpha
\beta }\left( 0\right) U\left( 0,x\right) \right\rangle =\left( \delta _{\mu
\alpha }\delta _{\nu \beta }-\delta _{\mu \beta }\delta _{\nu \alpha
}\right) g^2D\left( x^2\right) 
\]
\begin{equation}
+\frac 12\left( \delta _{\mu \alpha }\partial _\nu x_\beta -\delta _{\nu
\alpha }\partial _\mu x_\beta +\delta _{\nu \beta }\partial _\mu x_\alpha
-\delta _{\mu \beta }\partial _\nu x_\alpha \right) g^2D_1\left( x^2\right)
\label{expcorr}
\end{equation}
By comparing the expressions (\ref{fprop2}) and (\ref{expcorr}), the
Landau-Ginzburg model makes the following predictions, in the London limit,
for the functions $D\left( x^2\right) $ and $D_1\left( x^2\right) $: 
\[
g^2D\left( x^2\right) =\partial ^2S=\delta \left( x\right) -m_V^2L\left(
x\right) =\delta \left( x\right) -\frac{m_V^4}{\left( 2\pi \right) ^2}\frac{%
K_1\left( m_Vx\right) }{m_Vx} 
\]
\begin{equation}
=\frac{m_V^2}{2\pi ^2x^2}\left( \allowbreak \frac{2K_1\left( m_Vx\right) }{%
m_Vx}+K_0\left( m_Vx\right) \right)
\end{equation}
: 
\begin{equation}
g^2D_1\left( x^2\right) =-4\frac{dS}{dx^2}=\allowbreak \frac{m_V^2}{2\pi
^2x^2}\left( \allowbreak \frac{2K_1\left( m_Vx\right) }{m_Vx}+K_0\left(
m_Vx\right) \right)
\end{equation}

The cooled lattice data can be fitted, in the range $0.1\;fm\leq x\leq 1\;fm$
with the parametrization: 
\[
D\left( x^{2}\right) =Ae^{-\frac{x}{T_{g}}}+\frac{a}{x^{4}}e^{-\frac{x}{T_{p}%
}}\;\;\;\;\;\;D_{1}\left( x^{2}\right) =Be^{-\frac{x}{T_{g}}}+\frac{b}{x^{4}}%
e^{-\frac{x}{T_{p}}}\;\;\;\;\;\;x=\left| x\right| 
\]
\[
A=128\;GeV^{4}\;\;\;\;\;\;B=27\;GeV^{4}\;\;\;\;\;\;a=0.69\;\;\;\;\;\;b=0.46 
\]
\begin{equation}
T_{g}=0.22\;fm\;\;\;\;\;\;T_{p}=0.42\;fm
\end{equation}
The $\frac{1}{x^{4}}e^{-\frac{x}{T_{p}}}$ terms are negligible when $%
x>0.2\;fm$ and, since the London limit is unreliable at small $x$, we
neglect these terms for the comparison. Even so, the fit to the shape is
qualitative at best. The exponential decrease of the correlator allows us to
make the identification $m_{V}\simeq \frac{1}{T_{g}}=0.9\;GeV$, which is of
the same order of magnitude as the values quoted in Sect.\ref{sec:balifit}
and obtained by fitting the profiles of the electric field and magnetic
currents to lattice data. The reader is referred to the 1998 paper of Baker,
Brambilla, Dosch and Vairo \cite{Baker1998} for the improvements obtained
beyond the London limit.

\section{The London limit expressed in terms of a Kalb-Ramond field}

The confinement of static electric charges was modeled in Sect.\ref
{sec:london} in the London limit of the Landau-Ginzburg model. The same
confining force is obtained in terms of the following model action: 
\begin{equation}
I_{jJ}\left( A,\Phi \right) =\int d^4x\left( -\frac 12\left( \partial \cdot 
\bar{\Phi}\right) ^2-\frac 12\left( \partial \wedge A-m\Phi \right)
^2-j\cdot A-mG\cdot \Phi \right)  \label{kalbramond}
\end{equation}
which is expressed in terms of an antisymmetric tensor field $\Phi ^{\mu \nu
}=-\Phi ^{\nu \mu }$ and its dual $\bar{\Phi}$, often referred to as a
Kalb-Ramond field. In this action, $m$ is a mass parameter, $A^\mu $ is the
gauge field which couples to the electric current $j^\mu $ and $J^{\mu \nu
}=-J^{\nu \mu }$ is an antisymmetric source term for the field $\Phi $. The
latter is introduced so as to maintain the gauge invariance discussed below.
The action (\ref{kalbramond}) was studied in 1974 by Kalb and Ramond in the
context of interactions between strings \cite{Ramond1974}. The duality
transformation which leads to the use of Kalb-Ramond fields is proposed in
the 1984\ and 1994\ papers of Orland \cite{Orland1984,Orland1994}.\
Confining lagrangians are also expressed in terms of Kalb-Ramond fields in
the 1996 paper of Hosek \cite{Ripka96ter}. In 2001, Ellwanger and Wschebor
proposed a similar action to model low energy QCD and they showed that it
confines \cite{Wschebor2001,Wschebor2001bis}. A linear confining potential
was also derived from it in the 2002 paper of Deguchi and Kokubo \cite
{Deguchi2002}.

Let us show that this action is equivalent to the London limit of the
Landau-Ginzburg action. The kinetic term of the action (\ref{ijtheta}) is
often written in the literature in terms of the antisymmetric tensor: 
\begin{equation}
F_{\alpha \beta \gamma }=\partial _\alpha \Theta _{\beta \gamma }+\partial
_\beta \Theta _{\gamma \alpha }+\partial _\gamma \Theta _{\alpha \beta
}\;\;\;\;\;\;\;\;F^2\equiv \frac 16F_{\alpha \beta \gamma }F^{\alpha \beta
\gamma }
\end{equation}
It is simple to check that: 
\begin{equation}
-\frac 12\left( \partial \cdot \overline{\Theta }\right) ^2=\frac
1{12}F_{\alpha \beta \gamma }F^{\alpha \beta \gamma }\equiv \frac 12F^2
\end{equation}
so that the action (\ref{ijtheta}) is often written in the form \cite
{Ramond1974}: 
\begin{equation}
I_j\left( \Theta \right) =\int d^4x\left( \frac 12F^2-\frac{m^2}2\Theta
^2-mG\Theta \right)  \label{tthet}
\end{equation}

\subsection{The double gauge invariance}

The action (\ref{kalbramond}) is invariant under the usual abelian gauge
transformation: 
\begin{equation}
A\rightarrow A+\left( \partial \alpha \right)  \label{gauge1bis}
\end{equation}
It is however, also invariant under the joint gauge transformation: 
\begin{equation}
A\rightarrow A+L\;\;\;\;\;\;\Phi \rightarrow \Phi +\frac 1m\left( \partial
\wedge L\right)  \label{gauge2bis}
\end{equation}
because $\partial \cdot \overline{\partial \wedge L}=0$. In the presence of
the sources $j^\mu $ and $J^{\mu \nu }$, the double gauge invariance is
maintained, provided that the sources satisfy the (compatible) equations: 
\begin{equation}
\partial \cdot j=0\;\;\;\;\;\;\;\;\;\partial \cdot G=j  \label{sourcerel}
\end{equation}
The second gauge invariance relates the source term $J$ to the electric
current $j$. We can choose $L=-A$ so as to write the action in a form in
which the double gauge invariance is explicit: 
\begin{equation}
I_j\left( A,\Phi \right) =\int d^4x\left( -\frac 12\left( \partial \cdot 
\bar{\Phi}\right) ^2-\frac 12\left( \partial \wedge A-m\Phi \right)
^2+G\left( \partial \wedge A-m\Phi \right) \right)  \label{kbinvar}
\end{equation}
We can define an antisymmetric field $\Theta ^{\mu \nu }$: 
\begin{equation}
\partial \wedge A-m\Phi =-m\Theta \;\;\;\;\;\;\Theta =\Phi -\frac 1m\partial
\wedge A
\end{equation}
The field $\Theta ^{\mu \nu }$ has the property of being invariant under the
gauge transformation (\ref{gauge2bis}). Noting that $\partial \cdot \bar{\Phi%
}=\partial \cdot \overline{\Theta }$, the action (\ref{kbinvar}) can be
written as a functional of $\Theta $ alone, namely: 
\begin{equation}
I_j\left( \Theta \right) =\int d^4x\left( -\frac 12\left( \partial \cdot 
\overline{\Theta }\right) ^2-\frac{m^2}2\Theta ^2-mG\Theta \right)
\label{ijtheta}
\end{equation}
This is equivalent to a choice of gauge in which the gauge field $A^\mu $ is
absorbed by the field $\Theta ^{\mu \nu }$.

We can use (\ref{apvt:e}) to write the action in the form: 
\begin{equation}
I_j\left( \Theta \right) =\int d^4x\left( \frac 12\Theta \left( E\partial
^2-m^2\right) \Theta -mG\Theta \right) =\int d^4x\left( \frac 12\Theta
\left( E\left( \partial ^2+m^2\right) -m^2K\right) \Theta -mG\Theta \right)
\end{equation}
where $K$ and $E$ are the longitudinal and transverse projectors (\ref
{apvt:kedef}). The action is stationary with respect to variations of $%
\Theta $ if: 
\begin{equation}
\left( E\left( \partial ^2+m^2\right) -m^2K\right) \Theta
=mG\;\;\;\;\;\;\Theta =\left( \frac E{\partial ^2+m^2}-\frac K{m^2}\right) mG
\end{equation}
We can eliminate $\Theta $ from the action, to get: 
\[
I_j=-\int d^4x\frac 12G\left( \frac{m^2E}{\partial ^2+m^2}-K\right) G 
\]
\begin{equation}
=\int d^4x\left( -\frac 12\left( \partial \cdot \bar{G}\right) \frac
1{\left( \partial ^2+m^2\right) }\left( \partial \cdot \bar{G}\right) -\frac
12\bar{G}^2\right)  \label{ijgbar}
\end{equation}

This form is identical to the expression (\ref{ijgg}) obtained in the London
limit of the Landau-Ginzburg model, with $m=m_{V}$. The source term $G$
satisfies the equation $\partial \cdot G=j$ in both cases, and a straight
line string $G=\frac{1}{\left( n\cdot \partial \right) }\left( n\wedge
j\right) $ leads to the same confining potential (\ref{vlr1}). This is not a
coincidence, because we shall next display a so-called \emph{duality
transformation}, by means of which the Landau-Ginzburg model can be
expressed in terms of a Kalb-Ramond field (not only in the London limit).

\subsection{The duality transformation}

Let us show that the Landau-Ginzburg model can be expressed in terms of an
antisymmetric tensor field $\Theta $. For this purpose, we add to the
Landau-Ginzburg action (\ref{landsphi}) the term: 
\begin{equation}
\frac{m^2}2\left( \bar{\Theta}+\frac 1m\left( \partial \wedge B+\bar{G}%
\right) \right) ^2
\end{equation}
where $m$ is a constant mass. Adding such a term is permissible, because the
equation of motion of the field $\Theta $ simply makes the added term
vanish. The resulting action is: 
\[
I_j\left( \Theta ,B,S,\varphi \right) =\int d^4x\left\{ -\frac 12\left(
\partial \wedge B+\bar{G}\right) ^2+\frac{g^2S^2}2\left( B+\partial \varphi
\right) ^2+\frac 12\left( \partial S\right) ^2-\frac 12b\left(
S^2-v^2\right) ^2\right. 
\]
\begin{equation}
\left. +\frac{m^2}2\left( \bar{\Theta}+\frac 1m\left( \partial \wedge B+\bar{%
G}\right) \right) ^2\right\}
\end{equation}
The added term is chosen such that the term $-\frac 12\left( \partial \wedge
B+\bar{G}\right) ^2$ cancels.\ We are left with the action: 
\[
I_j\left( \Theta ,B,S,\varphi \right) =\int d^4x 
\]
\begin{equation}
\left\{ \frac{m^2}2\bar{\Theta}^2+m\bar{\Theta}\left( \partial \wedge
B\right) +m\bar{G}\bar{\Theta}+\frac{g^2S^2}2\left( B+\partial \varphi
\right) ^2+\frac 12\left( \partial S\right) ^2-\frac 12b\left(
S^2-v^2\right) ^2\right\}
\end{equation}
The identities (\ref{apvt:sda}) and (\ref{apvt:dds}) allow us to write: 
\begin{equation}
-m\int d^4xC\left( \partial \cdot \bar{\Theta}\right) =m\int d^4x\left(
B+\partial \varphi \right) \left( \partial \cdot \bar{\Theta}\right)
\end{equation}
so that the action can be cast into the form: 
\[
I_j\left( \Theta ,B,S,\varphi \right) =\int d^4x\left\{ \frac{g^2S^2}2\left(
B+\partial \varphi -m\frac{\left( \partial \cdot \bar{\Theta}\right) }{g^2S^2%
}\right) ^2\right. 
\]
\begin{equation}
\left. -\frac{m^2}{2g^2S^2}\left( \partial \cdot \bar{\Theta}\right) ^2+%
\frac{m^2}2\overline{\Theta }+m\bar{G}\overline{\Theta }^2+\frac 12\left(
\partial S\right) ^2-\frac 12b\left( S^2-v^2\right) ^2\right\}
\end{equation}
The first term of the action can be dropped because it vanishes when the
equation of motion for the field $B$ is satisfied. The remaining action is: 
\begin{equation}
I_j\left( \Theta ,S\right) =\int d^4x\left( -\frac{m^2}{2g^2S^2}\left(
\partial \cdot \bar{\Theta}\right) ^2+\frac{m^2}2\overline{\Theta }+m\bar{G}%
\overline{\Theta }^2+\frac 12\left( \partial S\right) ^2-\frac 12b\left(
S^2-v^2\right) ^2\right)
\end{equation}
This action, expressed in terms of the Kalb-Ramond field $\Theta $, is
identical to the action (\ref{landsphi}) of the Landau-Ginzburg model. In
the London limit, where $S=v$, we can choose $m=v$ and the action becomes
identical to the Kalb-Ramond action (\ref{ijtheta}).

The transformation of the action $I_j\left( \Theta ,S\right) $, expressed in
terms of a Kalb-Ramond field $\Theta ^{\mu \nu }$, into an action $I_j\left(
\Theta ,B,S,\varphi \right) $ which depends on the ''dual'' vector field $%
B^\mu $ is called a duality transformation. It is widely used in the
literature \cite{Antonov1998-2,Antonov1999},\cite
{Wschebor2001bis,Deguchi2002}. \cite{Wschebor2001,Wschebor2001bis}.

\subsection{The quantification of the massive Kalb-Ramond field}

To visualize the physical content of the action (\ref{ijtheta}), let us
express the Kalb Ramond field $\Theta ^{\mu \nu }$ in terms of two cartesian
3-dimensional fields $\vec{e}$ and $\vec{h}$: 
\begin{equation}
e^i=-\Theta ^{0i}=\frac 12\varepsilon ^{0ijk}\overline{\Theta }%
_{jk}\;\;\;\;\;\;\;\;\;\;\;h^i=-\overline{\Theta }^{0i}=-\frac 12\varepsilon
^{0ijk}\Theta _{jk}
\end{equation}

It is an easy exercise to check that: 
\begin{equation}
-\frac 12\left( \partial \cdot \overline{\Theta }\right) ^2=-\frac 12\left( 
\vec{\nabla}\cdot \vec{h}\right) ^2+\frac 12\left( \partial _t\vec{h}+\vec{%
\nabla}\times \vec{e}\right) ^2  \label{dphidual}
\end{equation}
The source term $G^{\mu \nu }$ can, in turn, be written in terms of the two
cartesian 3-dimensional vectors $\vec{E}^{st}$ and $\vec{H}^{st}$ defined in
(\ref{ehgstring}): 
\begin{equation}
E_{st}^i=-G^{0i}=\frac 12\varepsilon ^{0ijk}\bar{G}_{jk}\;\;\;\;\;%
\;H_{st}^i=-\frac 12\varepsilon ^{0ijk}G_{jk}=-\bar{G}^{0i}
\end{equation}
\begin{equation}
G\cdot \Phi =-\vec{e}\cdot \vec{E}^{st}+\vec{h}\cdot \vec{H}^{st}\;\;\;\;\;\;%
\vec{\nabla}\cdot \vec{E}^{st}=\rho \;\;\;\;\;\;-\partial _t\vec{E}^{st}-%
\vec{\nabla}\times \vec{H}^{st}=\vec{j}
\end{equation}
where $j^\mu =\left( \rho ,\vec{j}\right) $ is the electric current.

The action (\ref{ijtheta}) can thus be broken down to: 
\[
I_{\rho ,\vec{j}}\left( \vec{e},\vec{h}\right) =\int d^4x\left\{ \frac
12\left( \partial _t\vec{h}+\vec{\nabla}\times \vec{e}\right) ^2+\frac{m^2}2%
\vec{e}^2-\frac 12\left( \vec{\nabla}\cdot \vec{h}\right) ^2-\frac{m^2}2\vec{%
h}^2\right. 
\]
\begin{equation}
\left. +m\vec{e}\cdot \vec{E}^{st}-m\vec{h}\cdot \vec{H}^{st}\right\}
\label{irhoj}
\end{equation}
No time derivative acts on the field $\vec{e}$ so that it acts as the
constraint: 
\begin{equation}
\vec{\nabla}\times \left( \partial _t\vec{h}+\vec{\nabla}\times \vec{e}%
\right) +m^2\vec{e}=-m\vec{E}^{st}  \label{ehconstraint}
\end{equation}
The conjugate of the field $\vec{h}$ is: 
\begin{equation}
\vec{\pi}=\partial _t\vec{h}+\vec{\nabla}\times \vec{e}
\end{equation}
and the conjugate of the field $S$ is: 
\begin{equation}
P=\left( \partial _tS\right)
\end{equation}
Taking the constraint (\ref{ehconstraint}) into account, the hamiltonian, or
classical energy, is: 
\[
\mathcal{H}\left( \vec{\pi},\vec{h}\right) =\int d^3r\left( \vec{\pi}\cdot
\left( \partial _t\vec{h}\right) +P\left( \partial _tS\right) -L\right) 
\]
\begin{equation}
=\int d^3r\left[ \frac{\vec{\pi}^2}2+\frac 1{2m^2}\left( \vec{\nabla}\times 
\vec{\pi}+m\vec{E}^{st}\right) ^2+\frac 12\left( \vec{\nabla}\cdot \vec{h}%
\right) ^2+\frac 12m^2\vec{h}^2+m\vec{h}\cdot \vec{H}^{st}\right]
\label{pihps}
\end{equation}
In the absence of sources, the classical energy is a sum of positive terms.

\subsection{The elementary excitations}

Let us expand the hamiltonian (\ref{pihps}) to second order in the fields
around their vacuum values $S=v$, $\vec{\pi}=\vec{h}=0$, in the absence of
sources. The Hamiltonian reduces to: 
\[
\mathcal{H}\left( \vec{\pi},\vec{h}\right) =\int d^3r\left[ \frac{\vec{\pi}^2%
}2+\frac 1{2m^2}\left( \vec{\nabla}\times \vec{\pi}\right) ^2+\frac 12\left( 
\vec{\nabla}\cdot \vec{h}\right) ^2+\frac 12m^2\vec{h}^2\right] 
\]

\begin{equation}
=\int d^3r\left[ \frac 1{2m^2}\vec{\pi}_T\left( -\nabla ^2+m^2\right) \vec{%
\pi}_T+\frac 12m^2\vec{h}_T^2+\frac{\vec{\pi}_L^2}2+\frac 12\vec{h}_L\left(
-\nabla ^2+m^2\right) \vec{h}_L\right]  \label{hpih}
\end{equation}
where we used (\ref{apvt:dddc}).

As in Sect.\ref{sec:landexcite}, we expand the field $\vec{h}$ and its
conjugate $\vec{\pi}$ on the plane wave basis (\ref{rika}): 
\[
h^{i}\left( \vec{r}\right) =\frac{1}{\sqrt{2}}\sum_{ka}\alpha _{ka}\left(
\left\langle \vec{r}i\left| \vec{k}a\right. \right\rangle a_{\vec{k}a}+a_{%
\vec{k}a}^{\dagger }\left\langle \vec{k}a\left| \vec{r}i\right.
\right\rangle \right) =\vec{h}_{i}^{L}\left( \vec{r}\right) +\vec{h}%
_{i}^{T}\left( \vec{r}\right) 
\]
\begin{equation}
\pi ^{i}\left( \vec{r}\right) =\frac{1}{i\sqrt{2}}\sum_{ka}\frac{1}{\alpha
_{ka}}\left( \left\langle \vec{r}i\left| \vec{k}a\right. \right\rangle a_{%
\vec{k}a}-a_{\vec{k}a}^{\dagger }\left\langle \vec{k}a\left| \vec{r}i\right.
\right\rangle \right) =\vec{\pi}_{i}^{L}\left( \vec{r}\right) +\vec{\pi}%
_{i}^{T}\left( \vec{r}\right)  \label{hpiexp}
\end{equation}
where the transverse and longitudinal parts correspond respectively to the $%
a=1,2$ and $a=3$ contributions. If we assume boson commutation rules $\left[
a_{\vec{k}a},a_{\vec{k}^{\prime }b}^{\dagger }\right] =\delta _{\vec{k},\vec{%
k}^{\prime }}\delta _{ab}$, the fields $\vec{h}$ and $\vec{\pi}$ become
quantized: 
\begin{equation}
\left[ \pi ^{i}\left( \vec{r}\right) ,h^{j}\left( \vec{r}^{\prime }\right)
\right] =\delta ^{ij}\delta \left( \vec{r}-\vec{r}^{\prime }\right)
\end{equation}
Upon substitution of the expansion (\ref{hpiexp}) into the hamiltonian (\ref
{hpih}), we find that the following coefficients $\alpha _{ka}$ cancel the $%
\left( a_{\vec{k}a}a_{-\vec{k}a}+a_{\vec{k}a}^{\dagger }a_{-\vec{k}%
a}^{\dagger }\right) $ terms in the hamiltonian: 
\begin{equation}
\alpha _{k,a=1,2}=\frac{1}{\left( \vec{k}^{2}+m^{2}\right) ^{\frac{1}{4}}}%
\;\;\;\;\;\;\alpha _{k,a=3}=\frac{1}{m}\left( \vec{k}^{2}+m^{2}\right) ^{%
\frac{1}{4}}
\end{equation}
With these coefficients, the hamiltonian (\ref{hpih}) acquires the diagonal
form: 
\begin{equation}
\mathcal{H}=\sum_{\vec{k}a}\sqrt{\vec{k}^{2}+m^{2}}\left( a_{\vec{k}^{\prime
}b}^{\dagger }a_{\vec{k}a}+\frac{1}{2}\right)
\end{equation}
so that the elementary excitations of the massive Kalb-Ramond hamiltonian
consists of three vector particles of mass $m$, which is exactly the same as
the elementary excitations of the Landau-Ginzburg model in the London limit.
However, in the limit $m\rightarrow 0$ of vanishing mass, the transverse
field disappears and only one particle of mass $m$ subsists.

\begin{itemize}
\item  \textbf{Exercise}: Let $\bar{F}_\mu $ be the vector which is the dual
form of the tensor $F_{\alpha \beta \gamma }$: 
\begin{equation}
\bar{F}^\mu =\frac 16\varepsilon ^{\mu \alpha \beta \gamma }F_{\alpha \beta
\gamma }\;\;\;\;\;\;F^{\alpha \beta \gamma }=\varepsilon ^{\alpha \beta
\gamma \mu }\bar{F}_\mu \;\;\;\;\;\;\bar{F}^2=-F^2
\end{equation}
Note that the duality transformation $F\rightarrow \bar{F}$ of tensors with
an odd number of indices is reversible without a change in sign (App. \ref
{ap:dualvec}). Show that the components of the vector $\bar{F}^\mu $ are: 
\begin{equation}
\bar{F}^0=-\vec{\nabla}\cdot \vec{h}\;\;\;\;\;\;\bar{F}^i=\partial
_th^i+\left( \vec{\nabla}\times \vec{e}\right) ^i
\end{equation}
Show that, when $m=0$, the lagrangian (\ref{irhoj}) can be expressed in
terms of the four fields $\bar{F}^\mu $. The quantification of the fields in
this limit is discussed in the 1974 paper of Kalb and Ramond \cite
{Ramond1974}.
\end{itemize}

\subsection{The Nambu hierarchy of gauge potentials}

The kinetic term of the Kalb-Ramond action (\ref{tthet}) is written in terms
of an antisymmetric field tensor $F_{\alpha \beta \gamma }$ and its
associated gauge potential $\Theta _{\mu \nu }$. Nambu has displayed a
hierarchy of field tensors $F_{\mu \nu ,}\;F_{\alpha \beta \gamma },...$
which can be expressed in terms of gauge potentials $A_\mu ,\;A_{\mu \nu
},...$ \cite{Nambu1975}: 
\begin{equation}
F_{\mu \nu }=\partial _\mu A_\nu -\partial _\nu A_\mu \;\;\;\;\;F_{\mu \nu
\lambda }=\sum_{cycl}\partial _\mu A_{\nu \lambda }\;\;\;\;\;\;F_{\mu \nu
\lambda \rho }=\sum_{cycl}\partial _\mu A_{\nu \lambda \rho }
\end{equation}
The respective lagrangians: 
\[
L=-\frac 14F_{\mu \nu }F^{\mu \nu }+gA_\mu j^\mu 
\]
\[
L=-\frac 1{12m^2}F_{\mu \nu \lambda }F^{\mu \nu \lambda }+gA_{\mu \nu
}j^{\mu \nu } 
\]
\begin{equation}
L=-\frac 1{48m^2}F_{\mu \nu \lambda \rho }F^{\mu \nu \lambda \rho }+gA_{\mu
\nu \lambda }j^{\mu \nu \lambda }
\end{equation}
are invariant under the gauge transformations: 
\begin{equation}
A_\mu \rightarrow A_\mu +\left( \partial _\mu \Lambda \right)
\;\;\;\;\;\;A_{\mu \nu }\rightarrow A_{\mu \nu }+\left( \partial _\mu
\Lambda _\nu -\partial _\nu \Lambda _\mu \right) \;\;\;\;\;\;A_{\mu \nu
\lambda }\rightarrow A_{\mu \nu \lambda }+\sum_{cycl}\left( \partial _\mu
\Lambda _{\nu \lambda }\right)
\end{equation}
provided that the currents are conserved: 
\begin{equation}
\partial _\mu j^\mu =\partial _\mu j^{\mu \nu }=\partial _\mu j^{\mu \nu
\lambda }=0
\end{equation}

\section{The hamiltonian of the Landau-Ginzburg model}

Consider the action (\ref{actpsi}). In the unitary gauge, $\psi =S$ is a
real field and the action reduces to: 
\[
I_j\left( \vec{B},\chi ,S\right) =\int d^4x\left[ \frac 12\left( -\partial _t%
\vec{B}-\vec{\nabla}\chi +\vec{H}_{st}\right) ^2-\frac 12\left( -\vec{\nabla}%
\times \vec{B}+\vec{E}_{st}\right) ^2\right. 
\]
\begin{equation}
\left. -\frac{g^2S^2}2\vec{B}^2+\frac{g^2S^2}2\chi ^2+\frac 12\left(
\partial _tS\right) ^2-\frac 12\left( \vec{\nabla}S\right) ^2-\frac
12b\left( S^2-v^2\right) ^2\right]
\end{equation}
The constraint imposed by $\chi $ is: 
\begin{equation}
g^2S^2\chi +\vec{\nabla}\cdot \left( -\partial _t\vec{B}-\vec{\nabla}\chi +%
\vec{H}_{st}\right) =0  \label{eqchi}
\end{equation}
The conjugate momentum to the field $\vec{B}$: 
\begin{equation}
\frac{\delta I}{\delta \left( \partial _t\vec{B}\right) }=\left( \partial _t%
\vec{B}+\vec{\nabla}\chi -\vec{H}^{st}\right) =-\vec{H}
\end{equation}
The minus sign on the right hand side is inserted in order to conform with
the definition (\ref{ehbg}) of the magnetic field. The momentum conjugate of
the field $S$ is: 
\begin{equation}
P=\frac{\delta I}{\left( \partial _tS\right) }=\left( \partial _tS\right)
\end{equation}
Taking the constraint (\ref{eqchi}) into account, the hamiltonian, or
classical energy, is: 
\[
\mathcal{H}\left( \vec{H},\vec{B},P,S\right) =\int d^3r\;\left( -\vec{H}%
\left( \partial _t\vec{B}\right) +P\left( \partial _tS\right) \right) -I 
\]
\[
=\int d^3r\left[ \frac 12\vec{H}^2+\frac 12\left( -\vec{\nabla}\times \vec{B}%
+\vec{E}^{st}\right) ^2+\frac{g^2S^2}2\vec{B}^2+\frac 1{2g^2S^2}\left( \vec{%
\nabla}\cdot \vec{H}\right) ^2-\vec{H}\cdot \vec{H}^{st}\right. 
\]
\begin{equation}
\left. +\frac 12P^2+\frac 12\left( \vec{\nabla}S\right) ^2+\frac 12b\left(
S^2-v^2\right) ^2\right]  \label{hcps}
\end{equation}
In the absence of sources, the energy is a sum of positive terms.

\section{The elementary excitations of the Landau-Ginzburg model}

\label{sec:landexcite}

Vector fields can be expanded in the plane wave basis:

\begin{equation}
\left\langle \vec{r}i\left| \vec{k}a\right. \right\rangle =\frac 1{\sqrt{V}%
}e^{i\vec{k}\cdot \vec{r}}e_{\left( a\right) }^i\left( \vec{k}\right)
\;\;\;\;\;\;\left\langle \vec{k}a\left| \vec{r}i\right. \right\rangle =\frac
1{\sqrt{V}}e^{-i\vec{k}\cdot \vec{r}}e_{\left( a\right) }^i\left( \vec{k}%
\right)  \label{rika}
\end{equation}
where the vectors $\vec{e}_{\left( a=1,2,3\right) }$ are orthogonal unit
vectors with $\vec{e}_{\left( 3\right) }$ parallel to $\vec{k}$: 
\begin{equation}
\vec{e}_{\left( 3\right) }=\frac{\vec{k}}k\;\;\;\;\;\;\vec{e}_{\left(
a\right) }\times \vec{e}_{\left( b\right) }=\varepsilon _{abc}\vec{e}%
_{\left( c\right) }\;\;\;\;\;\;\left( \varepsilon _{123}=-\varepsilon
_{213}=1\right)
\end{equation}
and where $V$ is the volume in which the plane waves are normalized. The
basis (\ref{rika}) is complete and orthogonal: 
\begin{equation}
\sum_{\vec{k}a}\left\langle \vec{r}i\left| \vec{k}a\right. \right\rangle
\left\langle \vec{k}a\left| \vec{r}^{\prime }j\right. \right\rangle =\delta
\left( \vec{r}-\vec{r}^{\prime }\right) \delta _{ij}\;\;\;\;\;\;\int
d^3r\sum_i\left\langle \vec{k}a\left| \vec{r}i\right. \right\rangle
\left\langle \vec{r}i\left| \vec{k}^{\prime }b\right. \right\rangle =\delta
_{\vec{k}\vec{k}^{\prime }}\delta _{ab}
\end{equation}
The fields $\vec{B}$ and $\vec{H}$ can be expanded in this basis: 
\[
B^i\left( \vec{r}\right) =\frac 1{\sqrt{2}}\sum_{ka}\alpha _{ka}\left(
\left\langle \vec{r}i\left| \vec{k}a\right. \right\rangle a_{\vec{k}a}+a_{%
\vec{k}a}^{\dagger }\left\langle \vec{k}a\left| \vec{r}i\right.
\right\rangle \right) =C_L^i\left( \vec{r}\right) +C_T^i\left( \vec{r}%
\right) 
\]
\begin{equation}
H^i\left( \vec{r}\right) =\frac 1{\sqrt{2}}\sum_{ka}\frac i{\alpha
_{ka}}\left( \left\langle \vec{r}i\left| \vec{k}a\right. \right\rangle a_{%
\vec{k}a}-a_{\vec{k}a}^{\dagger }\left\langle \vec{k}a\left| \vec{r}i\right.
\right\rangle \right) =H_L^i\left( \vec{r}\right) +H_T^i\left( \vec{r}\right)
\label{chexp}
\end{equation}
The longitudinal and transverse parts $\vec{B}_L$ and $\vec{B}_T$ of $\vec{B}
$ correspond respectively to the contributions of $a=3$ and $a=1,2$. We have 
$\vec{\nabla}\times \vec{B}_L=0$ and $\vec{\nabla}\cdot \vec{B}_T=0$.
Similarly for the longitudinal and transverse parts of $\vec{H}$.

If we impose boson commutation rules on the $a_{ka}$ and $a_{ka}^{\dagger }$
coefficients: 
\begin{equation}
\left[ a_{ka},a_{k^{\prime }b}^{\dagger }\right] =\delta _{\vec{k}\vec{k}%
^{\prime }}\delta _{ab}
\end{equation}
the fields $\vec{B}$ and $\vec{H}$ become quantized: 
\begin{equation}
\left[ H^i\left( \vec{r}\right) ,B^j\left( \vec{r}^{\prime }\right) \right]
=i\left\langle \vec{r}i\left| \vec{r}^{\prime }j\right. \right\rangle
=i\delta ^{ij}\delta \left( \vec{r}-\vec{r}^{\prime }\right)
\end{equation}
The scalar field may be similarly quantized: 
\[
S\left( \vec{r}\right) =\frac 1{\sqrt{2}}\sum_k\alpha _k\left( \left\langle 
\vec{r}\left| \vec{k}\right. \right\rangle b_{\vec{k}}+b_{\vec{k}}^{\dagger
}\left\langle \vec{k}\left| \vec{r}\right. \right\rangle \right)
\;\;\;\;\;\;\;P\left( \vec{r}\right) =\frac 1{i\sqrt{2}}\sum_k\alpha
_k\left( \left\langle \vec{r}\left| \vec{k}\right. \right\rangle b_{\vec{k}%
}-b_{\vec{k}}^{\dagger }\left\langle \vec{k}\left| \vec{r}\right.
\right\rangle \right) 
\]
\begin{equation}
\left[ P\left( \vec{r}\right) ,S\left( \vec{r}^{\prime }\right) \right]
=\left\langle \left. \vec{r}\right| \vec{r}^{\prime }\right\rangle =\delta
\left( \vec{r}-\vec{r}^{\prime }\right)
\end{equation}

The elementary excitations of the vacuum, in the Landau-Ginzburg model, may
be obtained by expanding the fields to second order in the vicinity of their
vacuum values $S=v$, $\vec{B}=\vec{H}=0$ and in setting the sources $\vec{E}%
_{st}$ and $\vec{H}_{st}$ to zero. To second order in the fluctuating parts,
the hamiltonian can be written in the form: 
\[
\mathcal{H}\left( \vec{H},\vec{C},P,S\right) =\int d^{3}r\left[ +\frac{1}{2}%
\vec{H}_{T}^{2}+\frac{1}{2}\vec{B}^{T}\left( -\vec{\nabla}%
^{2}+m_{V}^{2}\right) \vec{B}^{T}+\frac{1}{2m_{V}^{2}}\vec{H}^{L}\left(
-\nabla ^{2}+m_{V}^{2}\right) \vec{H}^{L}+\frac{m_{V}^{2}}{2}\vec{B}%
_{L}^{2}\right. 
\]
\begin{equation}
\left. +\frac{1}{2}P^{2}+\frac{1}{2}\left( \vec{\nabla}S\right) ^{2}+\frac{1%
}{2}m_{H}^{2}\left( S-v\right) ^{2}\right]  \label{hlt}
\end{equation}
where we used (\ref{apvt:dddc}) and where $m_{H}$ and $m_{V}$ are the Higgs
and vector masses (\ref{mh}) and (\ref{mv}). If we substitute the expansions
(\ref{chexp}) into the hamiltonian (\ref{hlt}), we can choose the
coefficients $\alpha _{\vec{k}a}$ such that the terms $\left( a_{\vec{k}%
a}a_{-\vec{k}a}+a_{\vec{k}a}^{\dagger }a_{-\vec{k}a}^{\dagger }\right) $
vanish, and the hamiltonian reduces to: 
\begin{equation}
\mathcal{H}=\sum_{\vec{k}a}\sqrt{\vec{k}^{2}+m_{V}^{2}}\left( a_{\vec{k}%
a}^{\dagger }a_{\vec{k}a}+\frac{1}{2}\right) +\sum_{\vec{k}}\sqrt{\vec{k}%
^{2}+m_{H}^{2}}\left( b_{\vec{k}}^{\dagger }b_{\vec{k}}+\frac{1}{2}\right)
\end{equation}
The elementary excitations of the Landau-Ginzburg model consists of three
vector particles with mass $m_{V}$ and a scalar particle with mass $m_{H}$.

The appearance of massive vector particles with masses barely higher than $%
1\;GeV$ (see the values of the estimated vector and Higgs masses in Sect. 
\ref{sec:balifit}) has been invoked to criticize the dual superconductor
model. Indeed, as we shall see in Sect. \ref{sec:su3lg}, the Higgs field is
not a color singlet so that the model predicts the existence of freely
propagating particles with non-vanishing color.

\begin{itemize}
\item  \textbf{Exercise}: show that the coefficients $\alpha _{ka}$ are: 
\begin{equation}
\alpha _{k,a=1,2}=\left( \vec{k}^2+m_V^2\right) ^{\frac
14}\;\;\;\;\;\;\alpha _{k,a=3}=\frac 1m\left( \vec{k}^2+m_V^2\right) ^{\frac
14}
\end{equation}
Show that, in the limit $m_V\rightarrow 0$, the field $\vec{H}_L$ vanishes
and that $\vec{\nabla}\cdot \vec{B}=0$ so that only two massless vector
particles remain.
\end{itemize}

\section{The two-potential Zwanziger formalism}

\label{sec:zwanziger}

In sections \ref{sec:adynamics} and \ref{sec:bdynamics}, we saw that, in the
presence of both electric and magnetic currents, electrodynamics could be
expressed either in terms of a gauge potential $A^{\mu }$ associated to the
field strength $F=\partial \wedge A-\bar{G}$ or a gauge potential $B^{\mu }$
associated to the dual field strength $\bar{F}=\partial \wedge B+\bar{G}$.
The corresponding actions (\ref{iajj}) and (\ref{ibjj}) involve non-local
string terms. In 1971, Zwanziger wrote a beautiful paper in which he
proposed a \emph{local} lagrangian which uses two potentials $A^{\mu }$ and $%
B^{\mu }$ \cite{Zwanziger1971}. We shall apply his formalism to the
Landau-Ginzburg action of a dual superconductor. The theory was further
developed, in particular its Lorentz invariance, by several authors.\ For a
review see the 1979\ paper of Brandt, Neri and Zwanziger \cite{Zwanziger1979}%
. The Lorentz invariance is also discussed in the more recent 1998\ paper of
Gubarev, Polikarpov and Zakharov \cite{Polikarpov1998}. Other aspects of the
formalism are discussed in the 1976\ and 1979\ papers of Blagojevic and
Senjanovic \cite{Senjanovic1976,Senjanovic1979}. We shall show that, when
applied to a dual superconductor, the Zwanziger action leads to the form (%
\ref{landsphi})\ of the Landau-Ginzburg action.

\subsection{The field tensor $F^{\mu \nu }$ expressed in terms of two
potentials $A^{\mu }$ and $B^{\mu }$}

Zwanziger writes the components $n\cdot F$ and $n\cdot \bar{F}$ of the field
strength tensor, along a given fixed 4-vector $n^\mu $, in terms of two
potentials, namely: 
\begin{equation}
n\cdot F=n\cdot \left( \partial \wedge A\right) \;\;\;\;\;\;n\cdot \bar{F}%
=n\cdot \left( \partial \wedge B\right)  \label{nfab}
\end{equation}
He then expresses the field strength tensor $F$ in terms of these
components, using the identity (\ref{apvt:zwanident1}): 
\[
F=\frac 1{n^2}\left( n\wedge \left( n\cdot F\right) -\overline{n\wedge
\left( n\cdot \overline{F}\right) }\right) 
\]
\begin{equation}
=\frac 1{n^2}\left( n\wedge \left( n\cdot \left( \partial \wedge A\right)
\right) -\overline{n\wedge \left( n\cdot \left( \partial \wedge B\right)
\right) }\right)  \label{fab}
\end{equation}
and the dual tensor is: 
\begin{equation}
\bar{F}=\frac 1{n^2}\left( \overline{n\wedge \left( n\cdot \left( \partial
\wedge A\right) \right) }+n\wedge \left( n\cdot \left( \partial \wedge
B\right) \right) \right)  \label{fdab}
\end{equation}
The Maxwell equations $\partial \cdot F=j$ and $\partial \cdot \bar{F}%
=j_{mag}$ can be satisfied with potentials $A^\mu $ and $B^\mu $ which
satisfy the equations: 
\[
\frac 1{n^2}\left( \left( n\cdot \partial \right) ^2A-\left( n\cdot \partial
\right) \partial \left( n\cdot A\right) -n\left( n\cdot \partial \right)
\left( \partial \cdot A\right) +n\partial ^2\left( n\cdot A\right) +\left(
n\cdot \partial \right) \left( n\cdot \overline{\partial \wedge B}\right)
\right) =j 
\]
\begin{equation}
\frac 1{n^2}\left( \left( n\cdot \partial \right) ^2B-\left( n\cdot \partial
\right) \partial \left( n\cdot B\right) -n\left( n\cdot \partial \right)
\left( \partial \cdot B\right) +n\partial ^2\left( n\cdot B\right) -\left(
n\cdot \partial \right) \left( n\cdot \overline{\partial \wedge A}\right)
\right) =j_{mag}  \label{maxab}
\end{equation}
where we used successively (\ref{apvt:vec3}), (\ref{apvt:vec2}) and (\ref
{apvt:vec6}).

\begin{itemize}
\item  \textbf{Exercise}: Choose $n^\mu $ to be space-like and define: 
\begin{equation}
A^\mu =\left( \phi ,\vec{A}\right) \;\;\;\;\;\;\;\;B^\mu =\left( \chi ,\vec{B%
}\right) \;\;\;\;\;\;n^\mu =\left( 0,\vec{n}\right) \;\;\;\;\;\;\vec{n}\cdot 
\vec{n}=1
\end{equation}
Show that (\ref{fab}) and (\ref{fdab}) break down to: 
\[
-\left( n\cdot F\right) ^0=\vec{n}\cdot \vec{E}=-\left[ n\cdot \left(
\partial \wedge A\right) \right] ^0=\vec{n}\cdot \left( -\vec{\nabla}\phi
-\partial _t\vec{A}\right) 
\]
\[
\left( n\cdot F\right) ^i=-\left( \vec{n}\times \vec{H}\right) _i=\left[
n\cdot \left( \partial \wedge A\right) \right] ^i=-\left( \vec{n}\times
\left( \vec{\nabla}\times \vec{A}\right) \right) _i 
\]
\[
\left( n\cdot \overline{F}\right) ^0=-\vec{n}\cdot \vec{H}=\left[ n\cdot
\left( \partial \wedge B\right) \right] ^0=\vec{n}\cdot \left( -\vec{\nabla}%
\chi -\partial _t\vec{B}\right) 
\]
\begin{equation}
\left( n\cdot \overline{F}\right) ^i=\left( \vec{n}\times \vec{E}\right)
_i=\left[ n\cdot \left( \partial \wedge B\right) \right] ^i=\left( \vec{n}%
\times \left( \vec{\nabla}\times \vec{B}\right) \right) _i
\end{equation}
so that the potential $\vec{A}$ describes the longitudinal part of the
electric field and the transverse part of the magnetic field, whereas the
potential $\vec{B}$ describes the longitudinal part of the magnetic field
and the transverse part of the electric field. In this instance, the
longitudinal and transverse parts of the fields are defined relative to the
vector $\vec{n}$ and \emph{not} relative to $\vec{\nabla}$ as in (\ref
{apvt:dddc}).
\end{itemize}

\subsection{The Zwanziger action applied to a dual superconductor}

The Zwanziger action applied to a dual superconductor is: 
\[
I_{j}\left( A,B,S,\varphi \right) =\int d^{4}x\left[ -\frac{1}{2n^{2}}\left(
n\cdot \left( \partial \wedge A\right) \right) ^{2}-j\cdot A\right. 
\]
\[
-\frac{1}{2n^{2}}\left( n\cdot \partial \wedge A\right) \cdot \left( n\cdot 
\overline{\partial \wedge B}\right) +\frac{1}{2n^{2}}\left( n\cdot \partial
\wedge B\right) \cdot \left( n\cdot \overline{\partial \wedge A}\right) 
\]
\[
-\frac{1}{2n^{2}}\left( n\cdot \left( \partial \wedge B\right) \right) ^{2}+%
\frac{g^{2}S^{2}}{2}\left( B+\partial \varphi \right) ^{2} 
\]
\begin{equation}
\left. +\frac{1}{2}\left( \partial S\right) ^{2}-\frac{1}{2}b\left(
S^{2}-v^{2}\right) ^{2}\right]  \label{zwanlg}
\end{equation}
The first line describes the gauge field $A^{\mu }$ and its coupling to the
electric current $j^{\mu }$.\ The second line is an interaction between the
two gauge potentials $A^{\mu }$ and $B^{\mu }$. The third line describes the
interaction of the gauge field $B^{\mu }$ with the complex scalar field $%
Se^{ig\varphi }$ of the Landau-Ginzburg model.\ The last term describes the
dynamics of the order parameter $S$.\ 

The action (\ref{zwanlg}) is invariant under the following gauge
transformation of the field $A^{\mu }$: 
\begin{equation}
A\rightarrow A+\left( \partial \alpha \right)
\end{equation}
It is also invariant under the joint gauge transformation: 
\begin{equation}
B\rightarrow B+\left( \partial \beta \right) \;\;\;\;\;\;\varphi \rightarrow
\varphi -\beta
\end{equation}
The electric current $j^{\mu }$ might be caused by quarks, in which case the
term $-j\cdot A$ would be replaced by the term: 
\begin{equation}
\bar{q}\left[ i\left( \partial ^{\mu }+ieA\right) \gamma _{\mu }-m\right] q
\end{equation}
When this is done, the action becomes a local action for a system of
confined Dirac particles with electric charges.

The following are useful identities: 
\[
\int d^4x\;\left( n\cdot \partial \wedge A\right) \cdot \left( n\cdot 
\overline{\partial \wedge B}\right) =-\int d^4x\;\left( n\cdot \partial
\wedge B\right) \cdot \left( n\cdot \overline{\partial \wedge A}\right) 
\]
\[
=\int d^4x\;\varepsilon ^{\mu \nu \alpha \beta }A_\mu n_\nu \left( n\cdot
\partial \right) \partial _\alpha B_\beta 
\]
\begin{equation}
=-\int d^4x\;A\left( n\cdot \partial \right) \left[ n\cdot \overline{%
\partial \wedge B}\right] =+\int d^4x\;B\left( n\cdot \partial \right)
\left[ n\cdot \overline{\partial \wedge A}\right]  \label{idab}
\end{equation}
and: 
\[
-\int d^4x\;\left[ n\cdot \left( \partial \wedge A\right) \right] \left[
n\cdot \left( \partial \wedge A\right) \right] 
\]
\begin{equation}
=\int d^4x\left[ A\left( n\cdot \partial \right) ^2A-A\left( n\cdot \partial
\right) \partial \left( n\cdot A\right) -An\left( n\cdot \partial \right)
\left( \partial \cdot A\right) +\left( A\cdot n\right) \partial ^2\left(
n\cdot A\right) \right]  \label{idaa}
\end{equation}

\subsection{Elimination of the gauge potential $A^{\mu }$}

\label{sec:elima}

Let us show how to eliminate the field $A^\mu $ in order to reduce the
action (\ref{zwanlg}) to the form (\ref{landsphi})\ of the Landau-Ginzburg
action. Because the theory is invariant under the gauge transformation $%
A\rightarrow A+\left( \partial \alpha \right) $, we can, following
Zwanziger, add a gauge fixing term $\frac 1{2n^2}\left[ \partial \left(
n\cdot A\right) \right] ^2$ to the action. After doing this, and in view of
the identities (\ref{idab}) and (\ref{idaa}), the action (\ref{zwanlg}) can
be written in the form: 
\[
I_j\left( A,B,S,\varphi \right) =\int d^4x\left[ \frac 12A_\mu M^{\mu \nu
}A_\nu +A_\mu \left[ \frac 1{n^2}\left( n\cdot \partial \right) \left(
n\cdot \overline{\partial \wedge B}\right) ^\mu -j^\mu \right] -\frac
1{2n^2}\left( n\cdot \left( \partial \wedge B\right) \right) ^2\right. 
\]
\begin{equation}
\left. +\frac{g^2S^2}2\left( B+\partial \varphi \right) ^2+\frac 12\left(
\partial S\right) ^2-\frac 12b\left( S^2-v^2\right) ^2\right]
\end{equation}
where $M^{\mu \nu }$ is the matrix: 
\begin{equation}
M^{\mu \nu }=\frac 1{n^2}\left( n\cdot \partial \right) \left[ \left( n\cdot
\partial \right) g^{\mu \nu }-\partial ^\mu n^\nu -n^\mu \partial ^\nu
\right]
\end{equation}
The inverse matrix is: 
\begin{equation}
M_{\mu \nu }^{-1}=\frac{n^2}{\left( n\cdot \partial \right) ^2}\left( g_{\mu
\nu }-\frac 1{n^2}n_\mu n_\nu -\frac 1{\partial ^2}\partial _\mu \partial
_\nu \right)  \label{minv}
\end{equation}
The field $A^\mu $ satisfies the equation of motion: 
\begin{equation}
M^{\mu \nu }A_\nu =-\left[ \frac 1{n^2}\left( n\cdot \partial \right) \left(
n\cdot \overline{\partial \wedge B}\right) ^\mu -j^\mu \right]
\end{equation}
We can use the inverse matrix (\ref{minv}) to eliminate the field $A^\mu $
from the action, which becomes: 
\[
I_j\left( B,S,\varphi \right) =\int d^4x\left[ -\frac 12\left( \partial
\wedge B\right) ^2\right. 
\]
\[
\left. -\left( \partial \wedge B\right) \frac 1{\left( n\cdot \partial
\right) }\overline{n\wedge j}-\frac 12j^\mu \frac{n^2}{\left( n\cdot
\partial \right) ^2}\left( g_{\mu \nu }-\frac 1{n^2}n_\mu n_\nu \right)
j_\nu \right. 
\]
\begin{equation}
\left. +\frac{g^2S^2}2\left( B+\partial \varphi \right) ^2+\frac 12\left(
\partial S\right) ^2-\frac 12b\left( S^2-v^2\right) ^2\right]
\end{equation}
where we used the property: 
\begin{equation}
\partial _\nu \left( n\cdot \overline{\partial \wedge B}\right) ^\nu
=0\;\;\;\;\;\;n_\nu \left( n\cdot \overline{\partial \wedge B}\right) ^\nu =0
\end{equation}
as well as the second Zwanziger identity (\ref{apvt:zwanident2}) for the $%
\partial \wedge B$ terms. The action can finally be reduced to the form: 
\[
I_j\left( B,S,\varphi \right) =\int d^4x\left[ -\frac 12\left( \left(
\partial \wedge B\right) +\frac 1{\left( n\cdot \partial \right) }\overline{%
n\wedge j}\right) ^2\right. 
\]
\begin{equation}
\left. +\frac{g^2S^2}2\left( B+\partial \varphi \right) ^2+\frac 12\left(
\partial S\right) ^2-\frac 12b\left( S^2-v^2\right) ^2\right]
\end{equation}

This is precisely the form (\ref{landsphi}) of the Landau-Ginzburg action, 
\emph{with a straight-line string term} $\bar{G}=\frac 1{\left( n\cdot
\partial \right) }\overline{n\wedge j}$.\ 

There is however an apparent difference. The Landau-Ginzburg action (\ref
{landsphi}) contains a non-local string term $\bar{G}$ which we may choose
to be a string which stems from a positive charge and terminates on a
negative charge. In the local Zwanziger action (\ref{zwanlg}), the vector $%
n^{\mu }$ is fixed and independent of the position of the charges. This is a
source of difficulties, because, when the Zwanziger action is used with
classical fields, it breaks Lorentz invariance, as discussed at the
beginning of Sect.\ref{sec:zwanziger}.

\begin{itemize}
\item  \textbf{Exercise}: Consider projectors $K_{\mu \nu ,\alpha \beta }$
and $E_{\mu \nu ,\alpha \beta }$ defined, not in terms of $\partial ^\mu $
as in (\ref{apvt:kedef}), but in terms of the given vector $n^\mu $: 
\[
K_{\mu \nu ,\alpha \beta }=\frac 1{n^2}\left( g_{\mu \alpha }n_\nu n_\beta
-g_{\nu \alpha }n_\mu n_\beta +g_{\nu \beta }n_\mu n_\alpha -g_{\mu \beta
}n_\nu n_\alpha \right) 
\]
\begin{equation}
E=\varepsilon K\varepsilon \;\;\;\;\;\;E_{\mu \nu ,\alpha \beta }=\frac
14\varepsilon _{\mu \nu \sigma \rho }K^{\sigma \rho ,\gamma \delta
}\varepsilon _{\gamma \delta \alpha \beta }=\varepsilon _{\mu \nu \sigma
\rho }\frac 1{n^2}\left( g^{\sigma \gamma }n^\rho n^\delta \right)
\varepsilon _{\gamma \delta \alpha \beta }
\end{equation}
Check that the projectors $K$ and $E$ satisfy the relations: 
\begin{equation}
K^2=K\;\;\;\;\;\;E^2=-E\;\;\;\;\;\;KE=0\;\;\;\;\;\;K-E=G\;\;\;\;\;\;%
\varepsilon ^2=-G
\end{equation}
Check that: 
\begin{equation}
KF=\frac 1{n^2}n\wedge \left( n\cdot F\right) \;\;\;\;\;\;EF=\frac 1{n^2}%
\overline{n\wedge \left( n\cdot \overline{F}\right) }
\end{equation}
Show that the Zwanziger identity (\ref{apvt:zwanident1})\ can be expressed
in the form $F=\left( K-E\right) F$. Show that Zwanziger expresses the
''longitudinal'' part $KF$ of the field strength $F^{\mu \nu }$ in terms of
a vector potential $A^\mu $ and the ''transverse'' part $EF$ in terms of the
potential $B^\mu $: 
\begin{equation}
KF=K\left( \partial \wedge A\right) \;\;\;\;\;\;EF=-E\overline{\partial
\wedge B}
\end{equation}

\item  \textbf{Exercise}: Show that, if $S^{\mu \nu }$ and $T^{\mu \nu }$
are antisymmetric tensors , we have: 
\[
SKT=\frac 1{n^2}\left( n\cdot S\right) \left( n\cdot T\right)
\;\;\;\;\;\;SET=\bar{S}K\bar{T}=\frac 1{n^2}\left( n\cdot \bar{S}\right)
\left( n\cdot \bar{T}\right) 
\]
\begin{equation}
SK\bar{T}=\frac 1{n^2}\left( n\cdot S\right) \left( n\cdot \bar{T}\right) =-%
\bar{S}ET\;\;\;\;\;\;SE\bar{T}=-\bar{S}KT=-\frac 1{n^2}\left( n\cdot \bar{S}%
\right) \left( n\cdot T\right)  \label{skte}
\end{equation}
Use $\partial \cdot \overline{\partial \wedge A}=0$ to check that: 
\begin{equation}
\partial \cdot \left( K\overline{\partial \wedge A}\right) =\partial \cdot
\left( E\overline{\partial \wedge A}\right)  \label{dkdea}
\end{equation}
Show that the Zwanziger action (\ref{zwanlg}) can be written in the form: 
\[
I_{j,j_{mag}}\left( A,B\right) =\int d^4x\;\left\{ -\frac 12\left( \partial
\wedge A\right) K\left( \partial \wedge A\right) -\frac 12\left( \partial
\wedge B\right) K\left( \partial \wedge B\right) \right. 
\]
\begin{equation}
\left. -\frac 12\left( \partial \wedge A\right) K\overline{\partial \wedge B}%
+\frac 12\left( \partial \wedge B\right) K\overline{\partial \wedge A}%
-j\cdot A-j_{mag}\cdot B\right\}
\end{equation}
\end{itemize}

\chapter{Abelian gauge fixing}

\label{sec:abelgauge}

The formation of monopoles and their condensation in the QCD\ ground state
is a feature which is related to abelian gauge fixing, discussed in this
chapter. The gluon field acquires a singularity in the vicinity of points in
space where abelian gauge fixing fails and magnetic monopoles are formed
there. The ideas discussed in this chapter can be found in the 1974 and 1981
seminal papers of Polyakov \cite{Polyakov1974} and 't Hooft \cite
{tHooft1974b,tHooft81}. It is also very instructive to read the Sect.23.3
(vol.2) of Weinberg's Quantum Theory of Fields \cite{Weinberg1996}. The
formation of monopoles in QCD\ is still a subject of occasional debate \cite
{Kovner2002,Zakharov2002}. The choice of the abelian gauge is, of course,
not unique, and nor is the corresponding definition of the monopoles.\ The
recent gauge invariant definition of monopoles, proposed by Gubarev and
Zakharov \cite{Gubarev2002-1,Gubarev2002-2}, is not discussed in these
lectures.

The dynamical formation of monopoles in the QCD\ ground state is not
explained by the 't Hooft construction. A remarkable feature has however
been confirmed by lattice calculations 
\cite
{DiGiacomo2000-1,DiGiacomo2000-2,DiGiacomo2001-3,DiGiacomo2003},\cite
{Cosmai2000,Cosmai2001,Cosmai2003}, namely the previously surmised
condensation of monopoles in the QCD ground state. 
The lattice calculations evaluate the vacuum
expectation value of an operator which creates a magnetic monopole in an
Abelian gauge.\ It is found that the vacuum expectation value of this
operator is non-zero in the confining phase and zero in the deconfined
phase, thereby signaling the condensation of magnetic monopoles in the
confining phase. This is one of the main motivations for a phenomenological
description of the QCD\ ground state in terms of dual superconductors. The
lattice calculations suggest that the condensation of monopoles in the QCD\
ground state is remarkably independent of the gauge fixing condition.

The gluon field $A_a^\mu $ is a vector in color space. Of course, it would
be nice if we could make a gauge transformation at every space-time point $x$
which would rotate the gluon field so that only its diagonal components $A_3$
and $A_8$ would remain. This would reduce QCD to an abelian theory.\ The
trouble is, of course, that the gluon field has four components $A^\mu $ and
that it is only possible to align one component at a time. This is why a
scalar field is most often used to fix a gauge.

Let $\Phi \left( x\right) $ be a scalar field in the adjoint representation
of $SU\left( N_c\right) $, which means that the field is a vector in color
space with $N_c^2-1$ components $\Phi _a\left( x\right) $. The field can be
written in the form: 
\begin{equation}
\Phi \left( x\right) =\Phi _a\left( x\right) T_a  \label{phigauge}
\end{equation}
where $T_a$ are the $N_c^2-1$ generators of the $SU\left( N_c\right) $
group. They are equal to one half of the Pauli matrices in the case of $%
SU\left( 2\right) $ and to one half of the Gell-Mann matrices in the case of 
$SU\left( 3\right) $. The field $\Phi \left( x\right) $ does not have to be
one of the fields appearing in the model lagrangian. The choice of $\Phi $
is not innocent and will be discussed below. The orientation of the vector $%
\Phi \left( x\right) $ in color space, at the space-time point $x$, defines
a gauge. Different choices of $\Phi $ lead to different choices of the gauge.

Consider how this is done in practice. Local rotations in color space are
generated by operators of the form: 
\begin{equation}
\Omega \left( x\right) =e^{i\chi _{a}\left( x\right) T_{a}}  \label{omphi}
\end{equation}
The operators $\Omega \left( x\right) $ are elements of the color $SU\left(
N_{c}\right) $ group. A local rotation in color space is called a \emph{%
gauge transformation}. The generators $T_{a}$ are traceless hermitian $%
N_{c}\times N_{c}$ matrices, so that the field $\Phi \left( x\right) =\Phi
_{a}\left( x\right) T_{a}$ may be viewed as a traceless matrix in color
space. We can always perform a gauge transformation (a rotation of the
vector $\Phi $) so as to \emph{diagonalize} the matrix $\Phi \left( x\right) 
$. This means, that there always exists a rotation $\Omega \left( x\right) $%
, such that: 
\begin{equation}
\Omega \left( x\right) \Phi \left( x\right) \Omega ^{\dagger }\left(
x\right) =diag\left( \lambda _{1}\left( x\right) ,\lambda _{2}\left(
x\right) ,...,\lambda _{N_{c}}\left( x\right) \right)  \label{rot}
\end{equation}
The gauge in which $\Phi \left( x\right) $ is diagonal is called an \emph{%
abelian gauge}. The abelian gauge depends, of course, on the choice of the
scalar field $\Phi $.

When $N_c=2$, the abelian gauge is obtained by aligning the vector $\Phi
_a\left( x\right) $ along the color 3-axis: 
\begin{equation}
\Phi \left( x\right) \rightarrow \Omega \left( x\right) \Phi \left( x\right)
\Omega ^{\dagger }\left( x\right) =\Phi _3^{\prime }\left( x\right)
T_3=\frac 12\left( 
\begin{array}{ll}
\Phi _3^{\prime } & 0 \\ 
0 & -\Phi _3^{\prime }
\end{array}
\right)
\end{equation}

When $N_{c}=3$, two generators of the $SU\left( 3\right) $ group are
diagonal, namely: 
\begin{equation}
T_{3}=\frac{1}{2}\left( 
\begin{array}{ccc}
1 & 0 & 0 \\ 
0 & -1 & 0 \\ 
0 & 0 & 0
\end{array}
\right) \;\;\;\;\;\;\;T_{8}=\frac{1}{2\sqrt{3}}\left( 
\begin{array}{ccc}
1 & 0 & 0 \\ 
0 & 1 & 0 \\ 
0 & 0 & -2
\end{array}
\right)  \label{t3t8}
\end{equation}
and the abelian gauge is obtained by aligning the vector $\Phi _{a}\left(
x\right) $ along $T_{3}$ and $T_{8}$ axes: 
\[
\Phi \left( x\right) \rightarrow \Omega \left( x\right) \Phi \left( x\right)
\Omega ^{\dagger }\left( x\right) =\Phi _{3}^{\prime }\left( x\right)
T_{3}+\Phi _{8}^{\prime }\left( x\right) T_{8} 
\]
\begin{equation}
=\frac{1}{2}\left( 
\begin{array}{ccc}
\Phi _{3}^{\prime }+\frac{1}{\sqrt{3}}\Phi _{8}^{\prime } & 0 & 0 \\ 
0 & -\Phi _{3}^{\prime }+\frac{1}{\sqrt{3}}\Phi _{8}^{\prime } & 0 \\ 
0 & 0 & -\frac{2}{\sqrt{3}}\Phi _{8}^{\prime }
\end{array}
\right)
\end{equation}
Further gauge transformations, generated by two diagonal generators $T_{3}$
and $T_{8}$, leave the diagonal form of $\Phi \left( x\right) $ invariant.
Such rotations have the form $e^{i\chi _{3}T_{3}+i\chi _{8}T_{8}}=e^{i\chi
_{3}T_{3}}e^{i\chi _{8}T_{8}}$ and they belong to the residual $U\left(
1\right) \times U\left( 1\right) $ subgroup of $SU\left( 3\right) $, called
the \emph{maximal torus subgroup} of $SU\left( 3\right) $.

\section{The occurrence of monopoles in an abelian gauge}

\label{sec:occmonop}

There are points in space where the abelian gauge fixing becomes ill
defined. We shall see that such points, which are sometimes referred to as
topological defects, are sources of magnetic monopoles.

\subsection{The magnetic charge of a $SU\left( 2\right) $ monopole}

Consider the $SU\left( 2\right) $ case first. Let $\Omega \left( x\right) $
be the gauge transformation which brings the field $\Phi \left( x\right)
=\Phi _{a}\left( x\right) T_{a}$ into diagonal form: 
\begin{equation}
\Phi =\Phi _{a}T_{a}\rightarrow \Omega \Phi \Omega ^{\dagger }=\lambda
T_{3}=\left( 
\begin{array}{cc}
\lambda & 0 \\ 
0 & -\lambda
\end{array}
\right) \;\;\;\;\;\;\lambda =\sqrt{\Phi _{1}^{2}+\Phi _{2}^{2}+\Phi _{3}^{2}}
\end{equation}

The eigenvalues $\lambda \left( x\right) $ of the matrix $\Phi \left(
x\right) $ are, of course, gauge independent\footnote{%
They do, however, depend on the choice of the field $\Phi \left( x\right) $.}%
. A \emph{degeneracy} of the eigenvalues of $\Phi \left( x\right) $ occurs
when $\lambda =0$.\ At any one time, this implies that all three components $%
\Phi _{a=1,2,3}\left( \vec{r}\right) $ should vanish, and this can only
occur at specific \emph{points} $\vec{r}=\vec{r}_{0}$ in space such that: 
\begin{equation}
\Phi _{1}\left( \vec{r}_{0}\right) =0\;\;\;\;\;\;\Phi _{2}\left( \vec{r}%
_{0}\right) =0\;\;\;\;\;\;\Phi _{3}\left( \vec{r}_{0}\right) =0
\label{x123r}
\end{equation}

The three equations determine the three components $\left(
x_{0},y_{0},z_{0}\right) $ of the vector $\vec{r}_{0}$. At the point $\vec{r}%
_{0}$, defined by the equations (\ref{x123r}), it is not possible to define
the gauge and we shall that the gluon field develops a singularity at that
point.

In the vicinity of the point $\vec{r}_{0}$, we can express $\Phi \left( \vec{%
r}\right) $ in terms of a Taylor expansion: 
\begin{equation}
\Phi \left( \vec{r}\right) =\Phi _{a}\left( \vec{r}\right)
T_{a}=T_{a}C_{ab}\left( x_{b}-x_{0b}\right) \;\;\;\;\;\;C_{ab}=\left. \frac{%
\partial \Phi _{a}}{\partial x_{b}}\right| _{\vec{r}=\vec{r}_{0}}
\end{equation}
The matrix $C_{ab}$ defines a coordinate system in which the field $\Phi
\left( \vec{r}^{\prime }\right) $ has the form: 
\begin{equation}
\Phi \left( \vec{r}^{\prime }\right) =x_{a}^{\prime
}T_{a}\;\;\;\;\;\;x_{a}^{\prime }=C_{ab}\left( x_{b}-x_{0b}\right)
\label{shedge}
\end{equation}
In this coordinate system, the solution $\vec{r}_{0}$ of equation (\ref
{x123r}) is placed at the origin, and the field $\Phi \left( \vec{r}\right) $
has the\emph{\ hedgehog shape} displayed in Eq.(\ref{shedge}). In the
following, we work in this coordinate frame and drop the primes on $%
x^{\prime }$.

Let $\left( r,\theta ,\varphi \right) $ be the spherical coordinates of the
vector $\vec{r}$ (see App.\ref{ap:spherical}). In spherical coordinates, the
hedgehog field $\Phi \left( \vec{r}\right) =x_{a}T_{a}$ is represented by
the matrix: 
\[
\Phi \left( \vec{r}\right) =x_{a}T_{a}=T_{1}r\sin \theta \cos \varphi
+T_{2}r\sin \theta \sin \varphi +T_{3}r\cos \theta 
\]
\begin{equation}
=\frac{r}{2}\left( 
\begin{array}{cc}
\cos \theta & e^{-i\varphi }\sin \theta \\ 
e^{i\varphi }\sin \theta & -\cos \theta
\end{array}
\right)
\end{equation}
The matrix $\Omega $ which diagonalizes $\Phi $ is: 
\begin{equation}
\Omega \left( \theta ,\varphi \right) =\left( 
\begin{array}{cc}
e^{i\varphi }\cos \frac{\theta }{2} & \sin \frac{\theta }{2} \\ 
-\sin \frac{\theta }{2} & e^{-i\varphi }\cos \frac{\theta }{2}
\end{array}
\right) \;\;\;\;\;\;\Omega ^{\dagger }\left( \theta ,\varphi \right) =\left( 
\begin{array}{cc}
e^{-i\varphi }\cos \frac{\theta }{2} & -\sin \frac{\theta }{2} \\ 
\sin \frac{\theta }{2} & e^{i\varphi }\cos \frac{\theta }{2}
\end{array}
\right)  \label{omega}
\end{equation}
Indeed, we can check directly that: 
\begin{equation}
\Omega \Phi \Omega ^{\dagger }=\frac{r}{2}\left( 
\begin{array}{cc}
1 & 0 \\ 
0 & -1
\end{array}
\right) =rT_{3}
\end{equation}
\qquad \qquad \qquad \qquad

Now consider how the gluon field transforms under the \emph{same} gauge
transformation\footnote{%
The gauge transformation (\ref{gaginland}), defined in section \ref
{sec:landact}, corresponds to a rotation $\Omega =e^{ig\beta }$, in which
the angle $\beta $ is multiplied by the magnetic charge $g$. In the gauge
transformation (\ref{omphi}), the angles $\chi _a$ are not multiplied by the
coupling constant $e$. This is why a factor $\frac 1e$ appears in the gauge
transformation (\ref{aomega}) whereas no such factor appears in (\ref
{gaginland}).}:

\begin{equation}
A_\mu =A_{\mu a}T_a\rightarrow \Omega \left( A_\mu +\frac 1{ie}\partial _\mu
\right) \Omega ^{\dagger }\;\;\;\;\;\;\vec{A}=\vec{A}_aT_a\rightarrow \Omega
\left( \vec{A}+\frac 1{ie}\vec{\nabla}\right) \Omega ^{\dagger }
\label{aomega}
\end{equation}
The expression of the gradient $\vec{\nabla}$ in spherical coordinates is
given by (\ref{apvt:sph3}). The vector $\Omega \vec{\nabla}\Omega ^{\dagger
} $ is: 
\begin{equation}
\Omega \vec{\nabla}\Omega ^{\dagger }=\vec{e}_r\left( \Omega \frac \partial
{\partial r}\Omega ^{\dagger }\right) +\vec{e}_\theta \left( \Omega \frac
\partial {\partial \theta }\Omega ^{\dagger }\right) +\vec{e}_\varphi \frac
1{r\sin \theta }\left( \Omega \frac \partial {\partial \varphi }\Omega
^{\dagger }\right)
\end{equation}
where the unit vectors $\vec{e}$ are defined in (\ref{apvt:sph6}). From the
explicit expression (\ref{omega}) of $\Omega $, we find: 
\[
\Omega \frac \partial {\partial r}\Omega ^{\dagger }=0 
\]
\[
\Omega \frac \partial {\partial \theta }\Omega ^{\dagger }=\frac 12\left( 
\begin{array}{cc}
0 & -e^{i\varphi } \\ 
e^{-i\varphi } & 0
\end{array}
\right) =-ie^{i\varphi }T_2 
\]
\[
\Omega \frac \partial {\partial \varphi }\Omega ^{\dagger }=\allowbreak
\frac i2\left( 
\begin{array}{cc}
-\cos \theta -1 & e^{i\varphi }\sin \theta \\ 
e^{-i\varphi }\sin \theta & \cos \theta +1
\end{array}
\right) \allowbreak 
\]
\begin{equation}
=-i\left( 1+\cos \theta \right) T_3+i\sin \theta \cos \varphi T_1-i\sin
\theta \sin \varphi T_2
\end{equation}
so that: 
\begin{equation}
\frac 1{ie}\Omega \vec{\nabla}\Omega ^{\dagger }=\frac 1e\left( -\vec{e}%
_\theta T_2e^{i\varphi }-\vec{e}_\varphi \frac{1+\cos \theta }{r\sin \theta }%
T_3+\vec{e}_\varphi \frac 1r\left( \cos \varphi T_1-\sin \varphi T_2\right)
\right)
\end{equation}
$\allowbreak $The terms are all regular except for the term: 
\begin{equation}
\frac 1{ie}\left( \Omega \vec{\nabla}\Omega ^{\dagger }\right) _{sg}=-\frac
1e\vec{n}_\varphi \frac{1+\cos \theta }{r\sin \theta }T_3  \label{singa}
\end{equation}
which becomes singular when $\theta \rightarrow 0$, that is, on the positive 
$z$-axis.

Thus, in the abelian gauge, obtained by diagonalizing the field $\Phi \left(
x\right) $, the gluon field can be separated into a regular part $\vec{A}%
^{R} $ and the singular part (\ref{singa}): 
\begin{equation}
\vec{A}=\vec{A}_{a}T_{a}=\vec{A}_{a}^{R}T_{a}-\frac{1}{e}\vec{n}_{\varphi }%
\frac{1+\cos \theta }{r\sin \theta }T_{3}  \label{asph}
\end{equation}
Note that only the \emph{diagonal (abelian) part} of the gluon field
acquires a singular form. The singular part (\ref{singa}) has exactly the
form (\ref{amonop}) which a gauge field acquires in the vicinity of a
magnetic monopole situated at the origin, with a Dirac string running along
the positive $z$-axis. By comparing the expressions (\ref{asph}) and (\ref
{amonop}), we see that the \emph{magnetic charge} of the monopole is equal
to: 
\begin{equation}
g=-\frac{4\pi }{e}T_{3}  \label{gpie}
\end{equation}
Here $e$ is the color electric charge, that is, the QCD coupling constant.
This is another instance of the Dirac quantization condition (\ref
{chargequant}). In a way, the result (\ref{gpie}) is promising for low
energy phenomena because perturbation theory points to a divergence of the
QCD\ coupling constant $e$ at low energy and we may expect $g$ to be better
behaved. The way the running coupling constant $e$ and $g$ manage to
maintain a constant product $eg$ is discussed in 2001 papers of the Russian
group \cite{Polikarpov2001,Polikarpov2001ter}. In short, we have shown that,
in the vicinity of points where the eigenvalues of the matrix $\Phi \left(
x\right) $ are degenerate, that is, at points where the abelian gauge is ill
defined, the abelian part of the gluon field behaves as if a monopole with
magnetic charge $g=-\frac{4\pi }{e}T_{3}$ was sitting there.

\begin{itemize}
\item  \textbf{Exercise}: Consider the gauge transformation $\Omega ^{\prime
}=\Omega \left( \theta +\pi ,\varphi \right) $. Show that it also
diagonalizes $\Phi $, such that $\Omega ^{\prime }\Phi \Omega ^{^{\prime
}\dagger }=-rT_3$ and that the singular part of the transformed field $%
\Omega ^{\prime }A_\mu \Omega ^{\prime \dagger }$ is $\frac 1e\vec{n}%
_\varphi \frac{1-\cos \theta }{r\sin \theta }T_3$ which indicates the
presence of a magnetic charge $g=\frac{4\pi }eT_3$ with a string running
along the negative $z$-axis.
\end{itemize}

\subsection{The magnetic charges of $SU\left( 3\right) $ monopoles}

\label{sec:su3monop}

In the case of $SU\left( 3\right) $, there are two diagonal generators,
namely $T_{3}$ and $T_{8}$, given by (\ref{t3t8}). The abelian gauge is the
one in which the field $\Phi =\Phi _{a}T_{a}$ acquires the diagonal form: 
\begin{equation}
\Phi =\Phi _{a}T_{a}\rightarrow \Omega \Phi \Omega ^{\dagger }=diag\left(
\lambda _{1,}\lambda _{2,}\lambda _{3}\right) \;\;\;\;\; \;\lambda
_{1}+\lambda _{2}+\lambda _{3}=0
\end{equation}
Monopoles will occur at points in space where two eigenvalues become
degenerate.\ Indeed, consider the case where the first two eigenvalues are
degenerate: $\lambda _{1}=\lambda _{2}=\frac{\lambda }{2}$ and $\lambda
_{3}=-\lambda $. When the two eigenvalues $\lambda _{1}$ and $\lambda _{2}$
lie close to each other, the matrix $\Phi $ may be considered as diagonal in
all but the $SU\left( 2\right) $ subspace defined by the almost degenerate
eigenvalues: 
\begin{equation}
\Phi \simeq \frac{1}{2}\left( 
\begin{tabular}{cc}
$
\begin{array}{cc}
\lambda +\varepsilon _{3} & \varepsilon _{1}-i\varepsilon _{2} \\ 
\varepsilon _{1}+i\varepsilon _{2} & \lambda -\varepsilon _{3}
\end{array}
$ & $0$ \\ 
$0$ & $-2\lambda $%
\end{tabular}
\right) =\sum_{a=1}^{3}\Phi _{a}T_{a}+\Phi _{8}T_{8}
\end{equation}
\begin{equation}
\Phi _{1}=\varepsilon _{1}\;\;\;\;\;\;\Phi _{2}=\varepsilon
_{2}\;\;\;\;\;\;\Phi _{3}=\varepsilon _{3}\;\;\;\;\;\;\Phi _{8}=\sqrt{3}%
\lambda  \label{almost}
\end{equation}
Consider the rotation (or gauge transformation) $\Omega $ which brings this
almost diagonal matrix (\ref{almost}) into diagonal form: 
\begin{equation}
\Phi _{8}T_{8}+\sum_{a=1}^{3}\Phi _{a}T_{a}\rightarrow \Omega \left( \Phi
_{8}T_{8}+\sum_{a=1}^{3}\Phi _{a}T_{a}\right) \Omega ^{\dagger }=\Phi
_{8}T_{8}+\left( 
\begin{array}{lll}
\varepsilon & 0 & 0 \\ 
0 & -\varepsilon & 0 \\ 
0 & 0 & 0
\end{array}
\right)
\end{equation}
The rotation $\Omega $ simply orients the vector $\sum_{a=1}^{3}\Phi
_{a}T_{a}$ in the $T_{3}$ direction so that the eigenvalue $\varepsilon $
is: 
\begin{equation}
\varepsilon =\sqrt{\Phi _{1}^{2}+\Phi _{2}^{2}+\Phi _{3}^{2}}
\end{equation}
Degeneracies of the eigenvalues will occur at points $\vec{r}_{0}$ in space
where: 
\begin{equation}
\Phi _{1}\left( \vec{r}_{0}\right) =0\;\;\;\;\;\;\Phi _{2}\left( \vec{r}%
_{0}\right) =0\;\;\;\;\;\;\Phi _{3}\left( \vec{r}_{0}\right) =0
\end{equation}
These three equations define the three components $\left(
x_{0},y_{0},z_{0}\right) $ of the position vector $\vec{r}_{0}$, as in the $%
SU\left( 2\right) $ case. In the vicinity of the point $\vec{r}_{0}$, the
field $\Phi \left( \vec{r}\right) $ acquires the hedgehog shape (\ref{shedge}%
): 
\begin{equation}
\Phi \left( \vec{r}^{\prime }\right) =\sum_{a=1}^{3}x_{a}^{\prime
}T_{a}+\Phi _{8}\left( \vec{r}^{\prime }\right)
T_{8}\;\;\;\;\;\;x_{a}^{\prime }=\sum_{b=1}^{3}C_{ab}\left(
x_{b}-x_{0b}\right) \;\;\;\;\;\;C_{ab}=\left. \frac{\partial \Phi _{a}}{%
\partial x_{b}}\right| _{\vec{r}=\vec{r}_{0}}  \label{phisu3}
\end{equation}
We work in the coordinate frame $x_{a}^{\prime }$ and drop the primes. The
degeneracy point is then placed at the origin of coordinates. In spherical
coordinates, the field is:

\[
\Phi \left( \vec{r}\right) =x_aT_a+\Phi _8\left( \vec{r}\right) T_8=T_1r\sin
\theta \cos \varphi +T_2r\sin \theta \sin \varphi +T_3r\cos \theta +\Phi
_8\left( \vec{r}\right) T_8 
\]
\begin{equation}
=\frac r2\left( 
\begin{array}{cc}
\begin{array}{cc}
\cos \theta & e^{-i\varphi }\sin \theta \\ 
e^{i\varphi }\sin \theta & -\cos \theta
\end{array}
& 0 \\ 
0 & 0
\end{array}
\right) +\Phi _8\left( \vec{r}\right) T_8
\end{equation}
The gauge transformation, which brings this matrix into diagonal form is: 
\begin{equation}
\Omega \left( \theta ,\varphi \right) =\left( 
\begin{tabular}{cc}
$
\begin{array}{cc}
e^{i\varphi }\cos \frac \theta 2 & \sin \frac \theta 2 \\ 
-\sin \frac \theta 2 & e^{-i\varphi }\cos \frac \theta 2
\end{array}
$ & $0$ \\ 
$0$ & $1$%
\end{tabular}
\right)
\end{equation}
Under this gauge transformation, the gluon field becomes:

\begin{equation}
\vec{A}=\vec{A}_{a}T_{a}\rightarrow \Omega \left( \vec{A}+\frac{1}{ie}\vec{%
\nabla}\right) \Omega ^{\dagger }
\end{equation}
The calculation of $\Omega \vec{\nabla}\Omega ^{\dagger }$ proceeds exactly
as in the $SU\left( 2\right) $ case and we find: 
\begin{equation}
\frac{1}{ie}\Omega \vec{\nabla}\Omega ^{\dagger }=\frac{1}{e}\left( -\vec{e}%
_{\theta }T_{2}e^{i\varphi }-\vec{e}_{\varphi }\frac{1+\cos \theta }{r\sin
\theta }T_{3}+\vec{e}_{\varphi }\frac{1}{r}\left( \cos \varphi T_{1}-\sin
\varphi T_{2}\right) \right)
\end{equation}
The second term becomes singular in the vicinity of the positive $z$-axis.
The gauge transformed gluon field $\vec{A}$ can therefore be separated into
a regular part $\vec{A}^{R}$ and the singular part : 
\begin{equation}
\vec{A}=\vec{A}_{a}T_{a}=\vec{A}_{a}^{R}T_{a}-\frac{1}{e}\vec{e}_{\varphi }%
\frac{1+\cos \theta }{r\sin \theta }T_{3}  \label{asph31}
\end{equation}
Thus, in the vicinity of points where the first two eigenvalues coincide,
the diagonal gluon $\vec{A}_{3}$ feels the presence of a monopole with
magnetic charge $g=-\frac{4\pi }{e}T_{3}$.

Consider next the case where the last two eigenvalues are degenerate: $%
\lambda _{2}=\lambda _{3}=\lambda $ and $\lambda _{1}=-\lambda $. When the
two eigenvalues $\lambda _{2}$ and $\lambda _{3}$ lie close to each other,
the matrix $\Phi $ may be considered as diagonal in all but the $SU\left(
2\right) $ subspace defined by the almost degenerate eigenvalues: 
\begin{equation}
\Phi \simeq \frac{1}{2}\left( 
\begin{tabular}{cc}
$-2\lambda $ & $0$ \\ 
$0$ & $
\begin{array}{cc}
\lambda +\varepsilon _{3} & \varepsilon _{1}-i\varepsilon _{2} \\ 
\varepsilon _{1}+i\varepsilon _{2} & \lambda -\varepsilon _{3}
\end{array}
$%
\end{tabular}
\right) =\Phi _{8}t_{8}+\sum_{a=1}^{3}\varepsilon _{a}t_{a}
\end{equation}
\begin{equation}
\Phi _{1}=\varepsilon _{1}\;\;\;\;\;\;\Phi _{2}=\varepsilon
_{2}\;\;\;\;\;\;\Phi _{3}=\varepsilon _{3}\;\;\;\;\;\;\Phi _{8}=\sqrt{3}%
\lambda  \label{almost2}
\end{equation}
where we defined: 
\[
t_{1}=T_{6}=\frac{1}{2}\left( 
\begin{array}{ccc}
0 & 0 & 0 \\ 
0 & 0 & 1 \\ 
0 & 1 & 0
\end{array}
\right) ,\;\;t_{2}=T_{7}=\frac{1}{2}\left( 
\begin{array}{ccc}
0 & 0 & 0 \\ 
0 & 0 & -i \\ 
0 & i & 0
\end{array}
\right) 
\]
\begin{equation}
t_{3}=-\frac{1}{2}T_{3}+\frac{\sqrt{3}}{2}T_{8}=\frac{1}{2}\left( 
\begin{array}{ccc}
0 & 0 & 0 \\ 
0 & 1 & 0 \\ 
0 & 0 & -1
\end{array}
\right) \;\;\;\;\;\;t_{8}=-\frac{\sqrt{3}}{2}T_{3}-\frac{1}{2}T_{8}=\frac{1}{%
2\sqrt{3}}\left( 
\begin{array}{ccc}
-2 & 0 & 0 \\ 
0 & 1 & 0 \\ 
0 & 0 & 1
\end{array}
\right)
\end{equation}
From here we proceed as in the previous case, with the replacements $%
T_{a}\rightarrow t_{a}$ for $a=1,2,3$ and $8$. The rotation $\Omega $, which
brings the matrix (\ref{almost2}) to the diagonal form, is: 
\[
\Phi _{8}t_{8}+\sum_{a=1}^{3}\Phi _{a}t_{a}\rightarrow \Omega \left( \Phi
_{8}t_{8}+\sum_{a=1}^{3}\Phi _{a}t_{a}\right) \Omega ^{\dagger }=\Phi
_{8}t_{8}+\frac{1}{2}\left( 
\begin{array}{lll}
0 & 0 & 0 \\ 
0 & \varepsilon & 0 \\ 
0 & 0 & -\varepsilon
\end{array}
\right) 
\]
\begin{equation}
\varepsilon =\sqrt{\Phi _{1}^{2}+\Phi _{2}^{2}+\Phi _{3}^{2}}
\end{equation}
It transforms the gluon field to:

\[
\vec{A}=\vec{A}_{a}T_{a}\rightarrow \Omega \left( \vec{A}+\frac{1}{ie}\vec{%
\nabla}\right) \Omega ^{\dagger } 
\]
\begin{equation}
=\Omega \vec{A}\Omega ^{\dagger }+\frac{1}{e}\left( -\vec{e}_{\theta
}t_{2}e^{i\varphi }-\vec{e}_{\varphi }\frac{1+\cos \theta }{r\sin \theta }%
t_{3}+\vec{e}_{\varphi }\frac{1}{r}\left( \cos \varphi t_{1}-\sin \varphi
t_{2}\right) \right)
\end{equation}
The second term becomes singular in the vicinity of the positive $z$-axis.
The gauge transformed gluon field $\vec{A}$ can therefore be separated into
a regular part $\vec{A}^{R}$ and the singular part : 
\[
\vec{A}=\vec{A}_{a}T_{a}=\vec{A}_{a}^{R}T_{a}-\frac{1}{e}\vec{n}_{\varphi }%
\frac{1+\cos \theta }{r\sin \theta }t_{3} 
\]
\begin{equation}
=\vec{A}_{a}^{R}T_{a}-\frac{1}{e}\vec{n}_{\varphi }\frac{1+\cos \theta }{%
r\sin \theta }\left( -\frac{1}{2}T_{3}+\frac{\sqrt{3}}{2}T_{8}\right)
\end{equation}
where we expressed $t_{3}$ in terms of $T_{3}$ and $T_{8}$. Again, it is
only the diagonal gluons which become singular in the vicinity of the
positive $z$-axis. Thus, when the last two eigenvalues coincide, the
diagonal gluon $-\frac{1}{2}\vec{A}_{3}T_{3}+\frac{\sqrt{3}}{2}\vec{A}%
_{8}T_{8}$ feels the presence of a monopole with magnetic charge $g=-\frac{%
4\pi }{e}t_{3}=-\frac{4\pi }{e}\left( -\frac{1}{2}T_{3}+\frac{\sqrt{3}}{2}%
T_{8}\right) $.

In the final case where the first and third eigenvalues are degenerate, the
gluon field transforms to: 
\begin{equation}
\vec{A}=\vec{A}_{a}T_{a}\rightarrow \vec{A}_{a}^{R}T_{a}+\frac{1}{e}\vec{n}%
_{\varphi }\frac{1+\cos \theta }{r\sin \theta }\left( \frac{1}{2}T_{3}+\frac{%
\sqrt{3}}{2}T_{8}\right)
\end{equation}
and it is the diagonal gluon $-\frac{1}{2}\vec{A}_{3}T_{3}-\frac{\sqrt{3}}{2}%
\vec{A}_{8}T_{8}$ which feels the presence of a monopole with magnetic
charge $g=\frac{4\pi }{e}\left( \frac{1}{2}T_{3}+\frac{\sqrt{3}}{2}%
T_{8}\right) $.

These results may be summarized by saying that the topological defects of
abelian gauge fixing, in the case of $SU\left( 3\right) $, are sources of
magnetic monopoles, with magnetic charges equal to: 
\begin{equation}
g=\frac{4\pi }{e}\left( w_{a}\cdot T\right)
\end{equation}
where $w_{a=1,2,3}$ are the root vectors (\ref{apsu:weight}) of the color $%
SU\left( 3\right) $ group and $T$ is the vector $\left( T_{3},T_{8}\right) $%
. It is this observation which suggests the form of the $SU\left( 3\right) $
Landau-Ginzburg (Abelian Higgs) model presented in Sect.\ref{sec:su3lg}.

\section{The maximal abelian gauge and abelian projection}

\label{sec: maxabel}

The choice of the field $\Phi \left( x\right) $, used to fix the gauge, is
far from being a trivial problem and the reader is referred to the 1981
paper of 't Hooft for a discussion of some appropriate and inappropriate
choices \cite{tHooft81}. Many different choices have been used. Some are
defined in terms of Polyakov loops on a lattice \cite
{DiGiacomo2000-1,DiGiacomo2000-2,DiGiacomo2001-3},\cite{DiGiacomo2003} some
in terms of the lowest eigenvalue of the covariant laplacian operator \cite
{vanderSijs1998}, some in terms of a ''maximal abelian gauge'' \cite
{Kronfeld1987}. Most of these choices are defined on the lattice. A
discussion of abelian gauge fixing on the lattice is beyond the scope of
these lectures and we limit the discussion to a brief description of the
maximal abelian gauge, from which $92$ \% of the full string tension is
obtained in the $SU\left( 2\right) $ case \cite{Bali1996}. A useful account
of evidence for the occurrence of monopoles obtained from lattice
calculations in the maximal abelian projection, can be found in the 1997\
Cambridge lectures of Chernodub and Polikarpov \cite{Polikarpov1997}. The
reader is also referred to the Sect.4.10 of the extensive 2001 Physics
Report of Bali \cite{Bali2001} and he may find it instructive to consult the
1999 thesis of Ichie \cite{Ichie1999,Ichie2000} as well as the recent 2003
paper of Chernodub \cite{Chernodub2003-1}.

Consider color $SU\left( 3\right) $. The maximal abelian gauge attempts to
minimize the off-diagonal gluons. The gluon field $A^{\mu }$ can be
expressed thus: 
\begin{equation}
A^{\mu }=A_{a}^{\mu }T_{a}=A_{3}^{\mu }T_{3}+A_{8}^{\mu
}T_{8}+\sum_{a=1}^{3}C_{a}^{\mu \ast }E_{a}+C_{a}^{\mu }E_{-a}  \label{aac}
\end{equation}
In this form, the diagonal generators $T_{3}$ and $T_{8}$ are explicit and
the charged non-diagonal gluons $C_{a}$ and $C_{a}^{\ast }$ are expressed in
terms of the generators $E_{\pm a}$ defined in (\ref{apsu:evec}). Let us
represent the diagonal generators $T_{3}$ and $T_{8}$ by the two dimensional
vector $H=\left( T_{3},T_{8}\right) $. The commutator of the covariant
derivative $D^{\mu }=\partial ^{\mu }+ieA_{a}^{\mu }T_{a}$ with the diagonal
generators $H_{i=1,2}$ can be expressed in terms of the root vectors $\vec{w}%
_{a}$, defined in (\ref{apsu:weight}): 
\begin{equation}
\left[ D^{\mu },H_{i}\right] =ie\sum_{a=1}^{3}w_{ai}\left( C_{a}^{\ast \mu
}E_{a}-C_{a}^{\mu }E_{-a}\right)
\end{equation}
The commutator $\left[ D^{\mu },H_{i}\right] $ singles out the off-diagonal
part of the gluon field. Let us calculate the trace: 
\begin{equation}
R=tr\sum_{i=1}^{2}\left[ D_{\mu },H_{i}\right] \left[ D^{\mu },H_{i}\right]
=-e^{2}tr\sum_{i=1}^{2}\sum_{a,b=1}^{3}w_{ia}w_{ib}tr\left( C_{a\mu }^{\ast
}E_{a}-C_{a\mu }E_{-a}\right) \left( C_{b}^{\ast \mu }E_{b}-C_{b}^{\mu
}E_{-b}\right)
\end{equation}
It is easy to check that:

\begin{equation}
trE_aE_b=0\;\;\;\;\;\;trE_aE_{-b}=\delta _{ab}N_c\;\;\;\;\;\;\left(
a,b\right) >0\;\;\;\;\;\;
\end{equation}
so that, using (\ref{apsu:weg}), we find: 
\begin{equation}
R\left( x\right) =2e^2N_c\sum_{a=1}^3\left| C_a^\mu \left( x\right) \right|
^2
\end{equation}
We see that, in a gauge which minimizes the field $R\left( x\right) $, the
intensity of the charged gluons $C_a^{*\mu }\left( x\right) $ and $C_a^\mu
\left( x\right) $ is minimized.

-Let us seek this gauge. Let $R_{\Omega }$ be the field $R$ obtained by
performing a gauge transformation $D^{\mu }\rightarrow \Omega D^{\mu }\Omega
^{\dagger }$ of the covariant derivative: 
\begin{equation}
R_{\Omega }=tr\sum_{i=1}^{2}\left[ \Omega D_{\mu }\Omega ^{\dagger
},H_{i}\right] \left[ \Omega D^{\mu }\Omega ^{\dagger },H_{i}\right]
=tr\sum_{i=1}^{2}\left[ D_{\mu },\Omega ^{\dagger }H_{i}\Omega \right]
\left[ D^{\mu },\Omega ^{\dagger }H_{i}\Omega \right]
\end{equation}
We want $R\left( x\right) $ to be stationary with respect to infinitesimal
gauge transformations of the form $\Omega \simeq 1+i\chi $, where $\chi
=\chi _{a}T_{a}$. To first order in $\chi $, we have $\Omega ^{\dagger
}H_{i}\Omega =H_{i}-i\left[ \chi ,H_{i}\right] $ so that the first order
variation of $R_{\Omega }$ is: 
\begin{equation}
R_{\Omega }^{\left( 1\right) }=-2itr\sum_{i=1}^{2}\left[ D_{\mu },\left[
\chi ,H_{i}\right] \right] \left[ D^{\mu },H_{i}\right] =2itr\chi
\sum_{i=1}^{2}\left[ H_{i},\left[ D_{\mu },\left[ D^{\mu },H_{i}\right]
\right] \right]
\end{equation}
If this is to vanish for any $\chi =\chi _{a}T_{a}$, we must have: 
\begin{equation}
\left[ H_{i},\left[ D_{\mu },\left[ D^{\mu },H_{i}\right] \right] \right] =0
\label{hddh}
\end{equation}

In the $SU\left( 2\right) $ case, the vector $H_{i}$ has only one component $%
H_{1}=T_{3}$. The maximal abelian gauge is the one which aligns the vector: 
\begin{equation}
\Phi =\left[ D_{\mu },\left[ D^{\mu },T_{3}\right] \right]
\end{equation}
along the $T_{3}$ axis.

\emph{Abelian projection} in the continuum consists in making the
corresponding gauge transformation of the gluon field $A^{\mu }$ and in
retaining only the diagonal part. For example, in the expressions (\ref{asph}%
) and (\ref{asph31}), this means setting to zero the non-diagonal parts of $%
\vec{A}_{a}^{R}T_{a}$.\ The monopole singular part is retained in this
process.\ 

In the case of $SU\left( 3\right) $, the condition (\ref{hddh}) reads: 
\begin{equation}
\left[ T_{3},\left[ D_{\mu },\left[ D^{\mu },T_{3}\right] \right] \right]
+\left[ T_{8},\left[ D_{\mu },\left[ D^{\mu },T_{8}\right] \right] \right] =0
\end{equation}
Maximal abelian gauge fixing is more subtle in this case and the reader if
referred to the interesting 2002 paper by Stack, Tucker and Wensley \cite
{Tucker2002}.

\section{Abelian and center projection on the lattice.}

\label{sec:centerproj}

On the lattice the gluon field does not appear explicitly and, instead, the
action is expressed in terms of link variables. Abelian gauge fixing and
center projection on the lattice is usefully reviewed in the 1997\ paper of
Del Debbio, Faber, Greensite and Olejnik \cite{Greensite1997}. In $SU\left(
2\right) ,$ the maximal abelian gauge is the gauge which maximizes 
\begin{equation}
R=\sum_{\times }\sum_{\mu =1}^4Tr\left( \sigma _3U_{x,x+\mu }\sigma
_3U_{x,x+\mu }^{\dagger }\right)
\end{equation}
so as to make the link variables $U_{x,x+\mu }$ as diagonal as possible.
Under a gauge transformation, generated by the $SU\left( 2\right) $ element $%
\Omega \left( x\right) =e^{i\chi _a\left( x\right) T_a}$, the link variable
transforms as: 
\begin{equation}
U_{x,x+\mu }\rightarrow \Omega \left( x\right) U_{x,x+\mu }\Omega ^{\dagger
}\left( x+\mu \right)
\end{equation}
and $R$ transforms as: 
\begin{equation}
R_\Omega \rightarrow \sum_{\times }\sum_{\mu =1}^4Tr\left( \sigma _3\Omega
\left( x\right) U_{x,x+\mu }\Omega ^{\dagger }\left( x+\mu \right) \sigma
_3\Omega \left( x+\mu \right) U_{x,x+\mu }^{\dagger }\Omega ^{\dagger
}\left( x\right) \right)
\end{equation}
The maximal abelian gauge is then defined by the $SU\left( 2\right) $
element $\Omega \left( x\right) =e^{i\chi _a\left( x\right) T_a}$ in which
the angles $\alpha _a\left( x\right) $ are chosen so as to maximize $%
R_\Omega $.

Abelian projection means the replacement of the full link variables by
Abelian links $A$ according to the rule 
\begin{equation}
U=a_0I+i\vec{a}\cdot \vec{\sigma}\rightarrow A=\frac{a_0I+ia_3\sigma ^3}{%
\sqrt{a_0^2+a_3^2}}=\left( 
\begin{array}{cc}
e^{i\theta } & 0 \\ 
0 & e^{-i\theta }
\end{array}
\right)
\end{equation}
where $A$ stands for ''Abelian'' (and does not designate the gauge field).
In these expressions, we have omitted the induces $x$ and $\mu $ so that,
for example, $U$ stands for $U_{x,x+\mu }$ and $\theta $ stands for $\theta
_{x,x+\mu }$.

Abelian dominance, found by Suzuki and collaborators \cite
{Suzuki1990-2,Suzuki1991}, is essentially the fact that the confining string
tension can be extracted from the Abelian link variables alone.

The matrix $U$ remains abelian under $U\left( 1\right) $ gauge
transformations of the form 
\begin{equation}
A_{x,x+\mu }\rightarrow \left( 
\begin{array}{cc}
e^{i\alpha _{x}} & 0 \\ 
0 & e^{-i\alpha _{x}}
\end{array}
\right) \left( 
\begin{array}{cc}
e^{i\theta _{x,x+\mu }} & 0 \\ 
0 & e^{-i\theta _{x,x+\mu }}
\end{array}
\right) \left( 
\begin{array}{cc}
e^{i\alpha _{x+\mu }} & 0 \\ 
0 & e^{-i\alpha _{x+\mu }}
\end{array}
\right)
\end{equation}
We can proceed to make a further gauge fixing by choosing the angles $\alpha
\left( x\right) $ so as to maximize the quantity 
\begin{equation}
\sum_{x}\sum_{\mu =1}^{4}\cos ^{2}\left( \theta _{x,x+\mu }\right)
\end{equation}
This defines the \emph{maximal center gauge}. Of course, this still leaves a
remaining $Z_{2}$ symmetry because $\theta $ is only determined modulo $\pi $%
. The, at each link $\left( x,x+\mu \right) $ we can define a value of $%
Z_{x,x+\mu }$ as follows: 
\begin{equation}
Z_{x,x+\mu }=sign\left( \cos \theta _{x,x+\mu }\right)
\end{equation}
so that $Z_{x,x+\mu }$ takes the values $+1$ or $-1$. \emph{Center projection%
} means the replacement of the full link variables by Abelian links $\Lambda 
$ according to the rule 
\begin{equation}
U=a_{0}I+i\vec{a}\cdot \vec{\sigma}\rightarrow ZI=\left( 
\begin{array}{cc}
Z & 0 \\ 
0 & Z
\end{array}
\right)
\end{equation}
in the computation of observables and Polyakov lines.

\chapter{The confinement of $SU\left( 3\right) $ color charges}

\label{sec:su3model}

If we wish to describe color confinement in terms of the Meissner effect of
a dual superconductor, we need to adapt the Landau-Ginzburg model to the
dynamics of quarks and gluons, so as to accommodate their color quantum
numbers. The action (\ref{landginz}) of the Landau-Ginzburg model describes
a $U\left( 1\right) $ gauged self-interacting complex scalar field $\psi $.
Since the magnetic current of the dual superconductor is somehow related to
the monopoles which are formed by topological defects in a given gauge, as
described in Sect. \ref{sec:occmonop}, it might make sense to restrict the
covariant derivative $D_{\mu }$ to the corresponding abelian gauge. An
adaptation of the Landau-Ginzburg model to color $SU\left( 3\right) $, which
respects Weyl symmetry, was proposed in the 1989 paper of Maedan and Suzuki 
\cite{Suzuki89}.\ It was further developed in the 1993 paper of Kamizawa,
Matsubara, Shiba and Suzuki \cite{Suzuki1993} and the 1999 papers of
Chernodub and Komarov \cite{Chernodub1998,Chernodub1999}. We shall first
present the model in the absence of quark charges.\ The latter will be
introduced in Sect.\ref{sec:quarkcup}.

\section{An abelian $SU\left( 3\right) $ Landau-Ginzburg model}

\label{sec:su3lg}

\subsection{The model action and its abelian gauge invariance}

The model action, proposed by Maedan and Suzuki \cite{Suzuki1990}, has the
form: 
\[
I\left( B_{3},B_{8},\psi ,\psi ^{\ast }\right) =\int d^{4}x\left\{ -\frac{1}{%
2}\left( \partial \wedge B_{3}\right) ^{2}-\frac{1}{2}\left( \partial \wedge
B_{8}\right) ^{2}\right. 
\]
\begin{equation}
\left. +\sum_{a=1}^{3}\left[ \frac{1}{2}\left| \left( \partial _{\mu }\psi
_{a}+ig\left( w_{a}\cdot B_{\mu }\right) \psi _{a}\right) \right| ^{2}-\frac{%
1}{2}b\left( \psi _{a}\psi _{a}^{\ast }-v^{2}\right) ^{2}\right] \right\}
\label{suzmaed}
\end{equation}
The first term is the kinetic term of the abelian dual gauge fields $%
B_{3}^{\mu }$ and $B_{8}^{\mu }$ which can be grouped together to form the
vector $B^{\mu }=\left( B_{3}^{\mu },B_{8}^{\mu }\right) $. The model
involves three complex scalar fields $\psi _{a=1,2,3}$. Each scalar field $%
\psi _{a}$ is gauged with the \emph{abelian} covariant derivative $D^{\mu
}=\partial ^{\mu }+ig\left( w_{a}\cdot B_{\mu }\right) $ where $w_{a=1,2,3}$
are the three weight vectors (\ref{apsu:weight}) of the $SU\left( 3\right) $
group: 
\begin{equation}
w_{1}=\left( 1,0\right) \;\;\;\;\;\;w_{2}=\left( -\frac{1}{2},-\frac{\sqrt{3}%
}{2}\right) \;\;\;\;\;\;w_{3}=\left( -\frac{1}{2},\frac{\sqrt{3}}{2}\right)
\label{weight}
\end{equation}
Thus: 
\begin{equation}
\left( w_{1}\cdot B^{\mu }\right) =B_{3}^{\mu }\;\;\;\;\;\;\left( w_{2}\cdot
B^{\mu }\right) =-\frac{1}{2}B_{3}^{\mu }-\frac{\sqrt{3}}{2}B_{8}^{\mu
}\;\;\;\;\;\;\left( w_{3}\cdot B^{\mu }\right) =-\frac{1}{2}B_{3}^{\mu }+%
\frac{\sqrt{3}}{2}B_{8}^{\mu }  \label{wb}
\end{equation}
The magnetic charges appearing in the covariant derivative are assumed to be
proportional to the weight vectors $w_{a}$, as discussed in Sect. \ref
{sec:su3monop}. The explicit form of the second term of the action (\ref
{suzmaed}) is: 
\[
\frac{1}{2}\sum_{a=1}^{3}\left| \left( \partial _{\mu }\psi _{a}+ig\left(
w_{a}\cdot B_{\mu }\right) \psi _{a}\right) \right| ^{2}= 
\]
\[
=\frac{1}{2}\left| \partial \psi _{1}+igB_{3}\psi _{1}\right| ^{2}+\frac{1}{2%
}\left| \partial \psi _{2}+ig\left( -\frac{1}{2}B_{3}-\frac{\sqrt{3}}{2}%
B_{8}\right) \psi _{2}\right| ^{2} 
\]
\begin{equation}
+\frac{1}{2}\allowbreak \left| \partial \psi _{3}+ig\left( -\frac{1}{2}B_{3}+%
\frac{\sqrt{3}}{2}B_{8}\right) \psi _{3}\right| ^{2}
\end{equation}
The last term is a potential by means of which the three scalar fields
acquire non-vanishing values $\left| \psi _{a}\right| =v$ in the ground
state. Note that the last term does \emph{not} have the form $-\frac{1}{2}%
b\left( \sum_{a=1}^{3}\psi _{a}\psi _{a}^{\ast }-v^{2}\right) ^{2}$.

In this and the following sections, repeated indices $i,a,...$ are \emph{not}
assumed summed unless it is explicitly stated. Repeated indices of the
components $\mu ,\nu ,...$ of Lorentz vectors and tensors \emph{are} assumed
to be explicitly summed.

It is sometimes useful to use a polar representation of the scalar fields: 
\begin{equation}
\psi _a=S_ae^{ig\varphi _a}  \label{psipol}
\end{equation}
in which case the action (\ref{suzmaed}) acquires the form:

\[
I\left( B_3,B_8,S,\varphi \right) =\int d^4x\left[ -\frac 12\left( \partial
\wedge B_3\right) ^2-\frac 12\left( \partial \wedge B_8\right) ^2\right. 
\]
\begin{equation}
+\sum_{a=1}^3\left[ \frac{g^2S_a^2}2\left( \left( w_a\cdot B_\mu \right)
+\partial \varphi _a\right) ^2+\frac 12\left( \partial S_a\right) ^2-\frac
12b\left( S_a^2-v^2\right) ^2\right]  \label{suzphis}
\end{equation}

The actions (\ref{suzmaed}) and (\ref{suzphis}) have a double gauge
invariance. They are invariant with respect to the abelian gauge
transformation: 
\[
B_{3}\rightarrow B_{3}+\left( \partial \beta _{3}\right) 
\]
\[
\psi _{1}\rightarrow \psi _{1}e^{-ig\beta _{3}}\;\;\;\;\;\;\psi
_{2}\rightarrow \psi _{2}e^{i\frac{1}{2}g\beta _{3}}\;\;\;\;\;\psi
_{3}\rightarrow \psi _{3}e^{i\frac{1}{2}g\beta _{3}} 
\]
\begin{equation}
\varphi _{1}\rightarrow \varphi _{1}-\beta _{3}\;\;\;\;\;\varphi
_{2}\rightarrow \varphi _{2}+\frac{1}{2}\beta _{3}\;\;\;\;\varphi
_{3}\rightarrow \varphi _{3}+\frac{1}{2}\beta _{3}  \label{b38a}
\end{equation}
as well as to the abelian gauge transformation: 
\[
B_{8}\rightarrow B_{8}+\left( \partial \beta _{8}\right) 
\]
\[
\psi _{1}\rightarrow \psi _{1}\;\;\;\;\;\;\psi _{2}\rightarrow \psi _{2}e^{ig%
\frac{\sqrt{3}}{2}\beta _{8}}\;\;\;\;\;\;\psi _{3}\rightarrow \psi _{3}e^{-ig%
\frac{\sqrt{3}}{2}\beta _{8}} 
\]
\begin{equation}
\varphi _{1}\rightarrow \varphi _{1}\;\;\;\;\varphi _{2}\rightarrow \varphi
_{2}+\frac{\sqrt{3}}{2}\beta _{8}\;\;\;\;\varphi _{3}\rightarrow \varphi
_{3}-\frac{\sqrt{3}}{2}\beta _{8}  \label{b38b}
\end{equation}
We can impose a constraint on the phases $\varphi _{a}$ of the fields $\psi
_{a}$, namely: 
\begin{equation}
\varphi _{1}+\varphi _{2}+\varphi _{3}=0  \label{constrphi}
\end{equation}
The constraint (\ref{constrphi}) means that the degrees of freedom of the
system consist of the two gauge fields $B_{3}^{\mu }$ and $B_{8}^{\mu }$,
the three real fields $S_{a=1,2,3}$ and two phases chosen to be, for
example, $\varphi _{2}$ and $\varphi _{3}$, in which case $\varphi
_{1}=-\varphi _{2}-\varphi _{3}$.

The gauge transformations (\ref{b38a}) can then be expressed in the more
symmetric form: 
\[
\left( w_a\cdot B\right) \rightarrow \left( w_a\cdot B\right) +\left(
\partial \alpha _a\right) ,\;\;\;\;\psi _a\rightarrow \psi _ae^{ig\alpha
_a}\psi _a,\;\;\;\;\varphi _a\rightarrow \varphi _a-\alpha _a 
\]
\begin{equation}
\alpha _a=\left( w_a\cdot \beta \right) \;\;\;\;\;\;\;\;\left( \alpha
_1+\alpha _2+\alpha _3=0\right)  \label{wbgauge}
\end{equation}
where $\beta =\left( \beta _3,\beta _8\right) $.

Such gauge transformations are compatible with the abelian gauge fixing,
described in Sect.\ref{sec:occmonop}, which are only defined modulo residual 
$U\left( 1\right) $ transformations, to which the gauge transformations (\ref
{wbgauge}) belong.

The model action (\ref{suzmaed}), as all model actions so far, is QCD\
inspired but not derived. Variants to the form (\ref{suzphis}) have been
proposed and studied. For example, in reference \cite{Chernodub1999}, the
last term is replaced by $-\frac{1}{2}b\sum_{a=1}^{3}\left(
S_{a}^{2}-v_{a}^{2}\right) ^{2}$. In their 1990\ paper, Maedan, Matsubara
and Suzuki derive an effective Landau-Ginzburg model assuming the existence
of magnetic monopoles.\ The latter are assumed to interact minimally with
the gauge field $B^{\mu }$ to the dual field tensor $\bar{F}^{\mu \nu }$ and
a additional phenomenological interaction between the monopoles is
postulated \cite{Suzuki1990}. The partition function of the dual
superconductor can then be obtained by summing the trajectories of the
monopoles in a Feynman path integral, using a method developed by Bardakci
and Samuel \cite{Bardakci1978}. The summation is expressed in terms of an
action involving a gauged complex scalar field. This derivation of the
Landau-Ginzburg model is also considered in the 2003\ paper of Chernodub,
Ichiguro and Suzuki \cite{Chernodub2003-3}.

\section{The coupling of quarks to the gluon field}

\label{sec:quarkcup}

The color-$SU\left( 3\right) $ quarks may be regarded as color-electric
charges embedded in the dual superconductor, described by the action (\ref
{suzmaed}). They may be coupled to the system with the help of Dirac
strings, as done in Sect. \ref{sec:landact}. We generalize the expression (%
\ref{dotjbis}) by adding string sources $\left( \bar{G}_{3},\bar{G}%
_{8}\right) $ to the dual abelian field tensors $\left( \bar{F}_{3},\bar{F}%
_{8}\right) $: 
\begin{equation}
\bar{F}_{3}=\partial \wedge B_{3}+\bar{G}_{3}\;\;\;\;\;\;\;\bar{F}%
_{8}=\partial \wedge B_{8}+\bar{G}_{8}  \label{fcolg}
\end{equation}
The string sources $G_{3}$ and $G_{8}$ are antisymmetric tensors which
satisfy the equations: 
\begin{equation}
\partial _{\alpha }G_{3}^{\alpha \mu }=j_{3}^{\mu }\;\;\;\;\;\;\;\partial
_{\alpha }G_{8}^{\alpha \mu }=j_{8}^{\mu }  \label{maxsu3}
\end{equation}
where $\left( j_{3}^{\mu },j_{8}^{\mu }\right) $ are the color-electric
currents.\ If we assume that the quarks couple only to the abelian gluons $%
A_{3\mu }$ and $A_{8\mu }$, then they contribute a term to the lagrangian,
of the form: 
\begin{equation}
\bar{q}\left[ \gamma _{\mu }\left( i\partial ^{\mu }-eT_{3}A_{3\mu
}-eT_{8}A_{8\mu }\right) +m\right] q
\end{equation}
in which case the color-electric currents $j_{3}^{\mu }$ and $j_{8}^{\mu }$
would be: 
\begin{equation}
j_{3}^{\mu }=-e\bar{q}\gamma _{\mu }T_{3}q\;\;\;\;\;\;\;j_{8}^{\mu }=-e\bar{q%
}\gamma _{\mu }T_{8}q
\end{equation}
which agrees, of course, with the color charges (\ref{apsu:su3ch}) of the
quarks.

In the presence of color-electric charge, the action (\ref{suzmaed}) is
replaced by: 
\[
I\left( B_{3},B_{8},\psi ,\psi ^{\ast }\right) =\int d^{4}x\left\{ -\frac{1}{%
2}\left( \partial \wedge B_{3}+\bar{G}_{3}\right) ^{2}-\frac{1}{2}\left(
\partial \wedge B_{8}+\bar{G}_{8}\right) ^{2}\right. 
\]
\begin{equation}
\left. +\sum_{a=1}^{3}\left[ \frac{1}{2}\left| \left( \partial _{\mu }\psi
_{a}+ig\left( w_{a}\cdot B_{\mu }\right) \psi _{a}\right) \right| ^{2}-\frac{%
1}{2}b\left( \psi _{a}\psi _{a}^{\ast }-v^{2}\right) ^{2}\right] \right\}
\label{su3psi}
\end{equation}
Let us write:

\begin{equation}
B_3^\mu =\left( \chi _3,\vec{B}_3\right) \;\;\;\;\;\;\;B_8^\mu =\left( \chi
_8,\vec{B}_8\right)
\end{equation}
and, in analogy to (\ref{ehgstring}), let us express the string terms $%
G_{3,8}$ in terms of euclidean vectors $\vec{E}_{3,8}^{st}$ and $\vec{H}%
_{3,8}^{st}$: 
\[
E_{st,3}^i=-G_3^{0i}=\frac 12\varepsilon ^{0ijk}\bar{G}_{3,jk}\;\;\;\;\;%
\;H_{st,3}^i=-\bar{G}_3^{0i}=-\frac 12\varepsilon ^{0ijk}G_{3,jk} 
\]
\begin{equation}
E_{st,8}^i=-G_8^{0i}=\frac 12\varepsilon ^{0ijk}\bar{G}_{8,jk}\;\;\;\;\;%
\;H_{st,8}^i=-\bar{G}_8^{0i}=-\frac 12\varepsilon ^{0ijk}G_{8,jk}
\end{equation}
The action (\ref{su3psi}) can be expressed in terms of euclidean fields, as
in (\ref{epsi}):

\[
I_j\left( \psi ,\psi ^{*},\vec{B},\chi \right) =\int d^4x\left\{ +\frac
12\left( -\partial _t\vec{B}_3-\vec{\nabla}\chi _3+\vec{H}_3^{st}\right)
^2-\frac 12\left( -\vec{\nabla}\times \vec{B}_3+\vec{E}_3^{st}\right)
^2\right. 
\]
\[
+\frac 12\left( -\partial _t\vec{B}_8-\vec{\nabla}\chi _8+\vec{H}%
_8^{st}\right) ^2-\frac 12\left( -\vec{\nabla}\times \vec{B}_8+\vec{E}%
_8^{st}\right) ^2 
\]
\[
+\sum_{a=1}^3\left[ \frac 12\left| \left( \partial _t\psi _a+ig\left(
w_a\cdot \chi \right) \psi _a\right) \right| ^2-\frac 12\left| \left( \vec{%
\nabla}\psi _a-ig\left( w_a\cdot \vec{B}\right) \psi _a\right) \right|
^2\right. 
\]
\begin{equation}
\left. \left. -\frac 12b\left( \psi _a\psi _a^{*}-v^2\right) ^2\right]
\right\}  \label{eps38}
\end{equation}
The fields $\chi _3$ and $\chi _8$ act as constraints which we do not write
down.

\section{The energy of three static (quark) charges}

\begin{figure}[htbp]
\begin{center}
\includegraphics[width=10cm]{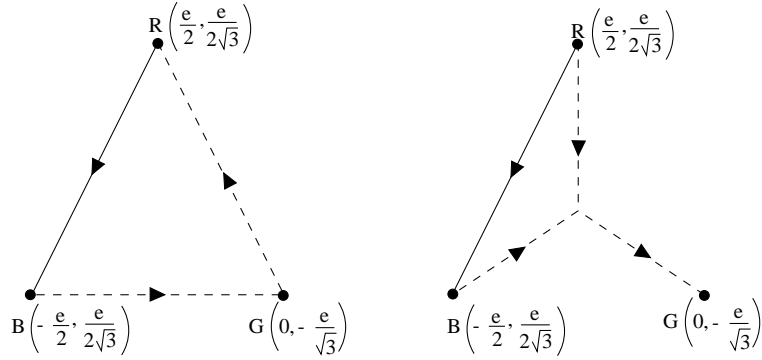}
\end{center}
\caption{The flux tubes formed by three static color-electric sources
denoted $R,B$ and $G$ (red,blue and green). The two color-electric charges
of each quark are denoted in parentheses. The full and dashed lines denote
respectively the flux tubes formed by the color-electric fields $\vec{E}_3$
and $\vec{E}_8 $. The left figure denotes the $\Delta $-shaped configuration
and the right figure denotes the Mercedes configuration. The side of the
triangle is $L$ and $M=\frac L{\protect\sqrt{3}}$ is the distance from a
summit to the center of the triangle.}
\label{fig:dbflux}
\end{figure}

Let us calculate the energy of three static color-electric charges, which,
for the sake of argument, we shall call quark charges. The quark charges are
listed in the table (\ref{apsu:su3ch}). Consider the case where three
quarks, red, blue and green, sit respectively at the points $\vec{R},\vec{B}$
and $\vec{G}$. Such a configuration is described by static color-charge
densities $\left( \rho _{3},\rho _{8}\right) $, with: 
\[
\rho _{3}\left( \vec{r}\right) =\frac{1}{2}e\delta \left( \vec{r}-\vec{R}%
\right) -\frac{1}{2}e\delta \left( \vec{r}-\vec{B}\right) 
\]
\begin{equation}
\rho _{8}\left( \vec{r}\right) =\frac{1}{2\sqrt{3}}e\delta \left( \vec{r}-%
\vec{R}\right) +\frac{1}{2\sqrt{3}}e\delta \left( \vec{r}-\vec{B}\right) -%
\frac{1}{\sqrt{3}}e\delta \left( \vec{r}-\vec{G}\right)  \label{su3charge}
\end{equation}
It is illustrated on Fig.\ref{fig:dbflux}. In Sect.\ref{sec:bdynamics}, we
showed that, in the presence of static charges, $\vec{H}^{st}=0$. The fields
are time-independent and the energy density is equal to minus the charge
density. It is simple to check that, when $\psi \psi ^{\ast }\neq 0$, the
constraints imposed by $\chi _{3}$ and $\chi _{8}$ are satisfied by $\chi
_{3}=\chi _{8}=0$. The energy obtained from (\ref{eps38}) reduces to: 
\[
\mathcal{E}\left( B,\psi ,\psi ^{\ast }\right) =\int d^{3}r\left\{ \frac{1}{2%
}\left( -\vec{\nabla}\times \vec{B}_{3}+\vec{E}_{st,3}\right) ^{2}+\frac{1}{2%
}\left( -\vec{\nabla}\times \vec{B}_{8}+\vec{E}_{st,8}\right) ^{2}\right. 
\]
\begin{equation}
\left. +\sum_{a=1}^{3}\left[ \frac{1}{2}\left| \left( \vec{\nabla}\psi
_{a}-ig\left( w_{a}\cdot \vec{B}\right) \psi _{a}\right) \right| ^{2}+\frac{1%
}{2}b\left( \psi _{a}\psi _{a}^{\ast }-v^{2}\right) ^{2}\right] \right\}
\label{ebpsi38}
\end{equation}
For the color-electric charges (\ref{su3charge}), the string terms $\vec{E}%
_{3}^{st}$ and $\vec{E}_{8}^{st}$ can be written in the form: 
\begin{equation}
\vec{E}_{st,3}\left( \vec{r}\right) =\frac{1}{2e}\int_{\vec{R}}^{\vec{B}}d%
\vec{Z}\;\delta \left( \vec{r}-\vec{Z}\right) \;\;\;\;\;\;\vec{E}%
_{st,8}\left( \vec{r}\right) =\frac{1}{2\sqrt{3}e}\left( \int_{\vec{B}}^{%
\vec{G}}-\int_{\vec{G}}^{\vec{R}}\right) d\vec{Z}\;\delta \left( \vec{r}-%
\vec{Z}\right)  \label{est38}
\end{equation}
where, for example, $\int_{\vec{R}}^{\vec{B}}d\vec{Z}$ is a line integral
along a path which stems from the point $\vec{R}$ and terminates at the
point $\vec{B}$. The expressions (\ref{est38}) satisfy the two equations (%
\ref{maxsu3}). Figure \ref{fig:dbflux} shows two examples of such strings.

We may proceed as in Sect. \ref{sec:ballcati}, and use the Ball-Caticha
trick \cite{Ball1988} which consists in expressing the string terms: 
\begin{equation}
\vec{E}_{st,3}\left( \vec{r}\right) =\vec{E}_3^0\left( \vec{r}\right) +\vec{B%
}_3^0\left( \vec{r}\right) \;\;\;\;\;\;\;\vec{E}_{st,8}\left( \vec{r}\right)
=\vec{E}_8^0\left( \vec{r}\right) +\vec{B}_8^0\left( \vec{r}\right)
\label{ebnor}
\end{equation}
in terms of the electric fields $\left( \vec{E}_3^0,\vec{E}_8^0\right) $ and
the gauge potentials $\left( \vec{B}_3^0,\vec{B}_8^0\right) $ which are
calculated in the \emph{normal} (non-superconducting) vacuum, in which $\psi
_a=\psi _a^{*}=0$. The electric fields in the normal vacuum are the usual
Coulomb fields produced by the color-electric charges (\ref{su3charge}) of
the quarks: 
\[
\vec{E}_3^0\left( \vec{r}\right) =-\frac e{8\pi }\vec{\nabla}\left( \frac
1{\left| \vec{r}-\vec{R}\right| }-\frac 1{\left| \vec{r}-\vec{B}\right|
}\right) 
\]
\begin{equation}
\vec{E}_8^0\left( \vec{r}\right) =-\frac e{8\pi \sqrt{3}}\vec{\nabla}\left(
\frac 1{\left| \vec{r}-\vec{R}\right| }+\frac 1{\left| \vec{r}-\vec{B}%
\right| }-\frac 2{\left| \vec{r}-\vec{G}\right| }\right)
\end{equation}
The gauge potentials $\left( \vec{B}_3^0,\vec{B}_8^0\right) $ in the normal
vacuum are given by expressions analogous to (\ref{botpath}): 
\[
\vec{B}_3^0\left( \vec{r}\right) =\frac e{4\pi }\vec{\nabla}_r\times \int_{%
\vec{R}}^{\vec{B}}d\vec{Z}\frac 1{\left| \vec{r}-\vec{Z}\right| } 
\]
\begin{equation}
\vec{B}_8^0\left( \vec{r}\right) =\frac e{4\pi \sqrt{3}}\vec{\nabla}_r\times
\left( \int_{\vec{B}}^{\vec{G}}-\int_{\vec{G}}^{\vec{R}}\right) d\vec{Z}%
\frac 1{\left| \vec{r}-\vec{Z}\right| }  \label{b38strings}
\end{equation}
When the forms (\ref{ebnor}) of the string terms are substituted into the
energy (\ref{ebpsi38}), an expression analogous to (\ref{ebbo}) is obtained: 
\[
\mathcal{E}\left( B,\psi ,\psi ^{*}\right) =-\frac{e^2}{8\pi \left| \vec{R}-%
\vec{B}\right| }-\frac{e^2}{24\pi \left| \vec{R}-\vec{G}\right| }-\frac{e^2}{%
24\pi \left| \vec{B}-\vec{G}\right| } 
\]
\[
+\int d^3r\left\{ \frac 12\left( -\vec{\nabla}\times \vec{B}_3+\vec{\nabla}%
\times \vec{B}_3^0\right) ^2+\frac 12\left( -\vec{\nabla}\times \vec{B}_8+%
\vec{\nabla}\times \vec{B}_8^0\right) ^2\right. 
\]
\begin{equation}
\left. +\sum_{a=1}^3\left[ \frac 12\left| \left( \vec{\nabla}\psi
_a-ig\left( w_a\cdot \vec{B}\right) \psi _a\right) \right| ^2+\frac
12b\left( \psi _a\psi _a^{*}-v^2\right) ^2\right] \right\}  \label{ebphis38}
\end{equation}

\section{Quantization of the electric and magnetic charges}

\label{sec:chqu38}

We can repeat the argument, given in Sect. \ref{sec:chargequant}, to show
that the energy (\ref{ebphis38}) does not depend on the shape of the strings
(\ref{b38strings}). A deformation of a 3-string (the string which defines $%
\vec{B}_3$) corresponds to the transformation $\vec{B}_3^0\rightarrow \vec{B}%
_3^0+\frac e{4\pi }\vec{\nabla}\Omega _3$ of the field $\vec{B}_3^0$.\ This
adds a term $\frac e{4\pi }\vec{\nabla}\times \left( \vec{\nabla}\Omega
_3\right) $ to the energy (\ref{ebphis38}) which can be compensated by a
singular gauge transformation of type (\ref{b38a}): 
\[
\vec{B}_3\rightarrow \vec{B}_3-\frac e{4\pi }\vec{\nabla}\Omega _3 
\]
\begin{equation}
\psi _1\rightarrow \psi _1e^{ieg\frac{\Omega _3}{4\pi }}\;\;\;\;\;\;\psi
_2\rightarrow \psi _2e^{-ieg\frac{\Omega _3}{8\pi }}\;\;\;\;\;\psi
_3\rightarrow \psi _3e^{-ieg\frac{\Omega _3}{8\pi }}  \label{b3gauge}
\end{equation}
thereby leaving the energy (\ref{ebphis38}) unchanged. As explained in Sect. 
\ref{sec:chargequant}, the solid angle $\Omega _3\left( \vec{r}\right) $ is
a discontinuous function of $\vec{r}$. We can, however, make the transformed
fields (\ref{b3gauge}) continuous and differentiable by imposing the
condition: 
\begin{equation}
eg=4n\pi  \label{charge4quant}
\end{equation}

Similarly, a deformation of an 8-string (the string which defines $\vec{B}_8$%
) corresponds to the transformation $\vec{B}_8^0\rightarrow \vec{B}%
_8^0+\frac e{4\pi \sqrt{3}}\vec{\nabla}\Omega _8$ of the field $\vec{B}_8^0$%
.\ The added gradient $\frac e{4\pi \sqrt{3}}\vec{\nabla}\Omega _8$ can be
compensated by a gauge transformation of type (\ref{b38b}): 
\[
B_8\rightarrow B_8-\frac e{4\pi \sqrt{3}}\vec{\nabla}\Omega _8 
\]
\begin{equation}
\psi _1\rightarrow \psi _1\;\;\;\;\;\;\psi _2\rightarrow \psi _2e^{-ig\frac
e{8\pi }\Omega _8}\;\;\;\;\;\;\psi _3\rightarrow \psi _3e^{ig\frac e{8\pi
}\Omega _8}  \label{b8gauge}
\end{equation}
thereby leaving the energy (\ref{ebphis38}) unchanged. Again, the
transformed fields (\ref{b8gauge}) become continuous and differentiable if
the charge quantization condition (\ref{charge4quant}) is satisfied. The
factor of 2, which distinguishes the quantization condition (\ref
{charge4quant}) from the quantization condition (\ref{chargequant}), is due
to the fact that electrons have electric charge $e$ whereas the quarks have
color-electric charge $\frac 12e$.

\section{Flux tubes formed by the electric and magnetic fields}

It is convenient to express the energy (\ref{ebphis38}) in terms of the
polar representation (\ref{psipol}) of the complex scalar fields: 
\[
\mathcal{E}\left( B,\varphi ,S\right) =-\frac{e^{2}}{8\pi \left| \vec{R}-%
\vec{B}\right| }-\frac{e^{2}}{24\pi \left| \vec{R}-\vec{G}\right| }-\frac{%
e^{2}}{24\pi \left| \vec{B}-\vec{G}\right| } 
\]
\[
\int d^{3}r\left\{ \frac{1}{2}\left( \vec{\nabla}\times \vec{B}_{3}+\vec{%
\nabla}\times \vec{B}_{3}^{0}\right) ^{2}+\frac{1}{2}\left( -\vec{\nabla}%
\times \vec{B}_{8}+\vec{\nabla}\times \vec{B}_{8}^{0}\right) ^{2}\right. 
\]
\begin{equation}
+\sum_{a=1}^{3}\left. \left[ \frac{g^{2}S_{a}^{2}}{2}\left( \left(
w_{a}\cdot \vec{B}\right) -\vec{\nabla}\varphi _{a}\right) ^{2}+\frac{1}{2}%
\left( \vec{\nabla}S_{a}\right) ^{2}+\frac{b}{2}\left(
S_{a}^{2}-v^{2}\right) ^{2}\right] \right\}  \label{eb38}
\end{equation}
We can choose to work in the gauge defined by $\alpha _{a}=\varphi _{a}$,
which is usually referred to as the unitary gauge. In this gauge, the energy
(\ref{eb38}) reduces to:

\[
\mathcal{E}\left( B,\varphi ,S\right) =-\frac{e^{2}}{8\pi \left| \vec{R}-%
\vec{B}\right| }-\frac{e^{2}}{24\pi \left| \vec{R}-\vec{G}\right| }-\frac{%
e^{2}}{24\pi \left| \vec{B}-\vec{G}\right| } 
\]
\[
\int d^{3}r\left\{ \frac{1}{2}\left( \vec{\nabla}\times \vec{B}_{3}+\vec{%
\nabla}\times \vec{B}_{3}^{0}\right) ^{2}+\frac{1}{2}\left( -\vec{\nabla}%
\times \vec{B}_{8}+\vec{\nabla}\times \vec{B}_{8}^{0}\right) ^{2}\right. 
\]
\begin{equation}
+\sum_{a=1}^{3}\left. \left[ \frac{g^{2}S_{a}^{2}}{2}\left( w_{a}\cdot \vec{B%
}\right) ^{2}+\frac{1}{2}\left( \vec{\nabla}S_{a}\right) ^{2}+\frac{b}{2}%
\left( S_{a}^{2}-v^{2}\right) ^{2}\right] \right\}  \label{eunit38}
\end{equation}
This expression of the energy, expressed in the unitary gauge, is not
independent of the shape of the strings, nor should it be, as explained in
Sect.\ref{sec:flux}. In the unitary gauge, the flux tubes develop around the
Dirac strings (\ref{b38strings}), which can be chosen to be straight lines
joining the charges. This choice presumably minimizes the energy.

The fields which minimize the energy (\ref{eunit38}) give rise to flux tubes
formed by the color-electric fields $\vec{E}_3$ and $\vec{E}_8$, as in the
Landau-Ginzburg model described in Sect. \ref{sec:fluxtube}. The electric
fields are given by the analogue of the expression (\ref{ehbg}): 
\begin{equation}
\vec{E}_\alpha =-\vec{\nabla}\times \vec{B}_\alpha +\vec{E}_{st,\alpha }=-%
\vec{\nabla}\times \vec{B}_\alpha +\vec{\nabla}\times \vec{B}_\alpha ^0+\vec{%
E}_\alpha ^0\;\;\;\;\;\left( \alpha =3,8\right)
\end{equation}
and the magnetic currents by the analogue of (\ref{gkmag}): 
\begin{equation}
\vec{j}_\alpha =-\vec{\nabla}\times \vec{E}_\alpha \;\;\;\;\;\left( \alpha
=3,8\right)
\end{equation}

Figure \ref{fig:dbflux} shows two possible equilibrium shapes of the flux
tubes.\ In the left figure, the flux tubes join the charges thereby forming
a $\Delta $-shaped pattern of electric fields. In the right figure, the $%
\vec{E}_{8}$ flux tubes converge first towards the center of the triangle,
thereby forming a Mercedes shaped pattern. Let us estimate the energy
contained in the flux tubes when the quarks are far apart. Each flux tube,
emanating from a quark gives rise to a linear confining potential, the
intensity of which is proportional to the squared charge of the quark it
stems from. It therefore contributes an energy which is proportional to the
square of the charge and to the length of the flux tube. For simplicity,
assume that the quarks are at the summit of an equilateral triangle of
length $L$, in which case the distance from a summit to the center of the
triangle is $M=\frac{L}{\sqrt{3}}$. In the $\Delta $-configuration, the
color-electric field $\vec{E}_{3}$ contributes an energy proportional to $%
\mathcal{E}_{3}\left( \Delta \right) =L\left( \frac{1}{2}e\right) ^{2}$ and
the color-electric field $\vec{E}_{8}$ an energy proportional to $\mathcal{E}%
_{8}\left( \Delta \right) =2L\left( \frac{1}{2\sqrt{3}}e\right) ^{2}$, so
that the total energy of the $\Delta $-shaped configuration is proportional
to: 
\begin{equation}
\mathcal{E}\left( \Delta \right) =\mathcal{E}_{3}\left( \Delta \right) +%
\mathcal{E}_{8}\left( \Delta \right) =L\left( \frac{1}{2}e\right)
^{2}+2L\left( \frac{1}{2\sqrt{3}}e\right) ^{2}=\frac{5}{12}%
Le^{2}=0.417\;Le^{2}
\end{equation}
In the Mercedes configuration, the color-electric field $\vec{E}_{3}$ still
forms a flux tube joining the red and blue quarks and contributes the same
energy as in the $\Delta $-configuration. The color-electric field $\vec{E}%
_{8}$ contributes an energy proportional to $\mathcal{E}_{8}\left( Y\right)
=2M\left( \frac{1}{2\sqrt{3}}e\right) ^{2}+M\left( \frac{1}{\sqrt{3}}%
e\right) ^{2}$, so that the total energy of the Y-shaped configuration is
proportional to: 
\begin{equation}
\mathcal{E}\left( Y\right) =\mathcal{E}_{3}\left( \Delta \right) +\mathcal{E}%
_{8}\left( Y\right) =\frac{1}{4}Le^{2}+\frac{1}{2}Me^{2}=\frac{1}{4}Le^{2}+%
\frac{1}{2}\frac{L}{\sqrt{3}}e^{2}=0.539\;Le^{2}
\end{equation}
Thus the $\Delta $-shaped configuration is energetically favored in this
estimate, by about 25\% of the total energy. However, in a type I
superconductor, flux tubes attract, so that this attraction could modify the
preceding estimate. A numerical calculation would be required to check this.
Lattice simulations appear to favor the $\Delta $-shaped configuration as
long as the distance $L$ between the quarks is less than $0.7\;fm$ \cite
{Forcrand2002}.

The actual flux tubes formed by the color-electric fields $\vec{E}_{3}$ and $%
\vec{E}_{8}$, obtained by minimizing the energy (\ref{ebphis38}) of the
Mercedes configuration are computed and displayed in the 1993 paper of
Kamizawa, Matsubara, Shiba and Suzuki \cite{Suzuki1993}.

\section{A Weyl symmetric form of the action}

\label{sec:weyl}

We are dealing with two gauge fields and three complex scalar fields and
this introduces an apparent asymmetry in the model, which, in fact, has the
virtue of respecting Weyl symmetry \footnote{%
The Weyl symmetry refers to the symmetry with respect to the exchange of the
three colors defined in the fundamental representation of the $SU\left(
3\right) $ group.}. We can obtain a form of the action, in which Weyl
symmetry is more explicit, by expressing the two abelian dual gauge fields $%
B^{\mu }=\left( B_{3}^{\mu },B_{8}^{\mu }\right) $in terms of the three
gauge fields $b_{\mu a}=\left( w_{a}\cdot B_{\mu }\right) $. Using the
completeness relations (\ref{apsu:wij}) or (\ref{apsu:wijbis}), we can write
the vector $B^{\mu }=\left( B_{3}^{\mu },B_{8}^{\mu }\right) $as follows: 
\begin{equation}
B^{\mu }=\frac{2}{3}\sum_{a=1}^{3}w_{a}\left( w_{a}\cdot B^{\mu }\right) =%
\frac{2}{3}\sum_{a=1}^{3}w_{a}b_{a}^{\mu }
\end{equation}
The three fields $b_{a\mu }$ are: 
\begin{equation}
b_{a}^{\mu }=\left( w_{a}\cdot B^{\mu }\right) \;\;\;\;\;\;b_{1}^{\mu
}+b_{2}^{\mu }+b_{3}^{\mu }=0  \label{awmu}
\end{equation}
They are not independent because they sum up to zero. They are, in fact, the
following linear combinations of the two fields $B_{3}^{\mu }$ and $%
B_{8}^{\mu }$: 
\[
b_{1}^{\mu }=\left( w_{1}\cdot B^{\mu }\right) =B_{3}^{\mu } 
\]
\begin{equation}
b_{2}^{\mu }=\left( w_{2}\cdot B^{\mu }\right) =-\frac{1}{2}B_{3}^{\mu }-%
\frac{\sqrt{3}}{2}B_{8}^{\mu }\;\;\;\;\;\;\;\;b_{3}^{\mu }=\left( w_{3}\cdot
B^{\mu }\right) =-\frac{1}{2}B_{3}^{\mu }+\frac{\sqrt{3}}{2}B_{8}^{\mu }
\label{bb}
\end{equation}
We have: 
\[
\left( B_{\mu }\cdot B^{\mu }\right) =B_{3\mu }B_{3}^{\mu }+B_{8\mu
}B_{8}^{\mu }=\frac{2}{3}\sum_{a=1}^{3}b_{a\mu }b_{a}^{\mu } 
\]
\[
\left( \partial \wedge B\right) ^{2}=\left( \partial \wedge B_{3}\right)
^{2}+\left( \partial \wedge B_{8}\right) ^{2}=\frac{2}{3}\sum_{a=1}^{3}%
\left( \partial \wedge b_{a}\right) ^{2} 
\]
\begin{equation}
\partial ^{\mu }\psi _{a}+ig\left( w_{a}\cdot B^{\mu }\right) \psi
_{a}=\partial ^{\mu }\psi _{a}+igb_{a}^{\mu }\psi _{a}  \label{bb23}
\end{equation}

Similarly, the two string terms $G=\left( G_3,G_8\right) $ can be expressed
in terms of three string terms $g_a=\left( w_a\cdot G\right) $ as follows: 
\begin{equation}
G=\frac 23\sum_{a=1}^3w_ag_a
\end{equation}
where: 
\begin{equation}
g_a=\left( w_a\cdot G\right) \;\;\;\;\;\;g_1+g_2+g_3=0
\end{equation}
The action (\ref{su3psi}) can then be written in the Weyl symmetric form: 
\[
I\left( b,\psi ,\psi ^{*}\right) =\int d^4x\sum_{a=1}^3\left[ -\frac
13\left( \partial \wedge b_a+\bar{g}_a\right) ^2+\frac 12\left| \partial
_\mu \psi _a+igb_{a\mu }\psi _a\right| ^2-\frac 12b\left( \psi _a\psi
_a^{*}-v^2\right) ^2\right] 
\]
\begin{equation}
=\int d^4x\sum_{a=1}^3\left[ -\frac 13\left( \partial \wedge b_a+\bar{g}%
_a\right) ^2+\frac 12\left( \partial S_a\right) ^2+\frac{g^2S_a^2}2\left(
b_a+\left( \partial \varphi _a\right) \right) ^2-\frac 12b\left(
S_a^2-v^2\right) ^2\right]  \label{ibsu3}
\end{equation}
When variations of the action are considered, the following constraints need
to be taken into account: 
\begin{equation}
\sum_{a=1}^3\varphi _a=0\;\;\;\;\;\;\sum_{a=1}^3b_a^\mu
=0\;\;\;\;\;\;\sum_{a=1}^3g_a^{\mu \nu }=0
\end{equation}
The possibly offending factors $\frac 13$ could be changed to $\frac 12$ by
redefining the fields $b_a$ and the source terms $\bar{g}_a$ but we do not
feel compelled to do this.

We can obtain a Weyl symmetric form of the action (\ref{eps38}) by defining: 
\begin{equation}
b_a^\mu =\left( \eta _a,\vec{b}_a\right) \;\;\;\;\;\;\;\;\eta _1+\eta
_2+\eta _3=0\;\;\;\;\;\;\;\;\;\vec{b}_1+\vec{b}_2+\vec{b}_3=0
\end{equation}
We obtain: 
\[
I_j\left( \psi ,\psi ^{*},\vec{b},\eta \right) =\int d^4x\sum_{a=1}^3\left\{
+\frac 13\left( -\partial _t\vec{b}_a-\vec{\nabla}\eta _a+\vec{h}%
_a^{st}\right) ^2-\frac 13\left( -\vec{\nabla}\times \vec{b}_a+\vec{e}%
_a^{st}\right) ^2\right. 
\]
\begin{equation}
\left. +\frac 12\left| \left( \partial _t\psi _a+ig\eta _a\psi _a\right)
\right| ^2-\frac 12\left| \left( \vec{\nabla}\psi _a-ig\vec{b}_a\psi
_a\right) \right| ^2-\frac 12b\left( \psi _a\psi _a^{*}-v^2\right) ^2\right\}
\label{iweyl}
\end{equation}
where: 
\begin{equation}
\vec{e}_{st,a}=\left( w_a\cdot \vec{E}_{st}\right) \;\;\;\;\;\;\;\;\;\;\;%
\vec{h}_{st,a}=\left( w_a\cdot \vec{H}_{st}\right)  \label{ehst}
\end{equation}

When the color-charge densities $\left( \rho _3,\rho _8\right) $of the
quarks are given by (\ref{su3charge}), the corresponding color-charge
densities $\rho _a=\left( w_a\cdot \rho \right) $ have the remarkable Weyl
symmetric form: 
\[
\rho _1\left( \vec{r}\right) =\vec{w}_1\cdot \vec{\rho}=\frac 12e\delta
\left( \vec{r}-\vec{R}\right) -\frac 12e\delta \left( \vec{r}-\vec{B}\right) 
\]
\[
\rho _2\left( \vec{r}\right) =\vec{w}_2\cdot \vec{\rho}=-\frac 12e\delta
\left( \vec{r}-\vec{R}\right) +\frac 12e\delta \left( \vec{r}-\vec{G}\right) 
\]
\begin{equation}
\rho _3\left( \vec{r}\right) =\vec{w}_3\cdot \vec{\rho}=\frac 12e\delta
\left( \vec{r}-\vec{B}\right) -\frac 12e\delta \left( \vec{r}-\vec{G}\right)
\label{rho123}
\end{equation}

The sources (\ref{ehst}) can be computed from the sources (\ref{est38}): 
\[
\vec{e}_{st,1}\left( \vec{r}\right) =\frac 1{2e}\int_{\vec{R}}^{\vec{B}}d%
\vec{Z}\;\delta \left( \vec{r}-\vec{Z}\right) 
\]
\begin{equation}
\vec{e}_{st,2}\left( \vec{r}\right) =\frac 1{2e}\int_{\vec{G}}^{\vec{R}}d%
\vec{Z}\;\delta \left( \vec{r}-\vec{Z}\right) \;\;\;\;\;\;\;\vec{e}%
_{st,3}\left( \vec{r}\right) =\frac 1{2e}\int_{\vec{B}}^{\vec{G}}d\vec{Z}%
\;\delta \left( \vec{r}-\vec{Z}\right)  \label{est123}
\end{equation}
which is compatible with the constraint $\vec{e}_{st,1}+\vec{e}_{st,2}+\vec{e%
}_{st,3}=0$. In the expressions above, $\int_{\vec{R}}^{\vec{B}}d\vec{Z}$
denotes a line integral along a path which begins at the point $\vec{R}$ and
ends at the point $\vec{B}$. The sources $g_a$ are strings which satisfy the
equations: 
\begin{equation}
\partial \cdot g_a=j_a  \label{su3dja}
\end{equation}
where: 
\begin{equation}
j_a^\mu =\left( w_a\cdot j^\mu \right) \;\;\;\;\;\;\;\;\;\;\;\;\left(
j_1^\mu +j_2^\mu +j_3^\mu \right) =0
\end{equation}
In the presence of static sources $\rho _a\left( \vec{r}\right) $, we have $%
\vec{h}_a^{st}=0$ and the fields become time-independent. The energy,
obtained fro; (\ref{iweyl}), is: 
\[
\mathcal{E}_j\left( \psi ,\psi ^{*},\vec{b},\eta \right) =\int
d^3r\sum_{a=1}^3\left\{ -\frac 13\left( \vec{\nabla}\eta _a\right) ^2-\frac
12g^2\eta _a^2\psi _a\psi _a^{*}\right. 
\]
\begin{equation}
\left. +\frac 13\left( -\vec{\nabla}\times \vec{b}_a+\vec{e}_{st,a}\right)
^2+\frac 12\left| \left( \vec{\nabla}\psi _a-ig\vec{b}_a\psi _a\right)
\right| ^2+\frac 12b\left( \psi _a\psi _a^{*}-v^2\right) ^2\right\}
\end{equation}
The constraints imposed by the $\eta _a$, or, more rigorously, by the
independent fields $\eta _3$ and $\eta _8$ are satisfied with $\eta _a=0$ so
that the energy is: 
\[
\mathcal{E}_j\left( \psi ,\psi ^{*},\vec{b}\right) 
\]
\begin{equation}
=\int d^3r\sum_{a=1}^3\left[ +\frac 13\left( -\vec{\nabla}\times \vec{b}_a+%
\vec{e}_{st,a}\right) ^2+\frac 12\left| \left( \vec{\nabla}\psi _a-ig\vec{b}%
_a\psi _a\right) \right| ^2+\frac 12b\left( \psi _a\psi _a^{*}-v^2\right)
^2\right]  \label{eweyl}
\end{equation}
When we attempt to minimize the energy (\ref{eweyl}) with respect to
variations of the fields, we must remember the constraints: 
\begin{equation}
\vec{b}_1\left( \vec{r}\right) +\vec{b}_2\left( \vec{r}\right) +\vec{b}%
_3\left( \vec{r}\right) =0\;\;\;\;\;\varphi _1\left( \vec{r}\right) +\varphi
_2\left( \vec{r}\right) +\varphi _2\left( \vec{r}\right)
=0\;\;\;\;\;\;\left( \psi _a=S_ae^{ig\varphi _a}\right)
\end{equation}

\begin{figure}[htbp]
\begin{center}
\includegraphics[width=10cm]{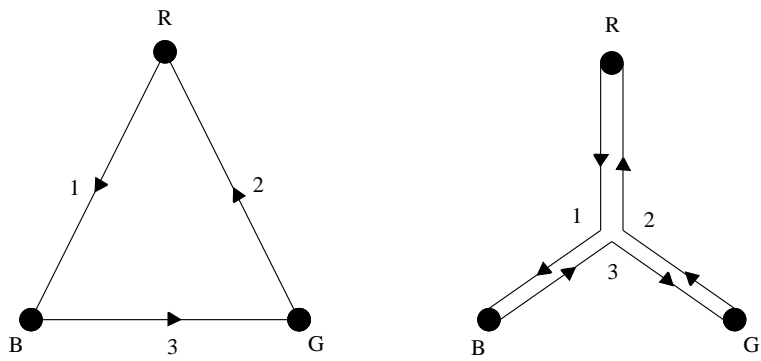}
\end{center}
\caption{The Weyl symmetric representation of the flux tubes formed by three
static color-electric sources denoted $R,B$ and $G$ (red,blue and green).
The left figure denotes the $\Delta $-shaped configuration and the right
figure denotes the Mercedes configuration.}
\label{fig:flux38}
\end{figure}

Figure (\ref{fig:flux38}) shows the flux tubes which are formed by the three
electric fields: 
\begin{equation}
\vec{e}_1=\vec{E}_3\;\;\;\;\;\vec{e}_2=-\frac 12\vec{E}_3-\frac{\sqrt{3}}2%
\vec{E}_8\;\;\;\;\;\;\vec{e}_3=-\frac 12\vec{E}_3+\frac{\sqrt{3}}2\vec{E}_8
\end{equation}
as well as the string terms (\ref{est123}). The figure shows two possible
equilibrium shapes of the flux tubes.\ In the $\Delta $-shaped pattern, the
flux tubes run in straight lines from one color source to the other, whereas
in the Mercedes configuration, they converge first toward the center of the
triangle. Flux tubes running in opposite directions attract each other and
this may lower the energy of the Mercedes configuration. Flux tubes in the
Weyl symmetric representation are discussed in the 1998 paper of Chernodub
and Komarov \cite{Chernodub1998}.

\appendix

\chapter{Vectors, tensors and their duality transformations}

\label{ap:apvt}

\section{Compact notation}

Scalar and vector products of two vectors $A^\mu $ and $B^\mu $ are written
as: 
\begin{equation}
A\cdot B=A_\mu B^\mu \;\;\;\;\;\;\;\;\left( A\wedge B\right) _{\mu \nu
}=A_\mu B_\nu -A_\nu B_\mu \;\;\;\;\;\;\left( \partial \wedge A\right) _{\mu
\nu }=\left( \partial _\mu A_\nu \right) -\left( \partial _\nu A_\mu \right)
\end{equation}
If $S^{\mu \nu }=-S^{\nu \mu }$ is an antisymmetric tensor, its contractions
with a vector $A^\mu $ are written in the form: 
\begin{equation}
\left( A\cdot S\right) ^\mu =A_\nu S^{\nu \mu }=-S^{\mu \nu }A_\nu =-\left(
S\cdot A\right) ^\mu
\end{equation}
If $S$ and $T$ are two antisymmetric tensors, we write: 
\begin{equation}
S\cdot T=ST=\frac 12S_{\mu \nu }T^{\mu \nu }\;\;\;\;\;\;S\cdot S\equiv
S^2=\frac 12S_{\mu \nu }S^{\mu \nu }
\end{equation}
The dot ''$\cdot $'' is not necessary when it is clear which symbols are
vectors, which are antisymmetric tensors, etc. In such cases we can write,
for example: 
\begin{equation}
\left( TA\right) _\mu =T_{\mu \nu }A^\nu \;\;\;\;\;\;ASB=A_\mu S^{\mu \nu
}B_\nu \;\;etc.
\end{equation}

\section{The metric $g^{\mu \nu }$ and the antisymmetric tensor $\varepsilon
^{\mu \nu \alpha \beta }$}

In Minkowski space, we use the metric: 
\begin{equation}
g^{\mu \nu }=g_{\mu \nu }=\left( 
\begin{array}{llll}
1 & 0 & 0 & 0 \\ 
0 & -1 & 0 & 0 \\ 
0 & 0 & -1 & 0 \\ 
0 & 0 & 0 & -1
\end{array}
\right) \;\;\;\;\;\;\det g=-1
\end{equation}
The antisymmetric tensor $\varepsilon ^{\mu \nu \alpha \beta }$ is: 
\begin{equation}
\varepsilon ^{0123}=-\varepsilon ^{1023}=...=1\;\;\;\;\;\;\;\;\;\varepsilon
_{\mu \nu \alpha \beta }=-\varepsilon ^{\mu \nu \alpha \beta }
\end{equation}
When indices of $\varepsilon _{\mu \nu \alpha \beta }$ are contracted, it is
useful to think in terms of scalar products of antisymmetric ''states'':

\begin{equation}
\frac 1{4!}\varepsilon ^{\alpha \beta \gamma \mu }\varepsilon _{\alpha \beta
\gamma {}\mu }=\frac 1{4!}\det g=-1
\end{equation}

\[
\varepsilon ^{\mu \alpha \beta \gamma }\varepsilon _{\mu \alpha ^{\prime
}\beta ^{\prime }\gamma ^{\prime }}=-\left\langle \alpha \beta \gamma \left|
\alpha ^{\prime }\beta ^{\prime }\gamma ^{\prime }\right. \right\rangle
=-\det \left( 
\begin{array}{lll}
g_{\alpha ^{\prime }}^\alpha & g_{\beta ^{\prime }}^\alpha & g_{\gamma
^{\prime }}^\alpha \\ 
g_{\alpha ^{\prime }}^\beta & g_{\beta ^{\prime }}^\beta & g_{\gamma
^{\prime }}^\beta \\ 
g_{\alpha ^{\prime }}^\gamma & g_{\beta ^{\prime }}^\gamma & g_{\beta
^{\prime }}^\gamma
\end{array}
\right) 
\]
\[
\frac 12\varepsilon ^{\mu \nu \alpha \beta }\varepsilon _{\mu \nu \alpha
^{\prime }\beta ^{\prime }}=-\left\langle \left. \alpha \beta \right| \alpha
^{\prime }\beta ^{\prime }\right\rangle =-\det \left( 
\begin{array}{ll}
g_{\alpha ^{\prime }}^\alpha & g_{\beta ^{\prime }}^\alpha \\ 
g_{\alpha ^{\prime }}^\beta & g_{\beta ^{\prime }}^\beta
\end{array}
\right) 
\]
\[
\frac 1{3!}\varepsilon ^{\alpha \beta \gamma \mu }\varepsilon _{\alpha \beta
\gamma {}\nu }=-\left\langle \mu \left| \nu \right. \right\rangle =-g_\nu
^\mu 
\]
\begin{equation}
\frac 1{4!}\varepsilon ^{\alpha \beta \gamma \mu }\varepsilon _{\alpha \beta
\gamma {}\mu }=-1  \label{apvt:epscontr}
\end{equation}
The rule is that upper indices appear in the bras, lower indices appear in
the kets.

\section{Vectors and their dual form}

The metric $g^{\mu \nu }$ acts as the unit operator acting on a vector $%
A^{\mu }$: 
\begin{equation}
gA=A\;\;\;\;\;\;g^{\mu \nu }A_{\nu }=A^{\mu }
\end{equation}

\subsection{Longitudinal and transverse components of vectors}

\label{apvt:vectors}

We define transverse and longitudinal projectors: 
\begin{equation}
T^{\mu \nu }=\left( g^{\mu \nu }-\frac{\partial ^\mu \partial ^\nu }{%
\partial ^2}\right) \;\;\;\;\;\;L^{\mu \nu }=\frac{\partial ^\mu \partial
^\nu }{\partial ^2}  \label{apvt:tldd}
\end{equation}
They have the properties: 
\begin{equation}
T^2=T\;\;\;L^2=L\;\;\;TL=LT=0\;\;\;T+L=g\;\;\;A=\left( T+L\right) A
\label{apvt:tlprop}
\end{equation}
if: 
\begin{equation}
A^T=TA\;\;\;\;\;\;A^L=LA\;\;\;\;\;\;A=A^T+A^L  \label{apvt:atl}
\end{equation}
Then: 
\begin{equation}
\partial \cdot A=\partial \cdot A^L\;\;\;\;\;\;\partial \wedge A=\partial
\wedge A^T\;\;\;\;\;\;\partial \cdot A^T=0\;\;\;\;\;\;\partial \wedge A^L=0
\end{equation}

\subsection{Identities involving vectors}

\label{apvt:sec:vecident}

In the following, $A^\mu \left( x\right) ,B^\mu \left( x\right) $ and $C^\mu
\left( x\right) $ are vector fields and $n^\mu $ is a $x$-independent vector:

\begin{equation}
A\cdot \left( B\wedge C\right) =\left( A\cdot B\right) C-\left( A\cdot
C\right) B  \label{apvt:vec1}
\end{equation}
\begin{equation}
n\cdot \left( \partial \wedge A\right) =\left( n\cdot \partial \right)
A-\partial \left( n\cdot A\right)  \label{apvt:vec2}
\end{equation}
\begin{equation}
\partial \cdot \left( n\wedge A\right) =\left( n\cdot \partial \right)
A-n\left( \partial \cdot A\right)  \label{apvt:vec3}
\end{equation}
\begin{equation}
\left( A\cdot \overline{B\wedge C}\right) ^\mu =-\varepsilon ^{\mu \alpha
\beta \gamma }A_\alpha B_\beta C_\gamma  \label{apvt:vec4}
\end{equation}
\begin{equation}
\left( n\cdot \overline{\partial \wedge A}\right) ^\mu =-\varepsilon ^{\mu
\alpha \beta \gamma }n_a\left( \partial _\beta A_\gamma \right)
\label{apvt:vec5}
\end{equation}
\begin{equation}
\left( \partial \cdot \overline{n\wedge A}\right) ^\mu =\varepsilon ^{\mu
\alpha \beta \gamma }n_\alpha \left( \partial _\beta A_\gamma \right)
=-n\cdot \left( \overline{\partial \wedge A}\right)  \label{apvt:vec6}
\end{equation}
\begin{equation}
\partial _\alpha \left( n\wedge \left( n\cdot \left( \partial \wedge
A\right) \right) \right) ^{\alpha \mu }=\left( n\cdot \partial \right)
^2A^\mu -\left( n\cdot \partial \right) \partial ^\mu \left( n\cdot A\right)
-n^\mu \left( n\cdot \partial \right) \left( \partial \cdot A\right)
+\partial ^2n^\mu \left( n\cdot A\right)  \label{apvt:vec7}
\end{equation}

\begin{equation}
\partial \cdot \left( \partial \wedge A\right) =\partial
^2TA\;\;\;\;\;\;\partial \cdot \overline{\partial \wedge A}=0
\label{apvt:vec8}
\end{equation}
\begin{equation}
\partial \cdot \left( TA\right) =0\;\;\;\;\;\;\partial \cdot \left(
LA\right) =\partial \cdot A\;\;\;\;\;\;\partial \wedge \left( TA\right)
=\partial \wedge A\;\;\;\;\;\;\partial \wedge \left( LA\right) =0
\label{apvt:vec9}
\end{equation}
In the following identities, surface terms are neglected: 
\[
\int d^4x\;\left( n\cdot \partial \wedge A\right) \cdot \left( n\cdot 
\overline{\partial \wedge B}\right) =-\int d^4x\;\left( n\cdot \partial
\wedge B\right) \cdot \left( n\cdot \overline{\partial \wedge A}\right) 
\]
\[
=\int d^4x\;\varepsilon ^{\mu \nu \alpha \beta }A_\mu n_\nu \left( n\cdot
\partial \right) \partial _\alpha B_\beta 
\]
\begin{equation}
=-\int d^4x\;A\left( n\cdot \partial \right) \left[ n\cdot \overline{%
\partial \wedge B}\right] =+\int d^4x\;B\left( n\cdot \partial \right)
\left[ n\cdot \overline{\partial \wedge A}\right]  \label{apvt:idab}
\end{equation}
\[
-\int d^4x\;\left[ n\cdot \left( \partial \wedge A\right) \right] \left[
n\cdot \left( \partial \wedge A\right) \right] 
\]
\begin{equation}
=\int d^4x\left[ A\left( n\cdot \partial \right) ^2A-A\left( n\cdot \partial
\right) \partial \left( n\cdot A\right) -An\left( n\cdot \partial \right)
\left( \partial \cdot A\right) +\left( A\cdot n\right) \partial ^2\left(
n\cdot A\right) \right]  \label{apvt:idaa}
\end{equation}
\[
\int d^4x\left( \partial \wedge A\right) \left( \partial \wedge B\right)
=-\int d^4xA\left[ \partial \cdot \left( \partial \wedge B\right) \right]
=-\int d^4xA\partial ^2TB 
\]
\begin{equation}
\int d^4x\left( \partial \cdot A\right) \left( \partial \cdot B\right)
=-\int d^4xA\partial \left( \partial \cdot B\right) =-\int d^4xA\partial ^2LB
\label{apvt:dadb}
\end{equation}

\subsection{Identities involving vectors and antisymmetric tensors}

Other useful identities, valid for an antisymmetric tensor $S$ and a vector $%
A$: 
\begin{equation}
A\cdot \left( A\cdot S\right) =0
\end{equation}
\begin{equation}
L\left( \partial \cdot S\right) =0\;\;\;\;\;\;T\left( \partial \cdot
S\right) =\left( \partial \cdot S\right)
\end{equation}
\begin{equation}
\int d^4x\;\left( \partial \wedge A\right) S=-\int d^4x\,A\cdot \left(
\partial \cdot S\right) \;\;\;\;\;\;\int d^4x\;\left( \overline{\partial
\wedge A}\right) S=-\int d^4x\,A\cdot \left( \partial \cdot \bar{S}\right)
\label{apvt:sda}
\end{equation}

\section{Antisymmetric tensors and their dual form}

The operator: 
\begin{equation}
G_{\mu \nu ,\alpha \beta }=g_{\mu \alpha }g_{\nu \beta }-g_{\mu \beta
}g_{\mu \beta }  \label{apvt:g}
\end{equation}
acts as the unit operator on an antisymmetric tensor $S$ : 
\begin{equation}
GS=S\;\;\;\;\;\;\frac 12G_{\mu \nu ,\alpha \beta }S^{\alpha \beta }=S_{\mu
\nu }
\end{equation}

\subsection{The dual of an antisymmetric tensor}

The dual of an antisymmetric tensor $S$ is denoted as $\bar{S}$: 
\begin{equation}
\bar{S}^{\mu \nu }=\frac 12\varepsilon ^{\mu \nu \alpha \beta }S_{\alpha
\beta }=\left( \varepsilon S\right) ^{\mu \nu }
\end{equation}
We can regard $\varepsilon $ as the operator $\varepsilon _{\mu \nu ,\alpha
\beta }=\varepsilon _{\mu \nu \alpha \beta }$ so as to write: 
\begin{equation}
\bar{S}_{\mu \nu }=\left( \varepsilon S\right) _{\mu \nu }=\varepsilon _{\mu
\nu \alpha \beta }S^{\alpha \beta }\;\;\;\;\left( S\varepsilon \right) _{\mu
\nu }=S^{\alpha \beta }\varepsilon _{\alpha \beta \mu \nu }=\bar{S}
\end{equation}
\begin{equation}
\varepsilon ^2=-G
\end{equation}
and the duality transformation of antisymmetric tensors is reversible with a
minus sign: 
\begin{equation}
\bar{S}=\varepsilon S\;\;\;\;\;S=-\left( \varepsilon \varepsilon S\right)
=-\varepsilon \bar{S}  \label{apvt:fdual}
\end{equation}
If $S$ and $T$ are two antisymmetric tensors, we have: 
\begin{equation}
TS=-T\varepsilon \varepsilon S=-\bar{T}\bar{S}\;\;\;\;\;\;\bar{T}%
S=T\varepsilon S=T\bar{S}
\end{equation}

\subsection{The Zwanziger identities}

An antisymmetric tensor $S^{\mu \nu }$ is entirely defined by the two
vectors $n\cdot S$ and $n\cdot \bar{S}$, where $n^\mu $ is a given vector.\
Let $S$ and $T$ be two antisymmetric tensors.\ If $n\cdot S=n\cdot T$ and $%
n\cdot \bar{S}=n\cdot \bar{T}$, then $S=T$. To understand why, let us choose 
$n^\mu $ space-like: 
\begin{equation}
n^\mu =\left( 0,\vec{n}\right) \;\;\;\;n^2=-\vec{n}\cdot \vec{n}
\end{equation}
Consider, for example, the six components of the electromagnetic field
tensor $F^{\mu \nu }=\left( \vec{E},\vec{H}\right) $ which can be written in
terms of two Euclidean 3-dimensional vectors $\vec{E}$ and $\vec{H}$ as in (%
\ref{fmunu}) and (\ref{fmunudual}).\ The argument is valid, of course, for
any other antisymmetric tensor. The components of the 4-vectors $n\cdot F$
and $n\cdot \bar{F}$ are: 
\[
\left( n\cdot F\right) ^0=-\vec{n}\cdot \vec{E}\;\;\;\;\;\;\left( n\cdot
F\right) ^i=-\left( \vec{n}\times \vec{H}\right) _i 
\]
\begin{equation}
\left( n\cdot \overline{F}\right) ^0=-\vec{n}\cdot \vec{H}\;\;\;\;\;\;\left(
n\cdot \overline{F}\right) ^i=\left( \vec{n}\times \vec{E}\right) _i
\label{apvt:nfeh}
\end{equation}
We see that $n\cdot F$ represents the longitudinal part of $\vec{E}$ and the
transverse part of $\vec{H}$, whereas $n\cdot \bar{F}$ represents the
transverse part of $\vec{E}$ and the longitudinal part of $\vec{H}$. The
vectors $\vec{n}\times \vec{E}$ and $\vec{n}\times \vec{H}$ have only two
components because they are orthogonal to the vector $\vec{n}$.\ Therefore
the eight components of $n\cdot F$ and $n\cdot \bar{F}$ represent, in fact,
the six independent components of the antisymmetric tensor $F^{\mu \nu }$.

The first Zwanziger identity \cite{Zwanziger1971} consists in writing any
antisymmetric tensor $F^{\mu \nu }$ in the form: 
\begin{equation}
F=\frac 1{n^2}\left( n\wedge \left( n\cdot F\right) -\overline{n\wedge
\left( n\cdot \overline{F}\right) }\right)  \label{apvt:zwanident1}
\end{equation}
which is correct because the left and right hand sides give the same value
for $n\cdot F$ and $n\cdot \bar{F}$.

The second Zwanziger identity reads: 
\begin{equation}
F^2=\frac 1{n^2}\left( \left( n\cdot F\right) ^2-\left( n\cdot \bar{F}%
\right) ^2\right)  \label{apvt:zwanident2}
\end{equation}
From (\ref{apvt:nfeh}), we see that:: 
\begin{equation}
\left( n\cdot F\right) ^2=\left( \vec{n}\cdot \vec{E}\right) ^2-\left( \vec{n%
}\times \vec{H}\right) ^2\;\;\;\;\;\;\left( n\cdot \bar{F}\right) ^2=\left( 
\vec{n}\cdot \vec{H}\right) ^2-\left( \vec{n}\times \vec{E}\right) ^2
\end{equation}
so that: 
\begin{equation}
\left( n\cdot F\right) ^2-\left( n\cdot \bar{F}\right) ^2=\left( \vec{n}%
\cdot \vec{E}\right) ^2+\left( \vec{n}\times \vec{E}\right) ^2-\left( \vec{n}%
\times \vec{H}\right) ^2-\left( \vec{n}\cdot \vec{H}\right) ^2=\vec{n}%
^2\left( \vec{E}^2-\vec{H}^2\right)
\end{equation}
Since $n^2=-\vec{n}\cdot \vec{n}$, the Zwanziger identity (\ref
{apvt:zwanident2}) is verified because $F^2=\vec{H}^2-\vec{E}^2$.

\subsection{Longitudinal and transverse components of antisymmetric tensors}

We define the projectors $K_{\mu \nu ,\alpha \beta }$ and $E_{\mu \nu
,\alpha \beta }$ which are the differential operators: 
\[
K_{\mu \nu ,\alpha \beta }=\frac 1{\partial ^2}\left( g_{\mu \alpha
}\partial _\nu \partial _\beta -g_{\nu \alpha }\partial _\mu \partial _\beta
+g_{\nu \beta }\partial _\mu \partial _\alpha -g_{\mu \beta }\partial _\nu
\partial _\alpha \right) 
\]
\begin{equation}
E=\varepsilon K\varepsilon \;\;\;\;\;\;E_{\mu \nu ,\alpha \beta }=\frac
14\varepsilon _{\mu \nu \sigma \rho }K^{\sigma \rho ,\gamma \delta
}\varepsilon _{\gamma \delta \alpha \beta }=\varepsilon _{\mu \nu \sigma
\rho }\frac 1{\partial ^2}\left( g^{\sigma \gamma }\partial ^\rho \partial
^\delta \right) \varepsilon _{\gamma \delta \alpha \beta }
\label{apvt:kedef}
\end{equation}
The projectors $K$ and $E$ are related by the equations: 
\begin{equation}
K^2=K\;\;\;\;\;\;E^2=-E\;\;\;\;\;\;KE=0\;\;\;\;\;\;K-E=G  \label{apvt:keid}
\end{equation}
We have: 
\[
KS=\frac 1{\partial ^2}\partial \wedge \left( \partial \cdot S\right) 
\]
\begin{equation}
ES=\varepsilon K\varepsilon S=\varepsilon K\bar{S}=\varepsilon \frac
1{\partial ^2}\left( \partial \wedge \left( \partial \cdot \overline{S}%
\right) \right) =\frac 1{\partial ^2}\overline{\partial \wedge \left(
\partial \cdot \overline{S}\right) }  \label{apvt:ddcs}
\end{equation}
so that $\left( K-E\right) S=S\equiv GS$ follows from (\ref{apvt:zwanident1}%
).

If $a$ and $b$ commute with $K$ and $E$, then: 
\begin{equation}
\left( \frac 1aK+\frac 1bE\right) \left( aK+bE\right) =K-E=G
\end{equation}
In the following, $S$ and $T$ are antisymmetric tensor fields, and we
neglect surface terms in the integrals:

\begin{equation}
\partial \cdot \left( \partial \cdot S\right) =0  \label{apvt:dds}
\end{equation}

\begin{equation}
\partial \cdot \left( ES\right) =0\;\;\;\;\;\;\partial \cdot \left(
KS\right) =T\left( \partial \cdot S\right) =\left( \partial \cdot S\right)
\;\;\;\;\;\;\;\partial \cdot S=\partial \cdot \left( KS\right)
\label{apvt:es}
\end{equation}
\begin{equation}
\int d^4x\left( \partial \cdot S\right) \left( \partial \cdot T\right)
=-\int d^4x\;S\partial ^2KT  \label{apvt:k}
\end{equation}
\begin{equation}
\int d^4x\left( \partial \cdot \bar{S}\right) \left( \partial \cdot \bar{T}%
\right) =-\int d^4x\;S\partial ^2ET  \label{apvt:e}
\end{equation}
\begin{equation}
K\left( \partial \wedge A\right) =\partial \wedge \left( TA\right) =\partial
\wedge A\;\;\;\;\;\;E\left( \partial \wedge A\right) =0
\end{equation}
\begin{equation}
K\left( \overline{\partial \wedge A}\right) =0\;\;\;\;\;\;E\left( \overline{%
\partial \wedge A}\right) =-\overline{\partial \wedge A}
\end{equation}

The projectors $K$ and $E$ can also be defined in terms of a given vector $%
n^\mu $. One simply replaces $\partial ^\mu $ by $n^\mu $. For example, the
projector $K$ can be defined thus: 
\[
K_{\mu \nu ,\alpha \beta }=\frac 1{n^2}\left( g_{\mu \alpha }n_\nu n_\beta
-g_{\nu \alpha }n_\mu n_\beta +g_{\nu \beta }n_\mu n_\alpha -g_{\mu \beta
}n_\nu n_\alpha \right) 
\]
and we still have: 
\begin{equation}
E=\varepsilon K\varepsilon
\;\;\;K^2=K\;\;\;\;\;\;E^2=-E\;\;\;\;\;\;KE=0\;\;\;\;\;\;K-E=G
\end{equation}

The identity $K-E=G$ is a statement of the Zwanziger identity (\ref
{apvt:zwanident1}).\ Indeed, we have: 
\[
KS=\frac 1{n^2}n\wedge \left( n\cdot S\right) 
\]
\begin{equation}
KS=\frac 1{n^2}n\wedge \left( n\cdot S\right) \;\;\;\;ES=\varepsilon K\bar{S}%
=\varepsilon \frac 1{n^2}\left( n\wedge \left( n\cdot \overline{S}\right)
\right) =\frac 1{n^2}\overline{n\wedge \left( n\cdot \overline{S}\right) }
\end{equation}
so that $\left( K-E\right) S=S\equiv GS$ is a statement of (\ref
{apvt:zwanident1}).

\section{Antisymmetric and dual 3-forms}

\label{ap:dualvec}

Let $T_{\alpha \beta \gamma }=-T_{\beta \alpha \gamma }=...$ be a tensor
which is completely antisymmetric with respect to the exchange of its
indices. We define: 
\begin{equation}
T^2=\frac 16T_{\alpha \beta \gamma }T^{\alpha \beta \gamma }
\end{equation}
there being $3!=6$ identical terms in the sum. The \emph{dual} of the tensor 
$T^{\alpha \beta \gamma }$ is the \emph{vector} $\bar{T}^\mu $ defined thus: 
\begin{equation}
\bar{T}^\mu =\frac 16\varepsilon ^{\mu \alpha \beta \gamma }T_{\alpha \beta
\gamma }
\end{equation}
and we have: 
\begin{equation}
\bar{T}^2=-T^2
\end{equation}
which, in explicit form, reads: 
\begin{equation}
\frac 16\varepsilon ^{\mu \alpha \beta \gamma }T_{\alpha \beta \gamma }\frac
16\varepsilon _{\mu \alpha ^{\prime }\beta ^{\prime }\gamma ^{\prime
}}T^{\alpha ^{\prime }\beta ^{\prime }\gamma ^{\prime }}=-\frac 16T_{\alpha
\beta \gamma }T^{\alpha \beta \gamma }
\end{equation}
This result can be checked using (\ref{apvt:epscontr}).

In the particular case where the tensor $T$ is defined in terms of the
derivative of an antisymmetric tensor $\Phi ^{\mu \nu }$: 
\begin{equation}
T_{\alpha \beta \gamma }=\partial _\alpha \Phi _{\beta \gamma }+\partial
_\beta \Phi _{\gamma \alpha }+\partial _\gamma \Phi _{\alpha \beta }
\end{equation}
we have: 
\begin{equation}
\bar{T}^\mu =\frac 16\varepsilon ^{\mu \alpha \beta \gamma }T_{\alpha \beta
\gamma }=-\left( \partial \cdot \bar{\Phi}\right) ^\mu
\end{equation}
so that: 
\begin{equation}
\frac 12T^2=-\frac 12\bar{T}^2=-\frac 12\left( \partial \cdot \bar{\Phi}%
\right) ^2
\end{equation}
Note also that, neglecting surface terms: 
\begin{equation}
\int d^4x\;\frac 12T^2=-\int d^4x\;\frac 12\left( \partial \cdot \bar{\Phi}%
\right) ^2=\frac 12\int d^4x\left( -\frac 12\left( \partial \cdot \Phi
\right) ^2-\frac 12\Phi \partial ^2\Phi \right)
\end{equation}

We can also define the \emph{dual} of a vector $V^\mu $ to be the \emph{%
antisymmetric tensor} $T^{\alpha \beta \gamma }$ defined as: 
\begin{equation}
\bar{V}^{\alpha \beta \gamma }=\varepsilon ^{\alpha \beta \gamma \mu }V_\mu
\end{equation}
and we have: 
\begin{equation}
\bar{V}^2=-V^2\;\;\;\;\;\;\frac 16\bar{V}^{\alpha \beta \gamma }\bar{V}%
_{\alpha \beta \gamma }=-V^\mu V_\mu
\end{equation}
Note also that, neglecting surface terms: 
\begin{equation}
\int d^4x\;\frac 12T^2=-\int d^4x\;\frac 12\left( \partial \cdot \bar{\Phi}%
\right) ^2=\frac 12\int d^4x\left( -\frac 12\left( \partial \cdot \Phi
\right) ^2-\frac 12\Phi \partial ^2\Phi \right)
\end{equation}
Note that, if: 
\begin{equation}
\Phi =\partial \wedge A
\end{equation}
then: 
\begin{equation}
T_{\alpha \beta \gamma }=0\;\;\;\;\;\;\bar{T}^\mu =0
\end{equation}
However, if: 
\begin{equation}
\Phi =\overline{\partial \wedge A}
\end{equation}
then: 
\[
T_{\alpha \beta \gamma }=\partial _\alpha \overline{\partial \wedge A}%
_{\beta \gamma }+\partial _\beta \overline{\partial \wedge A}_{\gamma \alpha
}+\partial _\gamma \overline{\partial \wedge A}_{\alpha \beta } 
\]
\begin{equation}
=\partial _\alpha \varepsilon _{\beta \gamma 12}\left( \partial ^1A^2\right)
+\partial _\beta \varepsilon _{\gamma \alpha 12}\left( \partial ^1A^2\right)
+\partial _\gamma \varepsilon _{\alpha \beta 12}\left( \partial ^1A^2\right)
\end{equation}
and the dual of $T$ is: 
\[
\bar{T}^\mu =\frac 16\varepsilon ^{\mu \alpha \beta \gamma }T_{\alpha \beta
\gamma }=\left( \partial ^2TA\right) ^\mu =\left( \partial \cdot \left(
\partial \wedge A\right) \right) ^\mu 
\]

\section{Three-dimensional euclidean vectors}

Formulas are taken from \cite{Jackson1975}.

The cartesian components are $i=\left( x,y,z\right) $ and the 3-dimensional
antisymmetric tensor is: 
\begin{equation}
\varepsilon _{123}=-\varepsilon _{213}=...=1
\end{equation}
The cartesian component $i=\left( x,y,z\right) $ of a three-dimensional
vector $\vec{a}$ is denoted by $a_i$. 
\begin{equation}
\vec{a}\cdot \left( \vec{b}\times \vec{c}\right) =\vec{b}\cdot \left( \vec{c}%
\times \vec{a}\right) =\vec{c}\cdot \left( \vec{a}\times \vec{b}\right)
=\varepsilon _{ijk}a_ib_jc_k
\end{equation}

\begin{equation}
\vec{a}\times \left( \vec{b}\times \vec{c}\right) =\left( \vec{a}\cdot \vec{c%
}\right) \vec{b}-\left( \vec{a}\cdot \vec{b}\right) \vec{c}
\label{apvt:abccross}
\end{equation}
\begin{equation}
\left( \vec{a}\times \vec{b}\right) \cdot \left( \vec{c}\times \vec{d}%
\right) =\left( \vec{a}\cdot \vec{c}\right) \left( \vec{b}\cdot \vec{d}%
\right) -\left( \vec{a}\cdot \vec{d}\right) \left( \vec{b}\cdot \vec{c}%
\right)
\end{equation}
\begin{equation}
\vec{\nabla}\times \vec{\nabla}\psi =0
\end{equation}
\begin{equation}
\vec{\nabla}\cdot \left( \vec{\nabla}\times \vec{a}\right) =0
\end{equation}
\begin{equation}
\vec{\nabla}\times \left( \vec{\nabla}\times \vec{a}\right) =\vec{\nabla}%
\left( \vec{\nabla}\cdot \vec{a}\right) -\vec{\nabla}^2\vec{a}
\label{apvt:crcr}
\end{equation}
\begin{equation}
\vec{\nabla}\cdot \left( \psi \vec{a}\right) =\vec{a}\cdot \vec{\nabla}\psi
+\psi \vec{\nabla}\cdot \vec{a}
\end{equation}
\begin{equation}
\vec{\nabla}\times \left( \psi \vec{a}\right) =\left( \vec{\nabla}\psi
\right) \times \vec{a}+\psi \vec{\nabla}\times \vec{a}
\end{equation}
\begin{equation}
\vec{\nabla}\left( \vec{a}\cdot \vec{b}\right) =\left( \vec{a}\cdot \vec{%
\nabla}\right) \vec{b}+\left( \vec{b}\cdot \vec{\nabla}\right) \vec{a}+\vec{a%
}\times \left( \vec{\nabla}\times \vec{b}\right) +\vec{b}\times \left( \vec{%
\nabla}\times \vec{a}\right)
\end{equation}
\begin{equation}
\vec{\nabla}\cdot \left( \vec{a}\times b\right) =\vec{b}\cdot \left( \vec{%
\nabla}\times \vec{a}\right) -\vec{a}\left( \vec{\nabla}\times \vec{b}\right)
\end{equation}
\begin{equation}
\vec{\nabla}\times \left( \vec{a}\times \vec{b}\right) =\vec{a}\left( \vec{%
\nabla}\cdot \vec{b}\right) -\vec{b}\left( \vec{\nabla}\cdot \vec{a}\right)
+\left( \vec{b}\cdot \vec{\nabla}\right) \vec{a}-\left( \vec{a}\cdot \vec{%
\nabla}\right) \vec{b}  \label{apvt:dab}
\end{equation}
We define longitudinal and transverse projectors: 
\begin{equation}
L_{ij}=\frac{\vec{\nabla}_i\vec{\nabla}_j}{\nabla ^2}\;\;\;\;\;\;\;T_{ij}=%
\delta _{ij}-\frac{\vec{\nabla}_i\vec{\nabla}_j}{\nabla ^2}%
\;\;\;\;\;\;L^2=L\;\;\;\;\;T^2=T\;\;\;\;\;LT=0
\end{equation}
A vector field may be decomposed into longitudinal and transverse parts: 
\begin{equation}
\vec{A}=\vec{A}_L+\vec{A}_T=\frac 1{\nabla ^2}\vec{\nabla}\left( \vec{\nabla}%
\cdot \vec{A}\right) -\frac 1{\nabla ^2}\vec{\nabla}\times \left( \vec{\nabla%
}\times \vec{A}\right)
\end{equation}
\begin{equation}
\vec{A}_L=\frac 1{\nabla ^2}\vec{\nabla}\left( \vec{\nabla}\cdot \vec{A}%
\right) \;\;\;\;\;\;\vec{A}_T=\vec{A}-\frac 1{\nabla ^2}\vec{\nabla}\left( 
\vec{\nabla}\cdot \vec{A}\right) =-\frac 1{\nabla ^2}\vec{\nabla}\times
\left( \vec{\nabla}\times \vec{A}\right)
\end{equation}
\begin{equation}
\int d^3r\;\left( \vec{\nabla}\cdot \vec{A}\right) ^2=-\int d^3r\;\vec{A}%
_L\nabla ^2\vec{A}_L\;\;\;\;\;\;\int d^3r\;\left( \vec{\nabla}\times \vec{A}%
\right) ^2=-\int d^3r\;\vec{A}_T\nabla ^2\vec{A}_T  \label{apvt:dddc}
\end{equation}
For the unit vector $\vec{e}=\frac{\vec{r}}r$, we have: 
\begin{equation}
\vec{\nabla}\cdot \vec{r}=3\;\;\;\;\;\;\vec{\nabla}\times \vec{r}=0
\end{equation}
\begin{equation}
\vec{\nabla}\cdot \vec{e}=\frac 2r\;\;\;\;\;\;\vec{\nabla}\times \vec{e}%
=0\;\;\;\;\;\;\left( \vec{e}=\frac{\vec{r}}r\right)
\end{equation}
\begin{equation}
\left( \vec{A}\cdot \vec{\nabla}\right) \vec{e}=\frac 1r\left( \vec{A}-\vec{e%
}\left( \vec{A}\cdot \vec{e}\right) \right) =\frac{\vec{A}_T}r
\end{equation}
\begin{equation}
\int d^3r\;\left( \vec{\nabla}\times \vec{A}\right) \cdot \vec{B}=\int d^3r\;%
\vec{A}\cdot \left( \vec{\nabla}\times \vec{B}\right)
\end{equation}

\begin{equation}
\int d^3r\left( \vec{\nabla}\times \vec{A}\right) \cdot \left( \vec{\nabla}%
\times \vec{A}\right) =\int d^3r\;\vec{A}\cdot \left[ \vec{\nabla}\times
\left( \vec{\nabla}\times \vec{A}\right) \right]
\end{equation}
\begin{equation}
\int_S\vec{A}\cdot d\vec{s}=\int_Vd^3r\;\vec{\nabla}\cdot \vec{A}%
\;\;\;\;\;\left( divergence\;theorem\right)  \label{apvt:divergence}
\end{equation}
\begin{equation}
\int_S\psi d\vec{s}=\int_Vd^3r\;\vec{\nabla}\psi
\end{equation}
\begin{equation}
\int_Sd\vec{s}\times \vec{A}=\int_Vd^3r\;\vec{\nabla}\times \vec{A}
\end{equation}
\begin{equation}
\int_S\left( \vec{\nabla}\times \vec{A}\right) \cdot d\vec{s}=\oint_C\vec{A}%
\cdot d\vec{l}\;\;\;\;\;\;\left( Stoke^{\prime }s\;theorem\right)
\label{apvt:stokes1}
\end{equation}
\begin{equation}
\int_Sd\vec{s}\times \vec{\nabla}\psi =\oint_C\psi d\vec{l}
\label{apvt:stokes2}
\end{equation}

Note that in four dimensions, $A\wedge B$ is a six-component antisymmetric
tensor, whereas in three dimensions, $\vec{a}\times \vec{b}$ is three
component vector.\ That is why the four-dimensional identity $A\cdot \left(
B\wedge C\right) =\left( A\cdot B\right) C-\left( A\cdot C\right) B$ plays
the role of the three-dimensional identity $\vec{a}\times \left( \vec{b}%
\times \vec{c}\right) =\left( \vec{a}\cdot \vec{c}\right) \vec{b}-\left( 
\vec{a}\cdot \vec{b}\right) \vec{c}$.\ That is also why a four-dimensional
vector $A^\mu $ is represented by the identity $A=\frac 1{n^2}\left[ n\cdot
\left( n\wedge A\right) +\left( n\cdot A\right) n\right] $, whereas a three
dimensional vector $\vec{A}$ is represented by the identity $\vec{A}=\frac
1{n^2}\left[ \left( \vec{n}\cdot \vec{A}\right) \vec{n}-\vec{n}\times \left( 
\vec{n}\times \vec{A}\right) \right] $.

\subsection{Cartesian coordinates}

In cartesian coordinates, the position vector is $\vec{r}=\left(
x,y,z\right) $.\ The three unit vectors $\vec{e}_{\left( i=x,y,z\right) }$
are: 
\begin{equation}
\vec{e}_{\left( i\right) }=\frac{x_i}r\;\;\;\;\;\;\vec{e}_{\left( i\right)
}\cdot \vec{e}_{\left( j\right) }=\delta _{ij}\;\;\;\;\;\;\vec{e}_{\left(
i\right) }\times \vec{e}_{\left( j\right) }=\varepsilon _{ijk}\vec{e}%
_{\left( k\right) }
\end{equation}
Any vector $\vec{A}$ can be expressed in terms of its cartesian components $%
A_{i=x,y,z}$: 
\begin{equation}
\vec{A}=A_x\vec{e}_{\left( x\right) }+A_y\vec{e}_{\left( y\right) }+A_z\vec{e%
}_{\left( z\right) }
\end{equation}
Then: 
\begin{equation}
\vec{\nabla}\psi =\vec{e}_{\left( x\right) }\frac{\partial \psi }{\partial x}%
+\vec{e}_{\left( y\right) }\frac{\partial \psi }{\partial y}+\vec{e}_{\left(
z\right) }\frac{\partial \psi }{\partial z}
\end{equation}

\begin{equation}
\vec{\nabla}\cdot \vec{A}=\frac{\partial A_x}{\partial x}+\frac{\partial A_y%
}{\partial y}+\frac{\partial A_z}{\partial z}
\end{equation}
\begin{equation}
\vec{\nabla}\times \vec{A}=\vec{e}_{\left( x\right) }\left( \frac{\partial
A_z}{\partial y}-\frac{\partial A_y}{\partial z}\right) +\vec{e}_{\left(
y\right) }\left( \frac{\partial A_x}{\partial z}-\frac{\partial A_z}{%
\partial x}\right) +\vec{e}_{\left( z\right) }\left( \frac{\partial A_y}{%
\partial x}-\frac{\partial A_x}{\partial y}\right)
\end{equation}
\begin{equation}
\vec{\nabla}^2\psi =\frac{\partial ^2\psi }{\partial x^2}+\frac{\partial
^2\psi }{\partial y^2}+\frac{\partial ^2\psi }{\partial z^2}
\end{equation}

\subsection{Cylindrical coordinates}

\label{apvt:cylcoord}

In cylindrical coordinates, the position vector is $\vec{r}=\left( \rho
,\theta ,z\right) $ where the cylindrical coordinates are expressed in terms
of the cartesian coordinates $\vec{r}=\left( x,y,z\right) $ as follows: 
\begin{equation}
x=\rho \cos \theta \;\;\;\;y=\rho \sin \theta \;\;\;\;z=z  \label{apvt:cyl1}
\end{equation}
The three unit vectors $\vec{e}_{\left( i=\rho ,\theta ,z\right) }$ are
defined by: 
\[
d\vec{r}=\vec{e}_{\left( \rho \right) }d\rho +\vec{e}_{\left( \theta \right)
}\rho d\theta +\vec{e}_{\left( z\right) }dz\;\;\;\;\int d^3r=\int_0^\infty
\rho d\rho \int_{-\infty }^\infty dz\int_0^{2\pi }d\theta 
\]
\begin{equation}
\vec{e}_{\left( i\right) }\cdot \vec{e}_{\left( j\right) }=\delta
_{ij}\;\;\;\;\;\;\vec{e}_{\left( i\right) }\times \vec{e}_{\left( j\right)
}=\varepsilon _{ijk}\vec{e}_{\left( k\right) }\;\;\;\;\;\left( i,j=\rho
,\theta ,z\right)  \label{apvt:cyl6}
\end{equation}
Any vector $\vec{A}$ can be expressed in terms of its cylindrical components 
$A_{i=\rho ,\theta ,z}$: 
\begin{equation}
\vec{A}=A_\rho \vec{e}_{\left( \rho \right) }+A_\theta \vec{e}_{\left(
\theta \right) }+A_z\vec{e}_{\left( z\right) }  \label{apvt:cyl2}
\end{equation}
Then: 
\begin{equation}
\vec{\nabla}\psi =\vec{e}_{\left( \rho \right) }\frac{\partial \psi }{%
\partial \rho }+\vec{e}_{\left( \theta \right) }\frac 1\rho \frac{\partial
\psi }{\partial \theta }+\vec{e}_{\left( z\right) }\frac{\partial \psi }{%
\partial z}  \label{apvt:cyl3}
\end{equation}

\begin{equation}
\vec{\nabla}\cdot \vec{A}=\frac 1\rho \frac \partial {\partial \rho }\left(
\rho A_\rho \right) +\frac 1\rho \frac{\partial A_\theta }{\partial \theta }+%
\frac{\partial A_z}{\partial z}  \label{apvt:cyl4}
\end{equation}
\begin{equation}
\vec{\nabla}\times \vec{A}=\vec{e}_{\left( \rho \right) }\left( \frac 1\rho 
\frac{\partial A_z}{\partial \theta }-\frac{\partial A_\theta }{\partial z}%
\right) +\vec{e}_{\left( \theta \right) }\left( \frac{\partial A_\rho }{%
\partial z}-\frac{\partial A_z}{\partial \rho }\right) +\vec{e}_{\left(
z\right) }\frac 1\rho \left( \frac \partial {\partial \rho }\left( \rho
A_\theta \right) -\frac{\partial A_\rho }{\partial \theta }\right)
\label{apvt:cyl5}
\end{equation}
\begin{equation}
\nabla ^2\psi =\frac 1\rho \frac \partial {\partial \rho }\left( \rho \frac{%
\partial \psi }{\partial \rho }\right) +\frac 1{\rho ^2}\frac{\partial
^2\psi }{\partial \theta ^2}+\frac{\partial ^2\psi }{\partial z^2}
\label{apvt:cyl7}
\end{equation}
\begin{equation}
\int_0^\infty \rho d\rho \int_0^{2\pi }d\theta e^{ik\rho \cos \theta
}f\left( \rho \right) =2\pi \int_0^\infty \rho d\rho J_0\left( k\rho \right)
f\left( \rho \right)
\end{equation}

\subsection{Spherical coordinates}

\label{ap:spherical}

In spherical coordinates, the position vector is $\vec{r}=\left( r,\theta
,\varphi \right) $ where the spherical coordinates are expressed in terms of
the cartesian coordinates $\vec{r}=\left( x,y,z\right) $ as follows: 
\begin{equation}
x=r\sin \theta \cos \varphi \;\;\;\;y=r\sin \theta \sin \varphi
\;\;\;\;z=r\cos \theta \;\;\;\;  \label{apvt:sph1}
\end{equation}
The three unit vectors $\vec{e}_{\left( i=\rho ,\theta ,\varphi \right) }$
are defined by: 
\[
d\vec{r}=\vec{e}_{\left( r\right) }dr+\vec{e}_{\left( \theta \right) }r\cos
\theta \;d\theta +\vec{e}_\varphi r\sin \theta \;d\varphi \;\;\;\;\;\int
d^3r=\int_0^\infty r^2dr\int_0^\pi \sin \theta \;d\theta \int_0^{2\pi
}d\varphi 
\]
\begin{equation}
\vec{e}_{\left( i\right) }\cdot \vec{e}_{\left( j\right) }=\delta
_{ij}\;\;\;\;\;\;\vec{e}_{\left( i\right) }\times \vec{e}_{\left( j\right)
}=\varepsilon _{ijk}\vec{e}_{\left( k\right) }\;\;\;\;\;\left( i,j=r,\theta
,\varphi \right)  \label{apvt:sph6}
\end{equation}
Any vector $\vec{A}$ can be expressed in terms of its spherical components $%
A_{i=r,\theta ,\varphi }$: 
\begin{equation}
\vec{A}=A_r\vec{e}_{\left( r\right) }+A_\theta \vec{e}_{\left( \theta
\right) }+A_\varphi \vec{e}_{\left( \varphi \right) }  \label{apvt:sph2}
\end{equation}
Then: 
\begin{equation}
\vec{\nabla}\psi =\vec{e}_{\left( r\right) }\frac{\partial \psi }{\partial r}%
+\vec{e}_{\left( \theta \right) }\frac 1r\frac{\partial \psi }{\partial
\theta }+\vec{e}_{\left( \varphi \right) }\frac 1{r\sin \theta }\frac{%
\partial \psi }{\partial \varphi }  \label{apvt:sph3}
\end{equation}

\begin{equation}
\vec{\nabla}\cdot \vec{A}=\frac{1}{r^{2}}\frac{\partial }{\partial r}\left(
r^{2}A_{r}\right) +\frac{1}{r\sin \theta }\frac{\partial }{\partial \theta }%
\left( \sin \theta \;A_{\theta }\right) +\frac{1}{r\sin \theta }\frac{%
\partial A_{\varphi }}{\partial \varphi }  \label{apvt:sph4}
\end{equation}
\begin{equation}
\vec{\nabla}\times \vec{A}=\vec{e}_{\left( r\right) }\frac{1}{r\sin \theta }%
\left( \frac{\partial }{\partial \theta }\left( \sin \theta \;A_{\varphi
}\right) -\frac{\partial A_{\theta }}{\partial \varphi }\right) +\vec{e}%
_{\left( \theta \right) }\left( \frac{1}{r\sin \theta }\frac{\partial A_{r}}{%
\partial \varphi }-\frac{1}{r}\frac{\partial }{\partial r}\left( rA_{\varphi
}\right) \right) +\vec{e}_{\left( \varphi \right) }\frac{1}{r}\left( \frac{%
\partial }{\partial r}\left( rA_{\theta }\right) -\frac{\partial A_{r}}{%
\partial \theta }\right)  \label{apvt:sph5}
\end{equation}
\begin{equation}
\vec{\nabla}^{2}\psi =\frac{1}{r^{2}}\frac{\partial }{\partial r}\left( r^{2}%
\frac{\partial \psi }{\partial r}\right) +\frac{1}{r^{2}\sin \theta }\frac{%
\partial }{\partial \theta }\left( \sin \theta \frac{\partial \psi }{%
\partial _{\theta }}\right) +\frac{1}{r^{2}\sin ^{2}\theta }\frac{\partial
^{2}\psi }{\partial \varphi ^{2}}
\end{equation}
\begin{equation}
\frac{1}{r^{2}}\frac{\partial }{\partial r}\left( r^{2}\frac{\partial \psi }{%
\partial r}\right) =\frac{1}{r}\frac{\partial ^{2}}{\partial r^{2}}\left(
r\psi \right)
\end{equation}

Note the identity: 
\begin{equation}
\vec{e}_{\left( r\right) }\frac 1{4\pi r^2}=-\vec{\nabla}\times \left( \frac{%
1+\cos \theta }{4\pi r\sin \theta }\vec{e}_{\left( \varphi \right) }\right)
\end{equation}
\begin{equation}
\int d^3re^{i\vec{k}\cdot \vec{r}}f\left( r\right) =4\pi \int_0^\infty
r^2drj_0\left( kr\right) f\left( r\right) =4\pi \int_0^\infty r^2dr\frac{%
\sin kr}{kr}f\left( r\right)
\end{equation}

\chapter{The relation between Minkowski and Euclidean actions}

\label{ap:minkeucl}

The Minkowski action leads to canonical quantization and it is used to
calculate matrix elements of the evolution operator $e^{-iHt}$.\ The
Euclidean action is used to calculate the partition function $tre^{-\beta H}$%
.\ Lattice calculations are formulated in terms of the Euclidean action. In
Minkowski space $g_{\mu \nu }=\left( 1,-1,-1,-1\right) $ and $\det g=-1$,
whereas in Euclidean space $g_{\mu \nu }=\left( 1,1,1,1\right) =\delta _{\mu
\nu }$ and $\det g=+1$. As a rule of the thumb, a Euclidean action can be
transformed into a Minkowski action by the following substitutions: 
\begin{equation}
\begin{tabular}{|c|c|c|}
\hline
Euclidean & $\rightarrow $ & Minkowski \\ \hline
$g_{\mu \nu }=diag\left( 1,1,1,1\right) =\delta _{\mu \nu }$ & $\rightarrow $
& $g_{\mu \nu }=diag\left( 1,-1,-1,-1\right) $ \\ \hline
$\left( t,\vec{r}\right) $ & $\rightarrow $ & $\left( it,\vec{r}\right) $ \\ 
\hline
$\left( \partial _t,\partial _i\right) =\left( \partial _t,\nabla _i\right) $
& $\rightarrow $ & $\left( -i\partial _t,\partial _i\right) =\left(
-i\partial _t,-\nabla _i\right) $ \\ \hline
$\left( A_0,A_i\right) $ & $\rightarrow $ & $\left( iA^0,A^i\right) $ \\ 
\hline
$\left( j_0,\vec{j}\right) $ & $\rightarrow $ & $\left( ij_0,\vec{j}\right) $
\\ \hline
$A_\mu A_\mu $ & $\rightarrow $ & $-A^\mu A_\mu $ \\ \hline
$\partial ^2=\partial _0^2+\partial _i^2=\partial _t^2+\partial _i^2$ & $%
\rightarrow $ & $-\partial ^2=-\partial _\mu \partial ^\mu $ \\ \hline
$\left( \partial \wedge A\right) _{0i}=\partial _0A_i-\partial _iA_0$ & $%
\rightarrow $ & $-i\left( \partial \wedge A\right) ^{0i}=-i\left( \partial
^0A^i-\partial ^iA^0\right) $ \\ \hline
$\left( \partial \wedge A\right) _{ij}=\partial _iA_j-\partial _jA_i$ & $%
\rightarrow $ & $-\left( \partial \wedge A\right) ^{ij}=-\partial
^iA^j+\partial ^jA^i$ \\ \hline
$\frac 14\left( \partial \wedge A\right) _{\mu \nu }^2$ & $\rightarrow $ & $%
\frac 14\left( \partial \wedge A\right) _{\mu \nu }\left( \partial \wedge
A\right) ^{\mu \nu }$ \\ \hline
$\varepsilon _{\mu \nu \alpha \beta }=\varepsilon ^{\mu \nu \alpha \beta }$
& $\rightarrow $ & $\varepsilon ^{\mu \nu \alpha \beta }=-\varepsilon _{\mu
\nu \alpha \beta }$ \\ \hline
$\varepsilon _{0123}=\varepsilon ^{0123}=1$ & $\rightarrow $ & $\varepsilon
^{0123}=-\varepsilon _{0123}=1$ \\ \hline
$\overline{\partial \wedge A}_{0i}=\frac 12\varepsilon _{0ijk}\left(
\partial \wedge A\right) _{jk}$ & $\rightarrow $ & $-\frac 12\varepsilon
^{0ijk}\left( \partial \wedge A\right) _{jk}=-\overline{\partial \wedge A}%
^{0i}$ \\ \hline
$\overline{\partial \wedge A}_{ij}=\varepsilon _{ij0k}\left( \partial \wedge
A\right) _{0k}$ & $\rightarrow $ & $i\overline{\partial \wedge A}%
^{ij}=i\varepsilon ^{ij0k}\left( \partial \wedge A\right) _{0k}$ \\ \hline
$\frac 14\overline{\partial \wedge A}_{\mu \nu }\overline{\partial \wedge A}%
_{\mu \nu }$ & $\rightarrow $ & $-\frac 14\overline{\partial \wedge A}_{\mu
\nu }\overline{\partial \wedge A}^{\mu \nu }$ \\ \hline
$\int d^4x=\int dtdxdydz$ & $\rightarrow $ & $\int d^4x=\int dtdxdydz$ \\ 
\hline
$\left( action\right) $ & $\rightarrow $ & $-\left( action\right) $ \\ \hline
$\vec{E}$ & $\rightarrow $ & $-i\vec{E}$ \\ \hline
$\vec{H}$ & $\rightarrow $ & $\vec{H}$ \\ \hline
\end{tabular}
\label{apme:trans}
\end{equation}
For example, the Minkowski Landau-Ginzburg action (\ref{landsphi}) of a dual
superconductor is: 
\begin{equation}
I_j\left( B,S,\varphi \right) =\int d^4x\left( -\frac 12\left( \partial
\wedge B+\bar{G}\right) ^2+\frac{g^2S^2}2\left( B+\partial \varphi \right)
^2+\frac 12\left( \partial S\right) ^2-\frac 12b\left( S^2-v^2\right)
^2\right)  \label{apme:lg}
\end{equation}
whereas the Euclidean action is: 
\begin{equation}
I_j\left( B,S,\varphi \right) =\int d^4x\left( \frac 12\left( \partial
\wedge B+\bar{G}\right) ^2+\frac{g^2S^2}2\left( B-\partial \varphi \right)
^2+\frac 12\left( \partial S\right) ^2+\frac 12b\left( S^2-v^2\right)
^2\right)  \label{apme:eu}
\end{equation}
The table (\ref{apme:trans}) can be used to recover the Minkowski action (%
\ref{apme:lg}) from the Euclidean action (\ref{apme:eu}).\ The change in
sign of the action is chosen such that the partition function can be written
in terms of a functional integral of the Euclidean action, in the form: 
\begin{equation}
Z=e^{-\beta H}=\int D\left( B,S,\varphi \right) e^{-I_j\left( B,S,\varphi
\right) }
\end{equation}
In general however, the functional integrals need to be adapted to the
acting constraints..

We can choose to represent the Euclidean field tensor $F^{\mu \nu }=F_{\mu
\nu }$ in terms of Euclidean electric and magnetic fields $\vec{E}$ and $%
\vec{H}$ thus: 
\begin{equation}
F^{\mu \nu }=\left( 
\begin{array}{cccc}
0 & -E_x & -E_y & -E_z \\ 
E_x & 0 & -H_z & H_y \\ 
E_y & H_z & 0 & -H_x \\ 
E_z & -H_y & H_x & 0
\end{array}
\right) \;\;\;\;\;\;\overline{F}_{\mu \nu }=\frac 12\varepsilon _{\mu \nu
\alpha \beta }F_{\alpha \beta }=\left( 
\begin{array}{cccc}
0 & -H_x & -H_y & -H_z \\ 
H_x & 0 & E_z & -E_y \\ 
H_y & -E_z & 0 & E_x \\ 
H_z & E_y & -E_x & 0
\end{array}
\right)
\end{equation}
\qquad If we want to express the Euclidean field tensor as $F=\partial
\wedge A$ then the relation between the Euclidean and Minkowski electric and
magnetic fields is the one given at the end of table \ref{apme:trans}.The
Euclidean electric and magnetic fields $\vec{E}$ and $\vec{H}$ are expressed
in terms of the Euclidean gauge potential $A_\mu =\left( \phi ,\vec{A}%
\right) $ as follows: 
\begin{equation}
\vec{E}=-\partial _t\vec{A}+\vec{\nabla}\phi \;\;\;\;\;\;\vec{H}=-\vec{\nabla%
}\times \vec{A}
\end{equation}

In the Euclidean formulation, $\varepsilon ^2=G$ and the duality
transformation of antisymmetric tensors is reversible without a change in
sign: 
\begin{equation}
\bar{S}_{\mu \nu }=\frac 12\varepsilon _{\mu \nu \alpha \beta }S_{\alpha
\beta }\;\;\;\;\;\;S_{\mu \nu }=\frac 12\varepsilon _{\mu \nu \alpha \beta }%
\bar{S}_{\alpha \beta }
\end{equation}
The projectors $K$ and $E$ are defined by (\ref{apvt:kedef}) with $g_{\mu
\nu }=\delta _{\mu \nu }$ and we have: 
\begin{equation}
K^2=K=\varepsilon K\varepsilon
\;\;\;\;\;\;E^2=E\;\;\;\;\;\;KE=0\;\;\;\;\;\;K+E=G  \label{apme:ke}
\end{equation}
with: 
\begin{equation}
G_{\mu \nu ,\alpha \beta }=\left( \delta _{\mu \alpha }\delta _{\nu \beta
}-\delta _{\mu \beta }\delta _{\nu \alpha }\right)
\end{equation}

\chapter{The generators of the $SU\left( 2\right) $ and $SU\left( 3\right) $
groups}

\label{ap:su}

\section{The $SU\left( 2\right) $ generators}

The three Pauli matrices are: 
\begin{equation}
\sigma ^1=\left( 
\begin{array}{ll}
0 & 1 \\ 
1 & 0
\end{array}
\right) \;\;\;\;\;\;\sigma ^2=\left( 
\begin{array}{ll}
0 & -i \\ 
i & 0
\end{array}
\right) \;\;\;\;\;\;\sigma ^3=\left( 
\begin{array}{ll}
1 & 0 \\ 
0 & -1
\end{array}
\right)  \label{apsu:pauli}
\end{equation}
\begin{equation}
\sigma _i\sigma _j=\delta _{ij}+i\varepsilon _{ijk}\sigma _l
\end{equation}
The three generators of the $SU\left( 2\right) $ group are $T_a=\frac
12\sigma _a$.

\section{The $SU\left( 3\right) $ generators and root vectors}

The eight Gell-Mann matrices are:

\[
\lambda _1=\left( 
\begin{array}{ccc}
0 & 1 & 0 \\ 
1 & 0 & 0 \\ 
0 & 0 & 0
\end{array}
\right) \;\;\;\;\;\;\lambda _2=\left( 
\begin{array}{ccc}
0 & -i & 0 \\ 
i & 0 & 0 \\ 
0 & 0 & 0
\end{array}
\right) \;\;\;\;\;\;\lambda _3=\left( 
\begin{array}{ccc}
1 & 0 & 0 \\ 
0 & -1 & 0 \\ 
0 & 0 & 0
\end{array}
\right) 
\]
\[
\lambda _4=\left( 
\begin{array}{ccc}
0 & 0 & 1 \\ 
0 & 0 & 0 \\ 
1 & 0 & 0
\end{array}
\right) \;\;\;\;\;\;\lambda _5=\left( 
\begin{array}{ccc}
0 & 0 & -i \\ 
0 & 0 & 0 \\ 
i & 0 & 0
\end{array}
\right) 
\]
\begin{equation}
\lambda _6=\left( 
\begin{array}{ccc}
0 & 0 & 0 \\ 
0 & 0 & 1 \\ 
0 & 1 & 0
\end{array}
\right) \;\;\;\;\;\;\lambda _7=\left( 
\begin{array}{ccc}
0 & 0 & 0 \\ 
0 & 0 & -i \\ 
0 & i & 0
\end{array}
\right) \;\;\;\;\;\;\lambda _8=\frac 1{\sqrt{3}}\left( 
\begin{array}{ccc}
1 & 0 & 0 \\ 
0 & 1 & 0 \\ 
0 & 0 & -2
\end{array}
\right)  \label{apsu:gelm}
\end{equation}
\begin{equation}
\lambda _a\lambda _b=\frac 23\delta _{ab}+\left( d_{abc}+if_{abc}\right)
\lambda _c\;\;\;\;\;\;T_aT_b=\frac 16\delta _{ab}+\frac 12\left(
d_{abc}+if_{abc}\right) T_c  \label{apsu:df}
\end{equation}
\begin{equation}
trT_aT_b=\frac 12\delta
_{ab}\,\,\,\,\,\,\;\;\;\;\;T_aT_b-T_bT_a=if_{abc}T_c\;\;\;\;\;%
\;T_aT_b+T_bT_a=\frac 13\delta _{ab}+d_{abc}T_c
\end{equation}
where $T_a=\frac 12\lambda _a$ are the generators of the $SU\left( 3\right) $
group. There are two diagonal generators, namely $T_3$ and $T_8$.\ They are
said to form a Cartan subalgebra in $SU\left( 3\right) $.

The $f_{abc}$ are antisymmetric in their indices $f_{123}=-f_{213}=...$ and
the non vanishing values are: 
\begin{equation}
f_{123}=1,\;\;\;f_{147}=f_{165}=f_{246}=f_{257}=f_{345}=f_{376}=\frac
12,\;\;\;f_{458}=f_{678}=\frac{\sqrt{3}}2  \label{apsu:fcoeff}
\end{equation}
The $d_{abc}$ are symmetric in their indices $d_{123}=d_{213}=...$ and the
non vanishing values are: 
\[
d_{118}=d_{228}=d_{338}=\frac 1{\sqrt{3}},\;\;\;\;%
\;d_{146}=d_{157}=d_{256}=d_{344}=d_{355}=\frac 12 
\]
\begin{equation}
d_{247}=d_{366}=d_{377}=-\frac
12,\;\;\;d_{448}=d_{558}=d_{668}=d_{778}=d_{888}=d_{118}=-\frac 1{2\sqrt{3}}
\label{apsu:dcoeff}
\end{equation}

The remaining six non-diagonal generators can be grouped together as
follows: 
\begin{equation}
E_{\pm 1}=\frac 1{\sqrt{2}}\left( T_1\pm iT_2\right) \;\;\;\;E_{\pm 2}=\frac
1{\sqrt{2}}\left( T_4\mp iT_5\right) \;\;\;E_{\pm 3}=\frac 1{\sqrt{2}}\left(
T_6\pm iT_7\right)  \label{apsu:evec}
\end{equation}
The commutators with the diagonal generators are: 
\[
\left[ T_3,E_{\pm 1}\right] =\pm E_{\pm 1}\;\;\;\;\left[ T_3,E_{\pm
2}\right] =\mp \frac 12E_{\pm 2}\;\;\;\left[ T_3,E_{\pm 3}\right] =\mp \frac
12E_{\pm 3} 
\]
\begin{equation}
\left[ T_8,E_{\pm 1}\right] =0\;\;\;\;\left[ T_8,E_{\pm 2}\right] =\mp \frac{%
\sqrt{3}}2E_{\pm 2}\;\;\;\left[ T_8,E_{\pm 3}\right] =\pm \frac{\sqrt{3}}%
2E_{\pm 3}  \label{apsu:tecom}
\end{equation}

The gluon field $A^\mu $ can be expressed thus: 
\begin{equation}
A^\mu =A_a^\mu T_a=A_3^\mu T_3+A_8^\mu T_8+\sum_{a=1}^3C_a^{\mu
*}E_a+C_a^\mu E_{-a}  \label{apsu:aac}
\end{equation}
where the non-diagonal gluon fields are: 
\[
C_1^\mu =\frac 1{\sqrt{2}}\left( A_1^\mu +iA_2^\mu \right) \;\;\;\;\;C_2^\mu
=\frac 1{\sqrt{2}}\left( A_4^\mu -iA_5^\mu \right) \;\;\;\;\;\;C_3^\mu
=\frac 1{\sqrt{2}}\left( A_6^\mu +iA_7^\mu \right) 
\]
\begin{equation}
C_1^{\mu *}=\frac 1{\sqrt{2}}\left( A_1^\mu -iA_2^\mu \right)
\;\;\;\;\;\;C_2^{\mu *}=\frac 1{\sqrt{2}}\left( A_4^\mu +iA_5^\mu \right)
\;\;\;\;\;\;C_3^{\mu *}=\frac 1{\sqrt{2}}\left( A_6^\mu -iA_7^\mu \right)
\label{apsu:ca}
\end{equation}

\section{Root vectors of $SU\left( 3\right) $}

\label{apsu:root}

We can represent the two diagonal generators $T_3$ and $T_8$ by a
two-dimensional vector $H$: 
\begin{equation}
H=\left( T_3,T_8\right)  \label{apsu:ht3t8}
\end{equation}
The commutators (\ref{apsu:tecom}) of $H$ with the non-diagonal generators
can be expressed in the form: 
\begin{equation}
\left[ H,E_{\pm a}\right] =\pm w_aE_{\pm a}\;\;\;\;\;\;\left( a=1,2,3\right)
\label{apsu:hecom}
\end{equation}
The vectors $w_a$ are called the \emph{root vectors} of the $SU\left(
3\right) $ group. Their components are: 
\begin{equation}
w_1=\left( 1,0\right) \;\;\;\;\;\;w_2=\left( -\frac 12,-\frac{\sqrt{3}}%
2\right) \;\;\;\;\;\;w_3=\left( -\frac 12,\frac{\sqrt{3}}2\right)
\label{apsu:weight}
\end{equation}
The root vectors have unit length. They are neither orthogonal nor linearly
independent, because they sum up to zero: 
\begin{equation}
\left( w_1\cdot w_1\right) =1\;\;\;\;\;\;\left( w_2\cdot w_2\right)
=1\;\;\;\;\;\;\left( w_3\cdot w_3\right) =1\;\;\;\;\;\;w_1+w_2+w_3=0
\label{apsu:weg}
\end{equation}
They form an over-complete set: 
\begin{equation}
\sum_{a=1}^3w_a^iw_a^j=\frac 32\delta _{ij}  \label{apsu:wij}
\end{equation}
where $w_a^i$ is the component $i$ of the root vector $w_a$. It can be
useful to think in terms of bras and kets and to write $w_a^i=\left\langle
\left. i\right| w_a\right\rangle =\left\langle \left. w_a\right|
i\right\rangle $ with $\left\langle \left. i\right| j\right\rangle =\delta
_{ij}$. In this notation, the completeness relation (\ref{apsu:wij}) reads: 
\begin{equation}
\frac 23\sum_{a=1}^3\left| w_a\right\rangle \left\langle w_a\right| =1
\label{apsu:wijbis}
\end{equation}

The \emph{abelian projection} of the gluon field is: 
\begin{equation}
A_\mu =A_{3\mu }T_3+A_{8\mu }T_8=\left( A_\mu \cdot H\right)
\;\;\;\;\;\;\;\;\;\;\;A_\mu \equiv \left( A_{3\mu },A_{8\mu }\right)
\end{equation}
In view of (\ref{apsu:wij}), we can write: 
\begin{equation}
A_\mu =\frac 23\sum_{a=1}^3w_a\left( w_a\cdot A_\mu \right)
\end{equation}
and the abelian projection of the gluon field is then: 
\begin{equation}
A_\mu \cdot H=\frac 23\sum_{a=1}^3\left( H\cdot w_a\right) \left( w_a\cdot
A_\mu \right) =\frac 23\sum_{a=1}^3a_{a\mu }t_a  \label{apsu:ahta}
\end{equation}
where the generators $t_a$ are: 
\[
t_a=\left( H\cdot w_a\right) \;\;\;\;\;\;\;\;\;\;t_1+t_2+t_3=0 
\]
\[
t_1=\left( H\cdot w_1\right) =T_3 
\]
\begin{equation}
t_2=\left( H\cdot w_2\right) =-\frac 12T_3-\frac{\sqrt{3}}%
2T_8\;\;\;\;\;\;t_3=\left( H\cdot w_3\right) =-\frac 12T_3-\frac{\sqrt{3}}%
2T_8  \label{apsu:atwa}
\end{equation}
and where the fields $a_{a\mu }$ are: 
\[
a_{\mu a}=\left( w_a\cdot A_\mu \right) \;\;\;\;\;\;a_{1\mu }+a_{2\mu
}+a_{3\mu }=0 
\]
\begin{equation}
a_{1\mu }=\frac 23\left( w_1\cdot A_\mu \right) =\frac 23A_{3\mu }
\end{equation}
\begin{equation}
a_{2\mu }=\frac 23\left( w_2\cdot A_\mu \right) =\frac 13\left( -A_{3\mu }-%
\sqrt{3}A_{8\mu }\right) \;\;\;\;\;\;\;\;a_{3\mu }=\frac 23\left( w_3\cdot
A_\mu \right) =\frac 13\left( -A_{3\mu }+\sqrt{3}A_{8\mu }\right)
\label{apsu:awmu}
\end{equation}

\chapter{Color charges of quarks and gluons}

\label{sec:colorcharges}

\section{$SU\left( 2\right) $ color charges}

In $SU\left( 2\right) $, the charge operator is: 
\begin{equation}
Q=eT_3=\frac 12\left( 
\begin{array}{cc}
e & 0 \\ 
0 & -e
\end{array}
\right)
\end{equation}
The quarks form a doublet in the fundamental representation of $SU\left(
2\right) $: 
\begin{equation}
q=\left( 
\begin{array}{c}
q_R \\ 
q_B
\end{array}
\right) \;\;\;\;\;\;Q\left( 
\begin{array}{c}
q_R \\ 
q_B
\end{array}
\right) =\frac 12e\left( 
\begin{array}{c}
q_R \\ 
-q_B
\end{array}
\right)
\end{equation}
The red and blue quarks have respectively color charges $\frac 12e$ and $%
-\frac 12e$.

The gluon field has the form (\ref{apsu:aac}): 
\begin{equation}
A^\mu =A_a^\mu T_a=A_3^\mu T_3+C_1^{\mu *}E_1+C_1^\mu E_{-1}
\end{equation}
and the color charges of the gluons are given by the commutator: 
\begin{equation}
\left[ Q,A^\mu \right] =e\left[ T_3,A^\mu \right] =eC_1^{\mu *}E_1-eC_1^\mu
E_{-1}
\end{equation}
The color charge of $A_3^\mu $ is zero and the fields $C_1^{\mu *}$ and $%
C_1^\mu $ have respectively color charges $e$ and $-e$.

\begin{equation}
\begin{tabular}{|c|c|c|c|c|}
\hline
$q_R$ & $q_B$ & $A_3^\mu $ & $C_1^{\mu *}$ & $C_{-1}^{\mu *}$ \\ \hline
$\frac e2$ & $-\frac e2$ & $0$ & $e$ & $-e$ \\ \hline
\end{tabular}
\label{apsu:su2ch}
\end{equation}

\section{$SU\left( 3\right) $ color charges}

\label{ap:su3charges}

In $SU\left( 3\right) $ there are two color charge operators, associated
respectively to the diagonal matrices $T_3$ and $T_8$: 
\begin{equation}
Q_3=eT_3=\frac 12\left( 
\begin{array}{ccc}
e & 0 & 0 \\ 
0 & -e & 0 \\ 
0 & 0 & 0
\end{array}
\right) \;\;\;\;\;\;\;\;Q_8=eT_8=\frac 1{2\sqrt{3}}\left( 
\begin{array}{ccc}
e & 0 & 0 \\ 
0 & e & 0 \\ 
0 & 0 & -2e
\end{array}
\right)  \label{apsu:q3q8}
\end{equation}
The charge operator can be represented by the vector $Q=\left(
Q_3,Q_8\right) =e\left( T_3,T_8\right) $.

The quarks form a triplet in the fundamental representation of $SU\left(
3\right) $: 
\begin{equation}
q=\left( 
\begin{array}{c}
q_{R} \\ 
q_{B} \\ 
q_{G}
\end{array}
\right) \;\;\;\;\;Q_{3}\left( 
\begin{array}{c}
q_{R} \\ 
q_{B} \\ 
q_{G}
\end{array}
\right) =\frac{1}{2}e\left( 
\begin{array}{c}
q_{R} \\ 
-q_{B} \\ 
0
\end{array}
\right) \;\;\;\;Q_{8}\left( 
\begin{array}{c}
q_{R} \\ 
q_{B} \\ 
q_{G}
\end{array}
\right) =\frac{1}{2\sqrt{3}}e\left( 
\begin{array}{c}
q_{R} \\ 
q_{B} \\ 
-2q_{G}
\end{array}
\right)
\end{equation}
The red, blue and green quarks have respectively color charges $Q_{3}$ equal
to $\frac{1}{2}e,-\frac{1}{2}e$ and $0$ and color charges $Q_{8}$ equal to $%
\frac{1}{2\sqrt{3}}e,\frac{1}{2\sqrt{3}}e$ and $-\frac{1}{\sqrt{3}}e$.

The gluon field has the form (\ref{apsu:aac}): 
\begin{equation}
A^\mu =A_a^\mu T_a=A_3^\mu T_3+\sum_{a=1}^3\left( C_a^{\mu *}E_a+C_a^\mu
E_{-a}\right)
\end{equation}
We calculate the commutators of the charge operator (\ref{apsu:q3q8})
written in the form (\ref{apsu:aac}): 
\begin{equation}
\left[ Q,A^\mu \right] =e\sum_{a=1}^3w_a\left( C_a^{\mu *}E_a-C_a^\mu
E_{-a}\right)
\end{equation}
where $w_a$ are the root vectors (\ref{apsu:hecom}). The color charges of $%
A_3^\mu $ and $A_8^\mu $ are zero and the fields $C_a^{\mu *}$ and $C_a^\mu $
have respectively color charges $ew_a$ and $-ew_a$.

\begin{equation}
\begin{tabular}{|c|c|c|c|c|c|c|c|c|c|c|c|}
\hline
& $q_R$ & $q_B$ & $q_G$ & $A_3$ & $A_8$ & $C_1^{\mu *}$ & $C_1^\mu $ & $%
C_2^{\mu *}$ & $C_2^\mu $ & $C_3^{\mu *}$ & $C_3^\mu $ \\ \hline
$Q_3$ & $\frac 12e$ & $-\frac 12e$ & $0$ & $0$ & $0$ & $e$ & $-e$ & $-\frac
12e$ & $\frac 12e$ & $-\frac 12e$ & $\frac 12e$ \\ \hline
$Q_8$ & $\frac 1{2\sqrt{3}}e$ & $\frac 1{2\sqrt{3}}e$ & $-\frac 1{\sqrt{3}}e$
& $0$ & $0$ & $0$ & $0$ & $-\frac{\sqrt{3}}2e$ & $\frac{\sqrt{3}}2e$ & $%
\frac{\sqrt{3}}2e$ & $-\frac{\sqrt{3}}2e$ \\ \hline
\end{tabular}
\label{apsu:su3ch}
\end{equation}

A model which confines only color charges will not confine the diagonal
gluons which have zero charge.

\section*{Acknowledgments}

I wish to thank professor Wolfram Weise for inviting me to write up these
lectures, for constructive criticism and for constant encouragement at times
when it was needed most. I thank Nicolas Wschebor for illuminating
discussions. I also benefited from numerous exchanges with the physicists at
ECT*, in particular from discussions with Pietro Faccioli and Luca Girlanda
and Evgeni Kolomeitsev.

\bibliographystyle{unsrt}
\bibliography{Njl}

\end{document}